\documentclass[onecolumn,longauth]{aa} 
\usepackage[T1]{fontenc}
\usepackage[utf8]{inputenc}
\usepackage{natbib}
\usepackage{dsfont}
\usepackage{multirow}
\usepackage{comment}
\usepackage{graphicx}	
\usepackage{amsmath}	
\usepackage{amssymb}	
\usepackage{xspace}        
\usepackage[dvipsnames]{xcolor}
\usepackage{newtxtext,newtxmath}
\usepackage{booktabs}
\usepackage{fancyhdr}
\usepackage{mathtools}
\usepackage{times}
\usepackage{listings}
\usepackage{dblfnote}
\DFNalwaysdouble
\usepackage{tikz}
\usetikzlibrary{shapes.geometric, arrows}
\usepackage[dvipsnames]{xcolor}
\tikzstyle{input} = [rectangle, rounded corners, minimum width=3cm, minimum height=1cm,text centered, draw=black, align=center, fill=SpringGreen!50]
\tikzstyle{classpy} = [rectangle,  rounded corners, minimum width=3cm, minimum height=1cm, text centered, draw=black, fill=Cerulean!50]

\DeclareMathOperator{\yellm}{\mathit{Y}}
\newcommand\yellmarb[3]{\, _{#1}\!\!\yellm^{#2}_{\!#3}}
\newcommand{\br}[1]{\left( #1 \right)}
\newcommand{\bc}[1]{\left\{ #1 \right\}}
\newcommand{\bb}[1]{\left[ #1 \right]}
\newcommand{\ba}[1]{\left\langle #1 \right\rangle}
\newcommand{\drfive}{{\sc dr5}}

\graphicspath{{/}}

\usepackage[linkbordercolor = green, urlbordercolor = green]{hyperref}
\usepackage{cleveref}

\newcommand{\mysc}[1]{{\sc #1}}

\newcommand{\review}[1]{{\color{black} #1}}
\newcommand{\restructured}[1]{{\color{black} #1}}

\begin{document}

   \title{KiDS-Legacy: Covariance validation and the unified \mysc{OneCovariance} framework for projected large-scale structure observables}

   \author{Robert Reischke\inst{1,2}\thanks{rreischke@astro.uni-bonn.de, reischke@posteo.net}, Sandra Unruh\inst{1}, Marika Asgari\inst{3,7,8}, Andrej Dvornik\inst{2}, Hendrik Hildebrandt\inst{2}, Benjamin Joachimi\inst{4}, Lucas Porth\inst{1}, Maximilian von Wietersheim-Kramsta\inst{4,5,6}, Jan Luca van den Busch\inst{2},  Benjamin St\"{o}lzner\inst{2}, Angus H. Wright\inst{2}, Ziang Yan\inst{2}, Maciej Bilicki\inst{9}, Pierre Burger\inst{10,1}, Nora Elisa Chisari\inst{11,15}, Joachim Harnois-D\'{e}raps\inst{3}, Christos Georgiou\inst{11}, Catherine Heymans\inst{2,12}, Priyanka Jalan\inst{9}, Shahab Joudaki\inst{13,14}, Konrad Kuijken\inst{15}, Shun-Sheng Li\inst{15,16}, Laila Linke\inst{17}, Constance Mahony \inst{18,2}, Davide Sciotti\inst{19,20}, Tilman Tr\"{o}ster\inst{21}, Mijin Yoon\inst{15}
          }
   \institute{
   Argelander-Institut für Astronomie, Universität Bonn, Auf dem Hügel 71, D-53121 Bonn, Germany\\
\email{rreischk@uni-bonn.de, reischke@posteo.net} \and
Ruhr University Bochum, Faculty of Physics and Astronomy, Astronomical Institute (AIRUB), German Centre for Cosmological Lensing, 44780 Bochum, Germany
\and School of Mathematics, Statistics and Physics, Newcastle University, Herschel Building, NE1 7RU, Newcastle-upon-Tyne, UK
\and Department of Physics and Astronomy, University College London, Gower Street, London WC1E 6BT, UK
\and
            Department of Physics, Institute for Computational  Cosmology, Durham University, South Road, Durham DH1 3LE, UK
        \and
             Department of Physics, Centre for Extragalactic Astronomy, Durham University, South Road, Durham DH1 3LE, UK
    \and  E. A. Milne Centre, University of Hull, Cottingham Road, Hull, HU6 7RX, UK
    \and  Centre of Excellence for Data Science, AI, and Modelling (DAIM), University of Hull, Cottingham Road, Hull, HU6 7RX, UK
    \and Center for Theoretical Physics, Polish Academy of Sciences, al. Lotnik\'ow 32/46, 02-668 Warsaw, Poland
    \and Department of Physics and Astronomy, University of Waterloo, 200 University Ave W, Waterloo, ON N2L 3G1, Canada
    \and Institute for Theoretical Physics, Utrecht University, Princetonplein 5, 3584CC Utrecht, The Netherlands
    \and Institute for Astronomy, University of Edinburgh, Blackford Hill, Edinburgh, EH9 3HJ, UK
    \and Centro de Investigaciones Energ\'{e}ticas, Medioambientales y Tecnol\'{o}gicas (CIEMAT), Av. Complutense 40, E-28040 Madrid, Spain
    \and Institute of Cosmology \& Gravitation, Dennis Sciama Building, University of Portsmouth, Portsmouth, PO1 3FX, United Kingdom
    \and Leiden Observatory, Leiden University, P.O. Box 9513, 2300 RA Leiden, the Netherlands,
    \and Aix-Marseille Universit\'{e}, CNRS, CNES, LAM, Marseille, France
    \and Universität Innsbruck, Institut für Astro- und Teilchenphysik, Technikerstr. 25/8, 6020 Innsbruck, Austria
    \and Donostia International Physics Center, Manuel Lardizabal Ibilbidea, 4, 20018 Donostia, Gipuzkoa, Spain
    \and INAF - Osservatorio Astronomico di Roma, via Frascati 33, 00040 Monteporzio Catone (Roma), Italy
    \and INFN - Sezione di Roma, Piazzale Aldo Moro, 2 - c/o Dipartimento di Fisica, Edificio G. Marconi, I-00185 Roma, Italy
    \and Institute for Particle Physics and Astrophysics, ETH Zürich Wolfgang-Pauli-Strasse 27, 8093 Zürich, Switzerland
}
    
   \date{\today}

\abstract{
We introduce {\mysc{OneCovariance}}, an open-source software designed to accurately compute covariance matrices for an arbitrary set of two-point summary statistics across a variety of large-scale structure tracers.  Utilising the halo model, we estimated the statistical properties of matter and biased tracer fields, incorporating all Gaussian, non-Gaussian, and super-sample covariance terms. The flexible configuration permits user-specific parameters, such as the complexity of survey geometry, the halo occupation distribution employed to define each galaxy sample, or the form of the real-space and/or Fourier space statistics to be analysed.

We illustrate the capabilities of \mysc{OneCovariance} within the context of a cosmic shear analysis of the final data release of the Kilo-Degree Survey (KiDS-Legacy). Upon comparing our estimated covariance with measurements from mock data and calculations from independent software, we ascertain that \mysc{OneCovariance} achieves accuracy at the per cent level. When assessing the impact of ignoring complex survey geometry in the cosmic shear covariance computation, we discover misestimations at approximately the $10\%$ level for cosmic variance terms. Nonetheless, these discrepancies do not significantly affect the KiDS-Legacy recovery of cosmological parameters. We derive the cross-covariance between real-space correlation functions, bandpowers, and COSEBIs, facilitating future consistency tests among these three cosmic shear statistics. Additionally, we calculate the covariance matrix of photometric-spectroscopic galaxy clustering measurements, validating the jackknife covariance estimates for calibrating KiDS-Legacy redshift distributions. The \mysc{OneCovariance} can be found on \href{https://github.com/rreischke/OneCovariance}{GitHub} together with  comprehensive documentation and examples.}

  \keywords{gravitational lensing: weak -- methods: statistical -- cosmological parameters -- large-scale structure of Universe.}

  \titlerunning{KiDS-Legacy: OneCovariance}
\authorrunning{R. Reischke et al.}

   \maketitle

\section{Introduction}
The current best practice for testing the cosmological model of the Universe using the large-scale structure (LSS) are two-point functions.  These measurements typically involve the coherent weak lensing distortion of galaxy images and the correlation of galaxy positions. All of them have been successfully measured and used for cosmological inference by the Kilo-Degree Survey 
\citep[KiDS, see e.g.][]{hildebrandt_kids-450_2017,asgari_kids-1000_2021,heymans_kids-1000_2020}\footnote{\href{https://kids.strw.leidenuniv.nl/}{https://kids.strw.leidenuniv.nl/}}, the Dark Energy Survey \citep[DES, see e.g.][]{abbott_dark_2018,amon_dark_2022,abbott_dark_2022},\footnote{\href{https://www.darkenergysurvey.org/}{https://www.darkenergysurvey.org/}} and the Subaru Hyper Suprime-Cam \citep[HSC, see e.g.][]{PhysRevD.105.123537, hamana_cosmological_2020, 2023PhRvD.108l3519D}.\footnote{\href{https://hsc.mtk.nao.ac.jp/ssp/}{https://hsc.mtk.nao.ac.jp/ssp/}}
These experiments have laid the groundwork for upcoming stage 4 surveys such as \textit{Euclid},\footnote{\href{https://www.euclid-ec.org/}{https://www.euclid-ec.org/}} The Legacy Survey of Space and Time (LSST),\footnote{\href{https://www.lsst.org/}{https://www.lsst.org/}} and \textit{Roman}.\footnote{\href{https://roman.gsfc.nasa.gov/}{https://roman.gsfc.nasa.gov/}} Leveraging galaxy positions and shapes enables the measurement of three unique two-point functions, shape-shape (cosmic shear, CS), shape-position (galaxy-galaxy-lensing, GGL), and position-position (galaxy clustering, GC). Aggregating these into a data vector establishes the standard $3\times2$-point analysis.  The main ingredient in this kind of analysis is the likelihood, which is commonly assumed to be Gaussian, so that it can be completely characterised by the data vector, model prediction, and covariance. 

There are various approaches to computing or estimating the covariance matrix. The most direct method involves utilising the data itself, generating pseudo-realisations through subsampling \citep{norberg_statistical_2009,friedrich_performance_2016,mohammad_creating_2022, trusov_two-point_2024}. Alternatively, simulation suites can be used to estimate the covariance if enough realisations are available and feasible \citep{hartlap_why_2007,2012MNRAS.426.1262H,dodelson_effect_2013,2013MNRAS.432.2433H,taylor_putting_2013,taylor_estimating_2014, joachimi_non-linear_2017, sellentin_insufficiency_2017, 2018MNRAS.481.1337H}. Lastly, theoretical modelling is usually the most efficient and arguably the most fundamental way to estimate the covariance matrix \citep{takada_cosmological_2003,eifler_dependence_2009,krause_cosmolike_2017,reischke_variations_2017,joachimi_kids-1000_2021,friedrich_dark_2020}. Each approach has its own advantages and disadvantages. 
Simulations tend to be relatively accurate, with systematic effects such as masking or variable survey depth easily accounted for. However, obtaining an unbiased covariance estimate necessitates numerous realisations, with at least as many simulations required as there are effective degrees of freedom in the data for the covariance matrix to be invertible. Current surveys typically involve around $10^2-10^3$ data points, while upcoming stage 4  surveys will boast $10^4$ data points for standard two-point statistics. Consequently, significant data compression \citep[e.g.][]{heavens_massive_2000,2015A&A...578A..50A,heavens_massive_2017} or running numerous computationally intensive simulations is necessary. 
Conversely, theoretical covariances are comparatively inexpensive to compute, although  modelling the aforementioned systematic effects is difficult and a simulation-based covariance matrix estimation might be more accurate.  While subsampling techniques offer a potential quick way to self-consistently estimate the covariance directly from the data, they tend to be biased on large scales where the subsamples are not independent due to cosmic variance.

Due to the symmetries of the Friedmann-Lema\^{i}tre-Robertson-Walker metric, theoretical computations are typically conducted in harmonic (Fourier) space. Rotational symmetry results in the decoupling of different angular multipoles, $\ell$, and allow  considerable data compression. However, observations of a masked sky compromise the orthogonality of the spherical harmonic basis, leading to mode coupling that necessitates careful modelling, known as pseudo-$C_\ell$ \citep[see e.g.][]{hikage_shear_2011,becker_cosmic_2016,hikage_cosmology_2019,alonso_unified_2019,nicola_cosmic_2020, loureiro_kids_2022, von_wietersheim-kramsta_kids-sbi_2024}. Unlike in harmonic space, standard real space correlation functions do not face these limitations to the first order in signal modelling, they encompass $w(\theta)$, $\gamma_\mathrm{t}(\theta)$, and $\xi_\pm(\theta)$ for GC, GGL, and CS respectively.\footnote{See \Cref{eq:gc_correlation_function}, \Cref{eq:ggl_correlation_function} and \Cref{eq:hankelshear1} for their respective definitions.} However, photometric galaxy surveys introduce a broad radial window function, mixing different physical scales and complicating efforts to disentangle them and mitigate astrophysical effects on small scales. Moreover, correlation functions integrate over the angular power spectra with an oscillatory but broad integration kernel, resulting in limited control over the physical scales influencing the measurements. Leakage from very low multipoles can alter the Gaussianity of the likelihood at large angular scales \citep{sellentin_skewed_2017,2020OJAp....3E..11L, Oehl_exact_2024}, while significant contributions from highly non-linear and theoretically uncertain scales occur at small angular separations. Importantly, the cosmological lensing signal does not produce $B$-modes to first order, as it represents the gradient of the lensing potential. Consequently, the $B$-mode signal serves as a consistency check for systematic effects. However, real space correlation functions mix $E$- and $B$-modes, complicating the interpretation of these tests \citep[see e.g.][]{schneider_ring_2007}.

The aforementioned problems of real space correlation functions are partially alleviated by considering different derived summary statistics. In weak lensing surveys and in particular KiDS, two alternatives to real space correlation functions used are band powers \citep{schneider_analysis_2002} and COSEBIs \citep{schneider_cosebis_2010}. While the former approximately separates $E$ and $B$ modes, the latter yields a complete separation by definition. The demonstration of consistent parameter constraints over a set of summary statistics adds significant confirmation to the robustness of the pipelines used, the systematics involved and the overall fidelity of the cosmological inference. Although many of these methods have been integrated within the community for a $3\times2$-point analysis and are publicly available for model predictions, a unified framework for the covariance matrix in the cosmological community for all kinds of summary statistics has been absent.

While some of the previously mentioned tools for calculating the covariance matrix are publicly available \citep[see e.g.][]{krause_cosmolike_2017,fang_2d-fftlog_2020}, they possess limited flexibility or lack comprehensive functionality when it comes to exchanging tracer fields, summary statistics, weight functions, bias, or halo occupation distribution prescriptions. Furthermore, it is not always straightforward to input files of pre-calculated quantities into these codes.
 Therefore, this paper presents the analytical covariance matrices alongside a \mysc{Python} code, the \mysc{OneCovariance} code, encompassing all functionalities used in the standard analysis of KiDS, ranging from CS \citep{asgari_kids-1000_2021} alone to a full $3\times2$-point analysis \citep{joachimi_kids-1000_2021, heymans_kids-1000_2020}, incorporating the stellar mass function \citep{dvornik_kids-1000_2023}, and employing all three types of summary statistics  (i.e. real space correlation functions, bandpowers, and COSEBIs/$\Psi$-stats\footnote{$\Psi$-stats are the COSEBIs-equivalent for GGL and GC equivalent.})
 used in KiDS for CS, galaxy clustering, and galaxy-galaxy lensing as well as arbitrary cross-correlations between them. 
 Moreover, it remains flexible enough to accommodate very general inputs and therefore can be used by any survey or collaboration to obtain quick and easy covariance matrices for photometric LSS observables.
 
 The goal of this paper is threefold: $(i)$ to furnish a comprehensive and instructive document encompassing all components required for theoretical computations of LSS covariances in photometric galaxy surveys, consolidating everything into a unified code, which we call \mysc{OneCovariance}, accessible to the cosmological community for generating covariance matrices effortlessly for any projected LSS parameter;
 $(ii)$ to introduce the mocks used for the validation of the covariance matrix in the final CS analysis of the Kilo-Degree Survey (KiDS-Legacy) featuring, among other things, a realistic footprint and variable number density; and $(iii)$ to employ the \mysc{OneCovariance} code in different scenarios. Our focus here, in addition to CS in KiDS-Legacy, includes the covariance matrix of clustering redshifts \citep{van_den_busch_testing_2020} and the cross-correlation between different summary statistics \citep{asgari_kids-1000_2021}. It should be highlighted that the \mysc{OneCovariance} code, while showcased for KiDS, is completely general and can be used for all  different types of analyses via simple file inputs to acquire fast covariance estimates. 
 
 Since the CS catalogue of KiDS-Legacy is not final at this stage, we   only performed simulations and comparisons for KiDS-Legacy-like settings. However, we do not anticipate any major changes that would influence the covariance validation. We   refer to data products that are final as of the writing of this paper as KiDS-Legacy, and to simulations or data products with nearly final settings as KiDS-Legacy-like. Furthermore, we generally refer to KiDS-Legacy as the entirety of the final analysis of the last data release of KiDS.

 The manuscript is organised as follows. In \Cref{sec:covariance_matrix} we discuss some general properties of covariance matrices in cosmology, its definition, the different contributions, and the flat sky approximation. In \Cref{sec:correlators_in_fourier_space} we discuss the prescription for non-linear structure formation. Here we   focus on a halo model-based description of structure formation, and we   discuss all its ingredients. The projection along the  line of sight and the corresponding covariance in harmonic space is discussed in \Cref{sec:harmonic_space}. We then discuss the projection to observables in \Cref{sec:harmonic_to_real} and describe the motivation and features of the \mysc{OneCovariance} code in \Cref{sec:basic_code}. The subsequent sections feature applications of the code to specific examples. We introduce a KiDS-Legacy-like CS sample in \Cref{sec:kids_legacy_sec} and describe the general characteristics of the covariance matrix as well as the influence of its modelling choices on parameter inference. Then, in \Cref{sec:sec_mock} we present the KiDS-Legacy-like mocks, incorporating variable depth, and we demonstrate the overall agreement between the analytical and numerical covariances. The results section concludes with \Cref{sec:examples}, which showcases two additional applications of the covariance matrix utilised in KiDS-Legacy: covariances between different summary statistics for consistency tests and the covariance for clustering redshift calibration. To facilitate swift access, we highlight key plots at the outset of each results section. Finally, in \Cref{sec:summary_onecov} and \Cref{sec:summary} we examine the current status and capabilities of the code, along with future directions and our conclusions.

\section{Covariance matrices in photometric galaxy surveys}
\label{sec:covariance_matrix}
In this section, we   outline the general theoretical modelling involved in constructing the covariance matrix. We  separate this discussion from the precise prescription of non-linear structure formation, which is described in more detail in \Cref{sec:correlators_in_fourier_space}. This separation allows us to distinguish observational definitions from the theoretical framework as clearly as possible.

\subsection{General considerations}
\label{sec:general_considerations}
Cosmological analyses of the LSS hinge on estimating a two-point function between any pair of observables (tracers), denoted as $t_i\big[\delta\big]$, which are functionals of the underlying matter field, $\delta(\boldsymbol{x},\chi)$. Here, $\boldsymbol{x}$ represents a spatial co-moving vector (i.e. a 3-dimensional vector in a spatial hyper-surface), and $\chi$ denotes the co-moving distance.
These two-point functions can be aggregated into a vector $\boldsymbol{\mathcal{O}}$ with components
\begin{equation}
\label{eq:observables_general}
     \mathcal{O}_\alpha\coloneqq \left\langle t_i\big[\delta\big] t_j\big[\delta\big]\right\rangle\; \quad \alpha\in\{\text{all unique combinations of $(t_i,t_j)$}\}\,.
\end{equation}
Here, the angular brackets, $\langle\cdot\rangle$, denote an ensemble average, akin to a spatial average \citep[for more details on this ergodicity argument see e.g.][]{desjacques_statistics_2020}. The number of unique combinations depends on the tracer used; if the underlying tracer is the same and $i$ simply labels a subset such as tomographic bins, $i\leq j$, otherwise, all combinations can exist.
The components of the covariance matrix are then defined as
\begin{equation}
\label{eq:covariance_general}
    (\boldsymbol{\tens{C}})_{\alpha\beta} \equiv\mathrm{Cov}(\boldsymbol{\mathcal{O}}_\alpha,\boldsymbol{\mathcal{O}}_\beta)\coloneqq \big\langle (\boldsymbol{\mathcal{O}}_\alpha -\langle\boldsymbol{\mathcal{O}}_\alpha\rangle)
    (\boldsymbol{\mathcal{O}}_\beta -\langle\boldsymbol{\mathcal{O}}_\beta\rangle)\big\rangle\,.
\end{equation}
Typically, we are provided with the statistical properties of the density field $\delta$ from theoretical calculations or numerical simulations, rather than those of its tracer. If the mapping from $\delta$ to $t_i$ is linear, all the statistical properties of $\delta$ extend accordingly to $t_i$. However, this is not always the case, such as in a perturbative bias expansion or the reduced shear beyond linear order.
For simplicity, we assume $\langle t_i[\delta]\rangle = 0$, resulting in the following breakdown of \Cref{eq:covariance_general} using \Cref{eq:observables_general} and Wick's theorem:
\begin{equation}
\label{eq:wichk}
     C_{ijmn} = \langle t_it_m\rangle \langle t_jt_n\rangle + \langle t_it_n\rangle \langle t_jt_m\rangle + \langle t_it_jt_mt_n\rangle_\mathrm{c}\,.
\end{equation}
Here, we have omitted the argument of the different tracers, $[\delta]$, for notational simplicity. The first two terms are known as Gaussian terms, deriving from products of the original observable, while the subscript `c' in the last term denotes the connected part of the correlator, i.e. the term originating from a genuine four-point function and not the product of two-point function. This quantity vanishes for Gaussian fields. For surveys with incomplete sky coverage, the connected part of the covariance can be decomposed into a part arising from modes within the survey, termed the non-Gaussian (nG) component, and a term generated by modes larger than the survey, known as the super-sample covariance \citep[SSC,][]{takada_power_2013}:
\begin{equation}    
\label{eq:connected_split}
 \langle t_it_jt_mt_n\rangle_\mathrm{c} = 
 \langle t_it_jt_mt_n\rangle_\mathrm{nG} + 
 \langle t_it_jt_mt_n\rangle_\mathrm{SSC}\,.
\end{equation}
It should be noted that both these terms are entirely arising from the connected term. Particularly, the SSC term stems from the cosmic variance's reaction to unobservable modes, manifested as the squeezed limit of a connected four-point function \citep{2024A&A...681A..33L}.

Real observables invariably serve as discrete tracers of the underlying smooth field, hence they are subject to stochastic noise due to the finite number of tracers, such as galaxies. Hence, one can schematically express
\begin{equation}
\label{eq:noise_add}
    \hat{t}_i = t_i + n_i\,,
\end{equation}
where $n_i$ denotes some stochastic noise component, this component emerges only when the same object correlates with itself and is thus also referred to as the shot or shape noise component for galaxy clustering or CS, respectively. With this understanding, assuming that the signal does not correlate with the noise, the Gaussian terms, i.e., the first two terms in \Cref{eq:wichk}, each decompose into three components,
\begin{equation}
\label{eq:Gaussian_split}
    \langle \hat t_i\hat t_m\rangle \langle \hat t_j\hat t_n\rangle \;\;\;\sim\;\;\;  \underbrace{\langle t_it_m\rangle \langle t_jt_n\rangle}_{\text{sample variance}}\;\;\; +\;\;\; \underbrace{ \langle t_it_m\rangle \langle n_{j}n_{n}\rangle +  \langle t_jt_n\rangle \langle n_{i}n_{m}\rangle}_{\text{mixed}} \;\;\;+\;\;\; \underbrace{\langle n_{j}n_n\rangle \langle n_{i}n_{m}\rangle }_{\text{shot/shape noise}}\,,
\end{equation}
that are distinguished by their scaling with the signal. We note that the sample variance term is often called cosmic variance in the literature.

\subsection{Angular polyspectra and Limber approximation}
Observables of (photometric) galaxy surveys are projected quantities of three-dimensional fields. Consider such a three-dimensional field $A(\boldsymbol{x} ,f_\mathrm{K}\chi(z)) \in \{t_i\}$ at position $\boldsymbol{x}$ and redshift $z$. Its projected version, $a(\boldsymbol{\hat{n}})$, is 
\begin{equation}
\label{eq:projection}
    a(\boldsymbol{\hat{n}}) = \int\mathrm{d}\chi\;W_A(\chi) A(f_\mathrm{K}(\chi)\boldsymbol{\hat{n}},\chi)\,,
\end{equation}
where $W_A(\chi)$ is some weighting function whose exact form is probe specific and will be discussed in \Cref{sec:harmonic_space}. $\boldsymbol{\hat{n}}$ is the unit vector on the sphere and  $f_\mathrm{K}(\chi)$ is the co-moving angular diameter distance, such that $\boldsymbol{x} = f_\mathrm{K}(\chi)\boldsymbol{\hat{n}}$ and $\chi$ the co-moving radial distance.
In \Cref{app:full_sky_fluctuations} the definitions for a spherical harmonic decomposition of the projected field, $a$, are given. Leveraging those relations, the two-point function defines the angular power spectrum, $\mathcal{P}_{a_1a_2}(\ell_1)$, of the two-dimensional fields $a_1$ and $a_2$ as follows:
\begin{equation}
\begin{split}
\label{eq:powerspectrum_general}
    \big\langle a^{\mathstrut}_{1,\ell_1 m_1} a^{\mathstrut}_{2,\ell_2 m_2}\big\rangle \coloneqq &\; \mathcal{P}_{a_1a_2}(\ell_1) \delta^\mathrm{K}_{\ell_1\ell_2}\delta^\mathrm{K}_{m_1m_2}  \\
    =&\; \frac{2}{\uppi}\int\mathrm{d}{\chi_1} W_{a_1}(\chi_1)\int\mathrm{d}{\chi_2} W_{a_2}(\chi_2)\int k^2\mathrm{d}k\;  P_{A_1A_2}(k, \chi_1,\chi_2)
j_{\ell}(kf_\mathrm{K}(\chi_1))j_{\ell}(kf_\mathrm{K}(\chi_2)) \;\delta^\mathrm{K}_{\ell_1\ell_2}\delta^\mathrm{K}_{m_1m_2}\,.
\end{split}
\end{equation}
Here the dependence on $\chi_1$ and $\chi_2$ of the three-dimensional power spectrum, $P_{A_1A_2}(k, \chi_1,\chi_2)$, was made explicit. $\delta^\mathrm{K}_{ij}$ is the Kronecker delta. The more general equation for higher-order correlators is provided in \Cref{app:higher_order_correlators_sphere}. Throughout this paper the geometric mean is used to approximate the unequal time correlator such that for the power spectrum
\begin{equation}
\label{eq:geometric_mean}
    P_{A_1A_2}(k, \chi_1,\chi_2) \approx \left[P_{A_1A_2}(k, \chi_1)P_{A_1A_2}(k, \chi_2)\right]^{1/2}\,,
\end{equation}
which is an excellent approximation for almost all practical purposes \citep{castro_weak_2005,kitching_unequal-time_2017,kilbinger_precision_2017,2019PhRvD.100b3543C,de_la_bella_unequal-time_2020}. 

In the flat sky limit (in which case the transform to spherical harmonics essentially becomes a two-dimensional Fourier transform, see \Cref{sec:flat_sky}) angular polyspectra can be approximated using the Limber approximation \citep{limber_analysis_1954,loverde_extended_2008,leonard_n5k_2022}:
\begin{equation}
\label{eq:limber}
    \langle a_{\ell_1 m_1} ... a_{\ell_n m_n} \rangle \sim \int \frac{\mathrm{d}\chi}{f_\mathrm{K}(\chi)^{2n-2}} W_{A_1}(\chi)\dots W_{A_n}(\chi) P_{A_1\dots A_n} \left(\frac{{\ell}_1+0.5}{f_\mathrm{K}(\chi)}, \dots, \frac{{\ell}_n + 0.5}{f_\mathrm{K}(\chi)}, \chi\right)\,.
\end{equation}
Here geometrical factors occurring in \Cref{eq:angular_n_point_function} are disregarded due to the flat sky approximation. For $n>3$ this expression assumes that the angular average over possible configurations has already been carried out on the flat sky.
The use of the Limber approximation introduces negligible biases in cosmological parameter inference when considering CS only \citep[ss e.g.][]{joachimi_kids-1000_2021, leonard_n5k_2022}. Therefore,  unless stated otherwise, we utilise \Cref{{eq:limber}} for projected fields.

\section{Non-linear evolution: Halo model}
\label{sec:correlators_in_fourier_space}
Due to the non-linearity of gravity and the complex interaction between cold dark matter and baryons, understanding structure formation in cosmology presents a formidable challenge. Various approaches have been developed to tackle this issue. These include standard perturbation theory \citep[SPT, for an exhaustive review see][]{bernardeau_large-scale_2002}, halo model approaches \citep[HM, see e.g.][]{seljak_analytic_2000,cooray_halo_2002, asgari_halo_2023},
effective field theory \citep[EFT,][]{carrasco_effective_2012}, or kinetic field theory \citep[KFT, e.g.][]{bartelmann_kinetic_2021}. 

In addition to modelling the non-linear evolution of the smooth dark matter density field, there are additional complications: $(i)$ The total matter power spectrum encapsulates the clustering of both cold dark matter and baryons. Processes such as star formation, supernovae, active galactic nuclei (AGN), and cooling can displace baryons relative to dark matter, leading to a spatial distribution of total matter that differs from that of cold dark matter alone. This phenomenon is collectively known as baryonic feedback \citep[see e.g.][]{2013MNRAS.434..148S, somerville_physical_2015,vogelsberger_cosmological_2020,chisari_modelling_2019,nicola_breaking_2022}. $(ii)$ The tracers discussed in Section \ref{sec:covariance_matrix} are biased representations of the underlying matter distribution. This bias is most prominent in galaxy surveys \citep[referred to as galaxy bias, see e.g.][for a review]{desjacques_large-scale_2018}, but also affects CS measurements through intrinsic alignment (IA) effects \citep[see e.g.][for reviews]{schaefer_review:_2009, joachimi_galaxy_2015, vlah_eft_2019,fortuna_halo_2020-1}, which can be treated similarly to galaxy bias using an effective field theory (EFT) approach.

While perturbative approaches such as EFT are technically more rigorous than non-perturbative (but phenomenological) halo model calculations, they do not allow progressing into the highly non-linear regime, which is crucial for signals with low signal-to-noise ratios such as CS. In other words, the significance of the CS measurement is not large enough to fit all the EFT nuisance parameters and all cosmological information would be lost. Hence, incorporating prior knowledge regarding the statistical properties of the matter distribution via a halo model is key if one wants to access highly non-linear scales. Although KFT offers, in theory, an alternative route, it still has to be shown to provide good fits for models including baryonic feedback and higher-order statistics necessary for the covariance. Since the halo model has been shown to be flexible enough to describe different cosmological observations into the non-linear regime\footnote{It should be noted that the vanilla halo model requires some more attention to accurately fit the transition region between the one-halo and two-halo term. Furthermore, for higher-order statistics these tweaks become more involved (see \citet{2020ApJ...895..113T} for the bispectrum). However, for the trispectrum contribution to the covariance, the accuracy of the halo model is, at least currently, sufficient.} \citep[see][and references therein]{mead_hydrodynamical_2020}, we   stick to a halo model approach in this paper, following the methodology of KiDS-1000 \citep{joachimi_kids-1000_2021}. However, it is worth noting that the code presented here is sufficiently flexible to accommodate any external power spectra, power spectrum response, and/or tri-spectra for subsequent calculations. The halo model assumes that the cosmic density field is entirely composed of dark matter halos, whose distribution is governed by a mass function. By assuming a density profile and a halo bias, we can calculate the statistical properties of the matter field. Populating each halo with galaxies using a halo occupation distribution (HOD) enables a versatile model that accurately predicts the statistical properties of CS, GC, GGL for the cross-correlation between the two galaxy samples.

\subsection{Halo model ingredients: Galaxy bias and halo occupation distribution}
\label{sec:halo_mod_ingredients}
Here, we briefly summarise the components involved in the halo model calculations of matter and galaxy polyspectra. Galaxy bias is addressed through an HOD prescription \citep{van_den_bosch_cosmological_2013,dvornik_unveiling_2018,lacasa_non-gaussianity_2014,lacasa_covariance_2018,reischke_information_2020,dvornik_kids-1000_2023}.
To achieve this, one employs a biasing function, $b(M)$,  depending on halo mass $M$, defined via the average number density of galaxies $\langle N|M\rangle$ in a halo of mass $M$
\begin{equation}
    \label{eq:bias_m}
    b(M) \coloneqq \frac{\bar\rho_\mathrm{m}(z)}{\bar n_\mathrm{g}(z)}\frac{\langle N|M\rangle}{M}\,,
\end{equation}
where $\bar n_\mathrm{g}$ is the average number density of galaxies and $\bar\rho_\mathrm{m}(z)$ is the mean matter density in the Universe at redshift $z$. Average quantities in the occupation distribution are defined as
\begin{equation}
\label{eq:occ_dist}
    \langle N|M\rangle \coloneqq \sum_{N=0}^{\infty}NP(N|M)\,,
\end{equation}
where $P(N|M)$ is the probability of a halo to be occupied with $N$ galaxies. Moments of the bias at a redshift $z$, or any halo-related quantity, $A(M,z)$, are calculated via the differential number density of halos of mass $M$: the halo mass function $n_\mathrm{h}(M,z)$
\begin{equation}
\label{eq:averages_hm}
    \langle A \rangle(z) \coloneqq \int\mathrm{d}M\;n_\mathrm{h}(M,z) A(M,z)\,.
\end{equation}
The halo model assumes that all matter in the Universe is bound in halos. We assume those halos to be spherically symmetric and to have an average density profile $\rho_\mathrm{h}(r|M) = M u_\mathrm{h}(r|M)$, such that $u_\mathrm{h}$ is normalised and can be used in \Cref{eq:averages_hm} without introducing additional normalisation factors. The galaxy populations can be split   into central, `c', and satellite, `s', galaxies with different halo profiles $u_{\mathrm{c/s}}$ and halo occupations $\langle N_\mathrm{c/s}|M\rangle$. If there is a central galaxy in a halo on average, $\langle N_\mathrm{c}|M\rangle = 1$ the distribution of additional galaxies, i.e. the satellites, is assumed to be Poisonnian. This allows us to calculate the expected number of $n$-tuples, $\langle N^{(n)}|M\rangle$ in a halo, as required for the $1$, ..., $4$-halo terms
\begin{equation}
\label{eq:tuples}
\begin{split}
   &  \langle N |M\rangle  = \;  \langle N_\mathrm{c} |M\rangle  +  \langle N_\mathrm{s} |M\rangle   \quad\quad\quad\quad\;\;\;\quad\quad\quad\quad\quad\quad\quad\;\; n=1\,,  \\
   &  \langle N(N-1) |M\rangle  =  \;\langle N_\mathrm{c} |M\rangle\left(\langle N_\mathrm{s} |M\rangle^2 + 2\langle N_\mathrm{s} |M\rangle\right)\quad\quad\quad\quad n = 2 \,,\\
     &\langle N(N-1)(N-2) |M\rangle  =  \;\langle N_\mathrm{c} |M\rangle\left(\langle N_\mathrm{s} |M\rangle^3 + 3\langle N_\mathrm{s} |M\rangle\right)\quad n = 3\,, \\
     &\dots\,, 
     \end{split}
\end{equation}
where the number of tuples for each individual type e.g. `s' and `c' can be read off. In the case of matter, `m', the HOD is set to the halo mass $M$ for each point and the normalisation is taken care of by $\bar\rho_\mathrm{m}$, i.e. $\langle N_\mathrm{m}|M\rangle = M$. It should be noted that sub-Poissonian behaviour was found in \citet{dvornik_kids-1000_2023} for the satellite population. This could, in principle, be accounted for by including an additional free parameter on the two-point level. For the connected non-Gaussian term, this might lead to more complications. However, we do not expect that the latter is dominating and would change cosmological results significantly. 
It is instructive to introduce the following integrals \citep{cooray_power_2001,takada_power_2013,lacasa_covariance_2018} for the species $X_1,\dots,X_\mu$:
\begin{equation}
\label{eq:halomod_integral}
    I^\alpha_{\mu,X_{\{\mu\}}}(k_1,\dots,k_\mu,z) \coloneqq\frac{1}{\bar n_{X_1}(z)\dots \bar n_{X_\mu}(z)} \int\mathrm{d}M\; n_\mathrm{h}(M,z) \langle N^{(\mu)}_{\{X\}}|M\rangle b^\alpha_\mathrm{h}(M,z) \tilde u_{X_1}(k_1|M,z)\dots \tilde u_{X_\mu}(k_\mu|M,z) \,.
\end{equation}
Here $b^\alpha_\mathrm{h}$ is the halo bias, for which we use the fitting function from \citet{tinker_what_2010}. $\Tilde{u}(k|M)$ is the Fourier transform of the normalised halo profile. \restructured{In \Cref{app:halo_profiles} we provide $\Tilde{u}(k|M)$ for an NFW profile. However, they are in general arbitrary.}
Lastly, the average number density $n_X(z)$ can be calculated using \Cref{eq:averages_hm}:
\begin{equation}
    \bar n_X (z) = \int\mathrm{d}M\; n_\mathrm{h}(M,z) \langle N_X|M\rangle\,.
\end{equation}
We note again that $\bar n_\mathrm{m}(z) = \bar\rho_\mathrm{m}(z)$ with this definition. 

\subsection{Conditional stellar mass function}
\label{sec:csmf}
It remains to specify the occupation statistics of a galaxy with halo mass $M$. As the fiducial model, we chose the conditional stellar mass function \citep[CSMF,][]{yang_galaxy_2009,cacciato_galaxy_2009,cacciato_cosmological_2013, van_uitert_stellar--halo_2016, dvornik_unveiling_2018,dvornik_kids-1000_2023}, specifying the average distribution of galaxies of stellar mass $M_\star$ given they occupy a halo with mass $M$. It is also split into centrals and satellites so that the total CSMF is
\begin{equation}
    \Phi(M_\star|M) =  \Phi_\mathrm{c}(M_\star|M) +  \Phi_\mathrm{s}(M_\star|M)\,.
\end{equation}
The two contributions are modelled as a log-normal and a modified Schechter function
\begin{equation}
\begin{split}
    \Phi_\mathrm{c}(M_\star|M) = &\ \frac{1}{\sqrt{2\uppi}\ln(10)\sigma_\mathrm{c}M_\star}\exp\left[-\frac{\log(M_\star/M^*_\mathrm{c})^2}{2\sigma^2_\mathrm{c}}\right]\,,\\
    \Phi_\mathrm{s}(M_\star|M) = &\ \frac{\phi^*_\mathrm{s}}{M^*_\mathrm{s}}\left(\frac{M_\star}{M^*_\mathrm{s}}\right)^{\alpha_\mathrm{s}}\exp\left[-\left(\frac{M_\star}{M^*_\mathrm{s}}\right)^2\right]\,,
\end{split}
\end{equation}   
for centrals and satellites respectively.\footnote{We refer to the natural logarithm as $\ln(x)$ and to the logarithm to base ten as $\log(x)$.} Here $\sigma_\mathrm{c}$ describes the scatter of stellar mass given a halo mass while $\alpha_\mathrm{s}$ describes the power law scaling of the satellite abundance. It should be noticed that if $\alpha_\mathrm{s} <0 $ the expression for $\Phi_\mathrm{s}(M_\star|M)$ would formally diverge for low masses. However, the existence of satellites in the HOD are always conditioned on the existence of a central galaxy, thus ensuring convergence at low masses due to the exponential cut-off of $\Phi_\mathrm{c}(M_\star|M)$.
All free parameters in this model can, in principle, be arbitrary functions of the halo mass, $M$. For this fiducial case, we follow \citet{yang_galaxy_2009,dvornik_unveiling_2018,dvornik_kids-1000_2023} and use the following parametrisation:
\begin{equation}
\begin{split}
    &M^*_\mathrm{c} =  \ M_0 \frac{(M/M_1)^{\gamma_1}}{(1+ [M/M_1])^{\gamma_1-\gamma_2}}\,,\\
    &M^*_\mathrm{s} =  \ 0.56M^*_\mathrm{c}(M)\,,\\
    &\log\phi^*_\mathrm{s} =  \ b_0 +b_1\log\left(\frac{M}{10^{13}M_\odot h^{-1}}\right)\,.
\end{split}
\end{equation}
Integrating the CSMF over the HMF yields the galaxy SMF:
\begin{equation}
\label{eq:smf}
    \Phi_X(M_\star) = \int\mathrm{d}M\; \Phi_X(M_\star|M)n_\mathrm{h}(M,z)\,.
\end{equation}
Likewise, the occupation number in a stellar mass bin $[M_{\star,1},M_{\star,2}]$ is given by
\begin{equation}
    \langle N_X|M\rangle = \int_{M_{\star,1}}^{M_{\star,2}}\mathrm{d}M_\star\Phi_X(M_\star|M)\,,
\end{equation}
completing the HOD prescription. It should be noted that the covariance code presented here is modular and can support other HOD prescriptions.

\subsection{Halo model polyspectra}
\label{sec:polysectra}
With the ingredients defined in the previous section, one can calculate the power spectrum between species $A_1$ and $A_2$, ignoring non-linear bias terms \citep{2021MNRAS.503.3095M}, as
\begin{equation}
\label{eq:powerspectrum_halomodel}
    P_{A_1A_2}(k,z) = P^{\mathrm{1h}}_{A_1A_2}(k,z) + P^{\mathrm{2h}}_{A_1A_2}(k,z) \equiv I^0_{2, A_1,A_2}(k,k,z) + I^1_{1,A_1}(k,z)I^1_{1,A_2}(k,z) P_\mathrm{lin}(k,z)\,,
\end{equation}
with the 1- and 2-halo term as well as the linear matter power spectrum $P_\mathrm{lin}$. The one-halo term will tend to be a constant, as $k\to 0$ due to the infinite support of the halo profile. Due to this, it will overcome the two-halo term on large scales. To remove this unphysical behaviour, we introduce a large-scale damping for the 1h term in \Cref{eq:powerspectrum_halomodel}
\begin{equation}
   P^{\mathrm{1h}}_{A_1A_2}(k,z)\;\to\;P^{\mathrm{1h}}_{A_1A_2}(k,z) \;\mathrm{erf}(k/k_\mathrm{damp}) \,,
\end{equation}
with a fiducial damping scale of $k_\mathrm{damp} = 0.1\;h/\mathrm{Mpc}$.
Likewise, the halo model trispectrum is split  into 1-4-halo terms
\begin{equation}
\begin{split}
    T_{X_1\dots A_4}(\boldsymbol{k}_1,\boldsymbol{k}_2,\boldsymbol{k}_3,\boldsymbol{k}_4,z) = &\; T^{\mathrm{1h}}_{A_1\dots A_4} +  T^{\mathrm{2h}}_{A_1\dots A_4} + T^{\mathrm{3h}}_{A_1\dots A_4} +  T^{\mathrm{4h}} _{A_1\dots A_4}\;,
\end{split}
\end{equation}
where it is understood that the right-hand side depends as well on $(\boldsymbol{k}_1,\boldsymbol{k}_2,\boldsymbol{k}_3,\boldsymbol{k}_4,z)$. The individual terms are given by
\begin{align}
        T^{\mathrm{1h}}_{A_1\dots A_4} = & \;I^0_{4,A_1\dots A_4}(k_1,k_2,k_3,k_4)\,,\\
        T^{\mathrm{2h}}_{A_1\dots A_4} = & P_\mathrm{lin}(k_1) I^1_{3,A_2A_3A_4} (k_2,k_3,k_4)I^1_{1,A_1}(k_1) + 3\;\mathrm{perm.} + P_\mathrm{lin}(k_{12}) I^1_{2, A_1A_2}(k_1,k_2) I^1_{2, A_3A_4}(k_3,k_4) +2\; \mathrm{perm.}\,,\\
        T^{\mathrm{3h}}_{A_1\dots A_4} =&\; B_{\mathrm{tree}}(\boldsymbol{k}_1,\boldsymbol{k}_2,\boldsymbol{k}_{34}) I^1_{2,A_3A_4} (k_3,k_4)I^1_{1,A_1}(k_1)I^1_{1,A_2}(k_2) + P_\mathrm{lin}(k_1) P_\mathrm{lin}(k_2)\\  & + I^1_{2, A_3A_4}(k_3,k_4)I^1_{1,A_1}(k_1)I^1_{1,A_2}(k_2) \nonumber +  5\;\mathrm{perm.}\,,\\
        T^{\mathrm{4h}}_{A_1\dots A_4}  = &\; \prod_\mu I^1_{1,A_\mu}(k_\mu)T_{\mathrm{tree}}(\boldsymbol{k}_1,\boldsymbol{k}_2,\boldsymbol{k}_3,\boldsymbol{k}_4)\,.
\end{align}
We neglect the explicit dependence on redshift here and refer to \citet{lacasa_covariance_2018} for an in-depth discussion.
The perturbation theory bispectrum and trispectrum is given by \citep{fry_galaxy_1984}
\begin{align}
    B_{\mathrm{tree}}(\boldsymbol{k}_1,\boldsymbol{k}_2,\boldsymbol{k}_{3}) = & \;2 F^\mathrm{s}_2(\boldsymbol{k}_1,\boldsymbol{k}_2)P_\mathrm{lin}(k_1)P_\mathrm{lin}(k_2) + 2\;\mathrm{perm.}\,,\\
   T_{\mathrm{tree}}(\boldsymbol{k}_1,\boldsymbol{k}_2,\boldsymbol{k}_3,\boldsymbol{k}_4) = & \; 4\left[F^\mathrm{s}_2(\boldsymbol{k}_{12},-\boldsymbol{k}_1)F^\mathrm{s}_2(\boldsymbol{k}_{12},\boldsymbol{k}_3)P_\mathrm{lin}(k_1)P_\mathrm{lin}(k_{12})P_\mathrm{lin}(k_3) + 12\;\mathrm{perm.}\right]\\ \nonumber & +\; 6\left[P_\mathrm{lin}(k_1)P_\mathrm{lin}(k_2)P_\mathrm{lin}(k_3)F^\mathrm{s}_3(\boldsymbol{k}_{1},\boldsymbol{k}_2,\boldsymbol{k}_3)+ 4\;\mathrm{perm.}\right]\,,
\end{align}
where we omitted the explicit dependence on redshift. For the symmetric perturbation theory kernels, $F^\mathrm{s}_i$, we refer to \citet{bernardeau_large-scale_2002}. The trispectrum relates to the nG term in \Cref{eq:connected_split} for $\boldsymbol{k}_1 = -\boldsymbol{k}_2$ and $\boldsymbol{k}_2 = -\boldsymbol{k}_4$.
Lastly, we specify the SSC \citep{takada_power_2013,krause_cosmolike_2017,barreira_complete_2018}, i.e. the second term in \Cref{eq:connected_split}:
\begin{equation}
    T^\mathrm{SSC}_{A_1\dots A_4}(k_1,k_2,z) = \frac{\partial P_{A_1A_2}}{\partial\delta_\mathrm{bg}}(k_1,z) \frac{\partial P_{A_3A_4}}{\partial\delta_\mathrm{bg}}(k_2,z) \sigma^2_{\mathrm{bg},(A_1A_2)(A_3A_4)}(z)\,.
\end{equation}
We note that $k_1$ corresponds to the correlated pair $A_1,A_2$ and $k_2$ to $A_3,A_4$. The responses are given as 
\begin{equation}
\label{eq:responses}
\begin{split}
    \frac{\partial P_{A_1A_2}}{\partial\delta_\mathrm{bg}}(k,z) = &\; \left[\frac{68}{21} - \frac{1}{3}\frac{\mathrm{d}\log k^3 P_{A_1A_2}(k,z)}{\mathrm{d}\log k} \right] I^1_{1,A_1}(k,z)I^1_{1,A_2}(k,z)P_\mathrm{lin}(k,z) \\ &  + I^1_{2,A_1A_2}(k,k,z) - \left([b_{A_1} + b_{A_2}] P_{A_1A_2}(k,z)\right)_{\mathrm{if\;} {A_1, A_2\neq\mathrm{m}}}\,.    
\end{split}
\end{equation}
We note that in the last term the contribution of $b_{A_i}$ vanishes if the mean of $A_i$ is not constructed in the survey window, such as for CS. The responses are defined with respect to fluctuations in the background, `bg', matter field, which can be calculated directly from the survey mask and its spherical harmonic decomposition:
\begin{equation}
\label{eq:sigma_bg}
    \sigma^2_{\mathrm{bg},(A_1A_2)(A_3A_4)}(z) = \frac{1}{A_{(A_1A_2)}A_{(A_3A_4)}} \sum_\ell P_\mathrm{lin}(\ell/\chi)\sum_m a^{*(A_1A_2)}_{\ell m}a^{(A_3A_4)}_{\ell m}\,.
\end{equation}
The bracket notation $(A_1A_2)$ denotes the footprint of the survey over which the summary statistic between the tracers $A_1$ and $A_2$ has been evaluated. As expected $\sigma^2_\mathrm{bg}\to 0$ as $A_{(A_1A_2)}\to\infty$ due to statistical homogeneity and isotropy. If the survey footprint is not available via a mask file, we   assume a circular footprint and follow \citet{li_super-sample_2014}
\begin{equation}
    \sigma^2_{\mathrm{bg},(A_1A_2)(A_3A_4)}(z) = \chi(z)^2\int\frac{k\mathrm{d}k}{(2\uppi)^2} P_\mathrm{lin}(k,z)\left[\frac{2{\rm J}_1(k\chi(z)\theta_\mathrm{s})}{k\chi(z)\theta_\mathrm{s}}\right]^2\,,
\end{equation}
with a cylindrical Bessel function, ${\rm J}_1$, and the survey size $\theta_\mathrm{s} = \sqrt{\mathrm{max}(A_{(A_1A_2)}A_{(A_3A_4)})/\uppi}$, such that the larger survey area is used.

\section{Harmonic space covariance}
\label{sec:harmonic_space}
 The components outlined in \Cref{sec:correlators_in_fourier_space} can now be translated to harmonic space through the Limber projection, as given in \Cref{eq:limber}. To achieve this, the line-of-sight weights, $W_{a_i}(\chi)$, still have to be specified.  The \mysc{OneCovariance} code typically employs two types of tracers, $a_i$, which can represent either the CS measured from a source galaxy sample, denoted $\mathrm{m}_i$ or the positions of a lens galaxy sample $\mathrm{g}_i$, respectively. In the literature, different terminologies exist. With a `source' we refer to a galaxy whose lensed ellipticity will be used. Conversely, a `lens' is a galaxy whose position will be used. To summarise, GC would be the `lens'-`lens' auto-correlation, CS the `source'-`source' auto-correlation and GGL the `source'-`lens' cross-correlation. 
 The index $i$ labels the tomographic bin of the corresponding sample. With this, the weight functions assume the following form
\begin{align}
\label{eq:weights_specific_g}
    W_{\mathrm{g}_i}(\chi) = &\; n^{(i)}_\mathrm{l}(\chi)\,, \\
    \label{eq:weights_specific_m}
    W_{\mathrm{m}_i}(\chi)  = &  \frac{3f_\mathrm{K}(\chi)\Omega_\mathrm{m}}{2\chi^2_\mathrm{H}a}\int_\chi^\mathrm{\chi_\mathrm{H}}\mathrm{d}\chi^\prime\;\frac{f_\mathrm{K}(\chi^\prime - \chi)}{f_\mathrm{K}(\chi^\prime)}n^{(i)}_\mathrm{s}(\chi) - A_\mathrm{IA}\left[\frac{1+z(\chi)}{1+z_\mathrm{pivot}}\right]^{\eta_\mathrm{IA}}\frac{C_1\rho_\mathrm{cr}\Omega_\mathrm{m}}{D_+[a(\chi)]}n^{(i)}_\mathrm{s}(\chi) \,.
\end{align}
Here $\chi_\mathrm{H}$ is the co-moving distance to the horizon, $n^{(i)}_\mathrm{l}(\chi)$ and $n^{(i)}_\mathrm{s}(\chi)$ are the normalised redshift distributions of the lens- and source sample respectively. $D_+(a)$ is the linear growth factor and $C_1\rho_\mathrm{cr} \approx 0.0134$. The second term in the $W_{\mathrm{m}_i}$ weight function corresponds to the non-linear linear alignment model \citep[NLA,][]{hirata_intrinsic_2004,2007NJPh....9..444B} with alignment amplitude $A_\mathrm{IA}$ and redshift dependence $\eta_\mathrm{IA}$ at a pivot redshift $z_\mathrm{pivot}$. There are two remarks in order here: $(i)$ The NLA model is the fiducial choice of the \mysc{OneCovariance} code. However, the user can include any alignment model (at least in the dominating Gaussian part) by providing the corresponding $C_\ell$ as input. Therefore, including, for example, the Tidal Alignment-Tidal Torquing \citep[TATT,][]{2019PhRvD.100j3506B} is straightforwardly done (see \Cref{sec:basic_code}). $(ii)$ For the fiducial NLA model, any lensing tracer automatically assumes a linear response to the non-linearly evolved density field. Consequently, the alignment is also modelled in the SSC and NG terms of the covariance matrix and not just in the Gaussian part.

In addition to the cosmological signal ($a^i_{\ell m}$), observed modes ($\hat{a}^{i}_{\ell m}$) contain a noise realisation ($n^i_{\ell m}$) since the continuous field is sampled by a finite number of tracers (galaxies),
\begin{equation}
    \hat{a}_{i,\ell m} = a_{i,\ell m} + n_{i,\ell m}\,,
\end{equation}
with $\langle a_{i,\ell m} n_{j,\ell^\prime m^\prime}\rangle = 0$ and noise statistic $\langle n_{i,\ell m} n_{j,\ell^\prime m^\prime}\rangle = \mathcal{N}^{\mathstrut}_{i} \delta^\mathrm{K}_{ij}$, as the noise in different maps is uncorrelated, and the noise is scale independent. Strictly speaking, this relation only holds for full sky coverage where all the modes are independent. Therefore, the noise component will be treated in real space directly to avoid this complication. We come back to this issue in \Cref{sec:harmonic_to_real}.
The factor $\mathcal{N}_{i}$ determines the overall noise level depending on the observable under consideration:
\begin{equation}
\label{eq:noise_levels}
    \mathcal{N}_{i} = \left\{
    \begin{array}{cl}
         \frac{\sigma_{\epsilon_1,\;i}^2}{\bar n_i}&  \text{if $i\in$  source}\\
         \frac{1}{\bar n_i}& \text{if $i\in$  lens}\,\\ 
    \end{array}.\right.
\end{equation}
Here $\bar n_i$ is the average number density of the tracers and $\sigma^2_{\epsilon_1}$ is the single component ellipticity dispersion of the sources.
The standard idealised angular power spectrum estimator at each multipole is defined as 
\begin{equation}
\label{eq:cell  _estimator}
    \hat{C}_{a_1a_2} (\ell) \coloneqq \frac{1}{(2\ell + 1)}\sum_{m = -\ell}^{\ell} \hat{a}^{1*}_{\ell m}\hat{a}^{2}_{\ell m}\,,
\end{equation}
such that
\begin{equation}
\label{eq:noise_free_noise_cell}
    \langle \hat{C}_{a_1a_2} (\ell)\rangle = \mathcal{P}_{a_1a_2}({\ell})  + \mathcal{N}_{a_1}\delta^\mathrm{K}_{a_1a_2} \equiv C_{a_1a_2}(\ell)\;.
\end{equation}
In the flat sky approximation, the same estimator assumes the  form
\begin{equation}
    \hat{C}_{a_1a_2} (\ell) \coloneqq \frac{1}{A_\mathrm{s}N_\mathrm{modes}(\ell)} \sum_{\boldsymbol{\ell}\in \ell_\mathrm{ring}} \hat{a}^1_{\boldsymbol{\ell}}\hat{a}^2_{\boldsymbol{-\ell}}\,,
\end{equation}
where we defined an annulus in Fourier space with volume $2\uppi \ell\Delta\ell$, so that the number of available modes over the survey area $A_\mathrm{s}$ is
\begin{equation}
    N_\mathrm{modes}(\ell) = \frac{2\uppi \ell\Delta\ell}{(2\uppi)^2/A_\mathrm{s}} = 2 \ell\Delta\ell f_\mathrm{sky}\,,
\end{equation}
where $\Delta\ell \ll \ell$ was assumed.
The covariance matrix between two estimators is then, again by noting that $\ell_1$ corresponds to the pair $a_1,a_2$ and $\ell_2$ corresponds to the pair $a_3,a_4$:
\begin{equation}
\label{eq:covariance_splitting_flat}
\begin{split}
    \mathrm{Cov}[\hat{C}_{a_1a_2}(\ell_1)\hat{C}_{a_3a_4}(\ell_2)] = & \;\frac{1}{A_{(12)}N_\mathrm{modes}(\ell_1)}\frac{1}{A_{(34)}N_\mathrm{modes}(\ell_2)}\sum_{\tilde{\boldsymbol{\ell}}_1\in\ell_{1,\mathrm{ring}}}\sum_{\tilde{\boldsymbol{\ell}}_2\in\ell_{2,\mathrm{ring}}}\left( \left\langle \hat{a}^1_{\boldsymbol{\ell}_1}\hat{a}^2_{-\boldsymbol{\ell}_1} \hat{a}^3_{\boldsymbol{\ell}_2}\hat{a}^4_{-\boldsymbol{\ell}_2}\right\rangle
    - \left\langle\hat{a}^1_{\boldsymbol{\ell}_1}\hat{a}^2_{-\boldsymbol{\ell}_1}\right\rangle\langle\hat{a}^3_{\boldsymbol{\ell}_2}\hat{a}^4_{-\boldsymbol{\ell}_2}\rangle\right)\\
    = & \;\frac{1}{N_\mathrm{modes}(\ell_1)}\frac{1}{N_\mathrm{modes}(\ell_2)}\sum_{\tilde{\boldsymbol{\ell}}_1\in\ell_{1,\mathrm{ring}}}\left( {C}_{a_1a_3}(\ell_1){C}_{a_2a_4}(\ell_2) + {C}_{a_1a_4}(\ell_1){C}_{a_2a_3}(\ell_2)\right) \delta^\mathrm{K}_{\ell_1\ell_2} \\
    & + \;\frac{1}{\mathrm{max}(A_{(12)}A_{(34)})N_\mathrm{modes}(\ell_1)N_\mathrm{modes}(\ell_2)}\sum_{\tilde{\boldsymbol{\ell}}_1\in\ell_{1,\mathrm{ring}}}\sum_{\tilde{\boldsymbol{\ell}}_2\in\ell_{2,\mathrm{ring}}}T^{(a_1a_2)(a_3a_4)}(\boldsymbol{\ell}_1, -\boldsymbol{\ell}_1,\boldsymbol{\ell}_2,-\boldsymbol{\ell}_2)\,.
\end{split}
\end{equation}
Here $A_{ij,\mathrm{s}}$ is the survey area over which the angular power spectrum of the two fields $a_i$ and $a_j$ is estimated. Furthermore, $T^{(a_1a_2)(a_3a_4)}(\boldsymbol{\ell}_1, -\boldsymbol{\ell}_1,\boldsymbol{\ell}_2,-\boldsymbol{\ell}_2)$ is the trispectrum, see \Cref{eq:trispectrum_proj} below. We note that the area cancels out in the Gaussian term due to the application of two correlators
(compare Equation~\ref{eq:flat_sky_power}). 

Assuming that the angular power spectra do not vary significantly over the bandwidth of the ring $\ell_\mathrm{ring}$, the Gaussian term in the flat sky approximation is given by the commonly used expression 
\begin{equation}
\label{eq:Gaussian_flat_sky}
    \mathrm{Cov}_\mathrm{G}[\hat{C}_{a_1a_2}(\ell_1)\hat{C}_{a_3a_4}(\ell_2)]  \approx \frac{4\uppi\delta^{\mathrm{K}}_{\ell_1\ell_2}}{\mathrm{max}(A_{(12)}A_{(34)})2\ell_1\Delta\ell_1} \left[ {C}_{a_1a_3}(\ell_1){C}_{a_2a_4}(\ell_2) + {C}_{a_1a_4}(\ell_1){C}_{a_2a_3}(\ell_2)\right]\,.
\end{equation}
This can be compared to the full sky version of the Gaussian term which, at each multipole, is given by
\begin{equation}
\label{eq:Gaussian_covariance}
    \mathrm{Cov}_\mathrm{G}[\hat{C}_{a_1a_2}(\ell_1)\hat{C}_{a_3a_4}(\ell_2)] = \frac{4\uppi\delta^{\mathrm{K}}_{\ell_1\ell_2}}{\mathrm{max}(A_{(12)}A_{(34)})(2\ell_1+1)}\left[{C}_{a_1a_3}(\ell_1){C}_{a_2a_4}(\ell_2) + {C}_{a_1a_4}(\ell_1){C}_{a_2a_3}(\ell_2)\right]\,,
\end{equation}
where we introduced the sky fraction $\mathrm{max}(A_{(12)}A_{(34)})/(4\uppi)$ to mimic incomplete sky coverage \citep{van_uitert_kidsgama_2018}. Averaging over multipole bands and assuming again that the spectra do not vary significantly across the bands yields
\begin{equation}
    \mathrm{Cov}_\mathrm{G}[\hat{C}_{a_1a_2}(\ell_1)\hat{C}_{a_3a_4}(\ell_2)] \approx \frac{4\uppi\delta^{\mathrm{K}}_{\ell_1\ell_2}}{\mathrm{max}(A_{(12)}A_{(34)})(2\ell_1+1)\Delta\ell_1}\left[{C}_{a_1a_3}(\ell_1){C}_{a_2a_4}(\ell_2) + {C}_{a_1a_4}(\ell_1){C}_{a_2a_3}(\ell_2)\right]\,,
\end{equation}
which is the same expression as \Cref{eq:Gaussian_flat_sky} for $\ell \gg 1$, as expected.
\Cref{eq:Gaussian_covariance} amounts to the splitting discussed in \Cref{eq:Gaussian_split}.
We now turn back to the connected term in \Cref{eq:covariance_splitting_flat}:
\begin{equation}
\label{eq:averaged_trispec}
    \mathrm{Cov}_\mathrm{nG}[\hat{C}_{a_1a_2}(\ell_1)\hat{C}_{a_3a_4}(\ell_2)] = \frac{1}{\mathrm{max}(A_{(12)}A_{(34)})}\int_{\tilde{\ell}_1\in\ell_1}\frac{\mathrm{d}^2\tilde{\ell}_1}{A(\ell_1)}\int_{\tilde{\ell}_2\in\ell_2}\frac{\mathrm{d}^2\tilde{\ell}_2}{A(\ell_2)}   T^{(a_1a_2)(a_3a_4)}(\tilde{\boldsymbol{\ell}}_1, -\tilde{\boldsymbol{\ell}}_1,\tilde{\boldsymbol{\ell}}_2,-\tilde{\boldsymbol{\ell}}_2)\,,
\end{equation}
where we moved to the continuous version of the two-dimensional Fourier transformation.
All trispectra are projected with the Limber projection and the angular average is carried out already in $k$-space:
\begin{equation}
T^{(a_1a_2)(a_3a_4),\; X}_{\ell_1\ell_2} = \int\frac{\mathrm{d}\chi}{f_\mathrm{K}(\chi)^6}W_{a_1}(\chi)\dots W_{a_4}(\chi) \bar T^X_{(A_1A_2)(A_3A_4)}\left(\frac{\ell_1 + 0.5}{f_\mathrm{K}(\chi)}, \frac{\ell_2 + 0.5}{f_\mathrm{K}(\chi)},\chi\right)\,,
\end{equation}
where
\begin{equation}
\label{eq:trispectrum_proj}
    \bar T^X_{(A_1A_2)(A_3A_4)}(k_1,k_2,z(\chi)) = 
    \left\{
    \arraycolsep=1.pt\def\arraystretch{1.9}
    \begin{array}{ll}
         T^\mathrm{SSC}_{(A_1A_2)(A_1A_2)}(k_1,k_2,z(\chi)) & \quad\text{if $X = \mathrm{SSC}$}\,, \\
         \int_0^\uppi\frac{\mathrm{d}\phi_\ell}{\uppi} T_{A_1\dots A_4}(\boldsymbol{k}_1,-\boldsymbol{k}_1,\boldsymbol{k}_2,-\boldsymbol{k}_2,z(\chi)) & \quad\text{if $X = \mathrm{nG}$}\,,
    \end{array}
    \right.
\end{equation}
where $\phi_\ell$ is the angle between the two wave vectors $\boldsymbol{k}_{1,2}$ in the plane perpendicular to the line of sight.
Altogether, the covariance in harmonic space is given by
\begin{equation}
    \label{eq:covariance_ell_space}
    \mathrm{Cov}[\hat{C}_{a_1a_2}(\ell_1)\hat{C}_{a_3a_4}(\ell_2)] = \mathrm{Cov}_\mathrm{G}\left[\hat{C}_{a_1a_2}(\ell_1)\hat{C}_{a_3a_4}(\ell_2)\right] + \frac{1}{\mathrm{max}(A_{(12)}A_{(34)})} T^{(a_1a_2)(a_3a_4),\; \mathrm{nG}}_{\ell_1\ell_2} + T^{(a_1a_2)(a_3a_4),\; \mathrm{SSC}}_{\ell_1\ell_2}\;
\end{equation}
at each multipole $\ell$. For an $\ell$-band averaged version, we use \Cref{eq:Gaussian_flat_sky} and \Cref{eq:averaged_trispec}. This is only important if a pure $\ell$-space covariance is required, since the averaging over scales is taken care of when mapping the theoretical $\ell$-space covariance to the observables in the next section.

\section{From harmonic space to observables}
\label{sec:harmonic_to_real}   
Whilst theoretical modelling is most straightforward in harmonic space, the theoretical power spectra, denoted as $C(\ell)$, are technically defined solely on the full sky. As real surveys typically cover only a portion of the sky, the advantage of harmonic space diminishes. This is because different $\ell$ modes become intermingled, and the partial sky $C_\ell$, often referred to as pseudo-$C_\ell$, are derived from the full-sky $C_\ell$ via convolution with the mode mixing matrix \citep[see e.g.][]{2002ApJ...567....2H, 2003ApJS..148..161K,2013A&A...554A.112R,2017MNRAS.465.1847E,alonso_unified_2019}. 

Deconvolution may not always be feasible for intricate sky masks, rendering the $C_\ell$ indirectly observable and serving primarily as an intermediary product.
Another challenge arises from the projection, as outlined in \Cref{eq:projection}, which combines various physical scales if the window function $W_A(\chi)$ possesses broad support in the co-moving distance. Additionally, the shear field's spin-2 nature permits its separation into E and B modes. 

To mitigate these difficulties, a variety of summary statistics are available, including the probes of the classical 3$\times$2-point analysis: GC, CS, and GGL. Accordingly, the \mysc{OneCovariance} code can accommodate diverse inputs for three-dimensional power spectra, angular power spectra, or weight functions for the projection, rendering the discussion applicable to all general tracers or summary statistics of two-point functions in the LSS.
In a broader context, any two-point summary statistic represents a linear transformation of the underlying two-point statistic in harmonic space, as nonlinear transformations would involve contributions from higher-order cumulants. Therefore, we write the summary statistic $\boldsymbol{\mathcal{O}}$ between two tracers $a_1$ and $a_2$ as
\begin{equation}
\label{eq:linearmap_ell_to_obs}
    \boldsymbol{\mathcal{O}}(L) = \int\frac{\ell\mathrm{d}\ell}{2\uppi}\; \boldsymbol{\tens{W}}_{L}(\ell) \mathrm{vec}(\boldsymbol{{\tens C}}_{a_1a_2})(\ell)\,.
\end{equation}
Here $L$ can be a discrete label or a continuous variable for some Fourier filter $\boldsymbol{\tens{W}}_L(\ell)$. Importantly, $\mathrm{vec}({\boldsymbol{\tens{C}}}_{a_1a_2})$, concatenates all unique $\boldsymbol{{\tens C}}_{a_1a_2}$ into a vector which is mapped to the new summary statistic $\boldsymbol{\mathcal{O}}$ via the aforementioned linear map.
Throughout this section, we   omit the formal integration limits from zero to infinity, but all integrals over $\ell$ are understood to run over this range if not specified differently. 
It should be noted that the internal structure of $\boldsymbol{{\tens C}}_{a_1a_2}$ depends on the underlying fields $a_1$ and $a_2$. If the spin of $a_1$ and $a_2$ is non-zero, such as for CS, they both consist of two complex components so that $\boldsymbol{\tens{C}}$ is a real $2\times 2$ matrix given by
\begin{equation}
    \boldsymbol{\tens{C}}_{a_1a_2}(\ell_1)\delta^\mathrm{K}_{\ell_1\ell_2}\delta^\mathrm{K}_{m_1m_2} = \big\langle \boldsymbol{a}^{\phantom{\dagger}}_{1,\ell_1 m_1}\boldsymbol{a}^\dagger_{2,\ell_2 m_2} \big\rangle\,,
\end{equation}
on the full sky for a homogeneous and isotropic random field. 
The CS submatrix of $\boldsymbol{\tens{W}}_L(\ell)$ is a symmetric and real $2\times 2$ matrix with two independent contributions: the curl-free E-mode signal and the divergence-free B-mode signal, where we assume that there is no measurable contribution to the EB spectrum. We thus bundle the combined $C(\ell)$ of all probes (for a standard $3\times 2$ analysis) in a vector by using the notation of Equation \ref{eq:weights_specific_g} and \ref{eq:weights_specific_m}
\begin{equation}
    [\mathrm{vec}(\boldsymbol{C})](\ell) \coloneqq(C_\mathrm{gg},C_\mathrm{gm},C_\mathrm{mm,EE},C_\mathrm{mm,BB})^\mathrm{T}(\ell)\,,
\end{equation}
where tomographic bin indices have been omitted for clarity. The resulting summary statistics are bundled in the same fashion:
\begin{equation}
    [\mathrm{vec}(\boldsymbol{\mathcal{O}})](L) \coloneqq ({\mathcal{O}}_\mathrm{gg},{\mathcal{O}}_\mathrm{gm}, {\mathcal{O}}_\mathrm{mm,EE}, {\mathcal{O}}_\mathrm{mm,BB})^\mathrm{T}(L)\,.
\end{equation}
In this manner, we can express the transformation from Fourier space to the summary statistic using \Cref{eq:linearmap_ell_to_obs}.
Real space correlation functions decouple the shot or shape noise contribution from different scales, $L$. Furthermore, the shot or shape noise contribution can be precisely estimated from the data itself, regardless of complex masks.
To leverage those properties the shot or shape noise levels defined in \Cref{eq:noise_levels}, which only apply to full-sky data, are improved by also providing a relation between real space correlation (see \Cref{sec:real_space_correlation_functions}) and the observables. Therefore, we also require the following correlation for some real space filter function $\boldsymbol{\tens{R}}_L(\theta)$
\begin{equation}
\label{eq:linearmap_theta_to_obs}
   \mathrm{vec}(\boldsymbol{\mathcal{O}}) (L) = \int\theta\mathrm{d}\theta\; \boldsymbol{\tens{R}}_{L}(\theta) \begin{pmatrix}
        w(\theta)\\
        \gamma_t(\theta)\\
        \xi_+(\theta)\\
        \xi_-(\theta)\\
    \end{pmatrix}\,,
\end{equation}
with the real space correlation functions defined via Hankel transformations of the angular power spectrum, \Cref{eq:powerspectrum_general}. 
For CS, this amounts to the two correlation functions \citep{bartelmann_weak_2001}
\begin{align}
\label{eq:hankelshear1}
\xi_+^{(ij)}(\theta) &= \int\frac{ \ell \mathrm{d}\ell }{2 \uppi} \; \left[\mathcal{P}_{\epsilon_i\epsilon_j,\mathrm{E}}(\ell) + \mathcal{P}_{\epsilon_i\epsilon_j,\mathrm{B}}(\ell)\right] \; {\mathrm{J}}_0(\ell \theta)\,, \\ 
\label{eq:hankelshear2}
\xi_-^{(ij)}(\theta) &= \int\frac{\ell \mathrm{d}\ell }{2 \uppi} \; \left[\mathcal{P}_{\epsilon_i\epsilon_j,\mathrm{E}}(\ell) - \mathcal{P}_{\epsilon_i\epsilon_j,\mathrm{B}}(\ell)\right] \; \mathrm{J}_4(\ell \theta)\,,
\end{align}
where $\mathrm{J}_\mu$ is a cylindrical Bessel function of the first kind of order $\mu$. The angular power spectrum $\mathcal{P}_{a_1a_2}(\ell)$ was decomposed into an E and B-mode component. Recall that we explicitly use the noise-free version of the angular power spectra here as defined in the estimator in \Cref{eq:noise_free_noise_cell}. For GGL one measures the tangential shear correlation function
\begin{equation}
\label{eq:ggl_correlation_function}
    \gamma^{(ij)}_\mathrm{t} (\theta) = \int\frac{\ell\mathrm{d}\ell}{2\uppi} \mathcal{P}_{\mathrm{n}_i\epsilon_j} (\ell)\; \mathrm{J}_2(\ell\theta)\,.
\end{equation}
The galaxy correlation function can be calculated as
\begin{equation}
\label{eq:gc_correlation_function}
    w^{(ij)}(\theta) = \int\frac{ \ell \mathrm{d}\ell }{2 \uppi} \; \mathcal{P}_{\mathrm{n}_i\mathrm{n}_j}(\ell)\;{\mathrm{J}}_0(\ell \theta)\,.
\end{equation}
We note that these are all the continuous versions and large multipole approximations of the discrete transformations. 
We   continue to review the definitions of the three most commonly used summary statistics (real space correlation functions, band powers and COSEBIs) and the corresponding covariances. A more detailed discussion of the general setting can be found in \Cref{app:mapping_to_summary}.

\subsection{Multiplicative shear bias}
Shear measurements for source samples are calibrated on image simulations \citep[see e.g.][]{kannawadi_towards_2019,2023A&A...670A.100L}. Residual biases are captured in a multiplicative and additive shear bias, which needs to be propagated into the cosmological inference. We assume that there are no residual spatial patterns in the additive contribution (e.g. from point-spread function leakage) and hence no correlation with the cosmological signal. Thus, the only source of error which has to be propagated in the covariance is the multiplicative $m$-correction.  The residual error on the multiplicative shear bias after calibration is labelled $\sigma^{a_1}_\mathrm{m}$ for source bin $a_1$. Considering the multiplicative shear bias uncertainty as an additive contribution to the covariance matrix \citep{joachimi_kids-1000_2021}, so that
\begin{equation}
    \mathrm{Cov}_\mathrm{mult}\left[ \mathcal{O}_{a_1a_2} (L_1), \mathcal{O}_{a_3a_4} (L_2)  \right] = \mathcal{O}_{a_1a_2} (L_1)\mathcal{O}_{a_3a_4} (L_2)\left[\sigma^{a_1}_\mathrm{m}\sigma^{a_3}_\mathrm{m} + \sigma^{a_1}_\mathrm{m}\sigma^{a_4}_\mathrm{m}+ \sigma^{a_2}_\mathrm{m}\sigma^{a_3}_\mathrm{m}+\sigma^{a_2}_\mathrm{m}\sigma^{a_4}_\mathrm{m}\right]\,.
\end{equation}
This approximation holds for $\sigma^{a_1}_\mathrm{m} \ll 1 $, for $\langle m\rangle = 0\rangle$ after correction and fully correlated $m$-bias corrections across tomographic bins. We note that only source samples have a $\sigma^{a_1}_\mathrm{m}$. by definition. The assumption $\langle m\rangle = 0\rangle$ is not crucial for the covariance, as this can be addressed by  directly shifting the data vector while keeping the covariance unchanged \citep{2023A&A...670A.100L}.

\subsection{Real space correlation functions}
\label{sec:real_space_correlation_functions}
When propagating the covariance matrix from the angular power spectra to the real space correlation functions in \Cref{eq:hankelshear1,eq:hankelshear2,eq:ggl_correlation_function,eq:gc_correlation_function} we take into account that the correlation functions are measured over finite angular separation bins centred at $\bar\theta_i$ with boundaries $[\theta_{\mathrm{l},i}, \theta_{\mathrm{u},i}]$, the bin average needs to be carried out explicitly. To capture the different weight functions, we   use the notation from \citet{joachimi_kids-1000_2021}
\begin{equation}
\label{eq:bin_average_theta}
    {\cal K}_\mu \left(\ell \bar{\theta}_i\right) := \frac{2}{\theta_{{\rm u},i}^2 - \theta_{{\rm l},i}^2 }\, \int_{\theta_{{\rm l},i}}^{\theta_{{\rm u},i}} \mathrm{d} \theta'\, \theta' {\rm J}_\mu(\ell \theta') =  \frac{2}{\br{ \theta_{{\rm u},i}^2 - \theta_{{\rm l},i}^2 } \ell^2}\, \times\, \left\{ \begin{array}{ll}
    \bb{x {\rm J}_1(x)}_{\ell \theta_{{\rm l},i}}^{\ell \theta_{{\rm u},i}}  & \mu=0\,, \\ \bb{ -x {\rm J}_1(x) - 2 {\rm J}_0(x) }_{\ell \theta_{{\rm l},i}}^{\ell \theta_{{\rm u},i}}  & \mu=2\,, \\ \bb{ \br{x -\frac{8}{x}} {\rm J}_1(x) - 8 {\rm J}_2(x) }_{\ell \theta_{{\rm l},i}}^{\ell \theta_{{\rm u},i}}  & \mu=4\,.  
    \end{array} \right.\;,
\end{equation}
ignoring the weights of the pair counts \citep[][]{2019A&A...624A.134A}.
The set of possible correlation functions is denoted accordingly in a compact form \citep[see the notation used in][]{joachimi_kids-1000_2021}:
\begin{equation*}
\bc{w, \ba{\gamma_{\rm t}}, \xi_+, \xi_-} \leftrightarrow \bc{\Xi_0, \Xi_2, \Xi_0, \Xi_4}\,.
\end{equation*}
We note that $w$ and $\xi_+$ have the same weight function. 
As discussed, the Gaussian covariance is split into three contributions as in \Cref{eq:Gaussian_split}. For the pure shot/shape noise (sn) term, this procedure is a numerical necessity, as an $\ell$-independent integrand will lead to a $\delta_\mathrm{D}$-contribution to the covariance.  Since we reconsider the idealised term in \Cref{app:mixed_term} for real space correlation functions, this splitting is also necessary here as the ‘mix’ term is the most important while also still being a somewhat uncertain term in the analytical covariance modelling. Numerically, it is, however, usually advantageous to combine the sample variance (‘sva’) and mixed terms to speed up the convergence of the integrals.

Using \Cref{eq:linearmap_ell_to_obs}, the definitions of the weights above and \Cref{eq:covariance_ell_space}, the ‘sva’ term is given by
\begin{equation}
    \label{eq:cov_g_sva}
{\rm Cov}_{\rm G, sva} \bb{\Xi_\mu^{(ij)}\left(\bar{\theta}_p\right);\, \Xi_\nu^{(mn\vphantom{j})}\left(\bar{\theta}_q\right)} = \frac{1}{2 \uppi\; \mathrm{max}(A_{(ij)}A_{(mn)}) } \int \!\mathrm{d}\ell\, \ell\; {\cal K}_\mu \left(\ell \bar{\theta}_p\right) {\cal K}_\nu \left(\ell \bar{\theta}_q\right) \bb{ \mathcal{P}_{im}(\ell) \mathcal{P}_{jn}(\ell) + \mathcal{P}_{in}(\ell) \mathcal{P}_{jm}(\ell) } \,,
\end{equation}
where the tomographic indices $i,j,m,n$ label either a source or a lens sample.
We measure all areas from a binary \textsc{healpix} mask down to a certain angular scale so that masked stars and other features only reduce the effective number density and not the survey area. This assumption is valid as long as this dilution scale is smaller than the smallest scale over which the cosmological signal is measured. Typically, one assumes $N_\mathrm{side}= 4096$ for the \textsc{healpix} \citep{2005ApJ...622..759G} evaluation, corresponding to a pixel size of just below one arcmin. 
 
 In an ideal survey, the pure noise contribution is just the product of the two individual noise contributions, \Cref{eq:noise_levels}, to the measurement of the two-point statistic amounting to a term proportional to the number of (non-unique) pairs in bin $\bar\theta_p$:
\begin{equation}
    N^{(ij)}_{\mathrm{pair,\;ideal}}\left(\bar{\theta}_p\right) = \uppi  \left(\theta_{\mathrm{u},p}^2 -  \theta_{\mathrm{l},p}^2\right) \bar n_i \bar n_j A_{(ij)}\,.
\end{equation}
This equation is accurate in the absence of survey boundaries. For finite areas, however, the number of pairs as a function of angular scale differs from the expected $\propto\theta^2$ scaling. To obtain a more accurate prescription of the pure shot noise contribution, the number of pairs is directly measured at the catalogue level, providing $N^{(ij)}_{\mathrm{pair}}\left(\bar{\theta}_p\right)$. Altogether,
\begin{equation}
\label{eq:noise_covariance}
{\rm Cov}_{\rm G, sn} \bb{\Xi_\mu^{(ij)}\left(\bar{\theta}_p\right);\, \Xi_\nu^{(mn\vphantom{j})}\left(\bar{\theta}_q\right)} =\delta^\mathrm{K}_{\mu\nu} \delta^\mathrm{K}_{\bar{\theta}_p \bar{\theta}_q} \br{\delta^\mathrm{K}_{im} \delta^\mathrm{K}_{jn} + \delta^\mathrm{K}_{in} \delta^\mathrm{K}_{jm} } \frac{{\cal T}_{(ij)(mn)}^{\rm sn} }{N_{\rm pair}^{(ij)}\left(\bar{\theta}_p\right) } \,, 
\end{equation}
where we defined the noise levels
\begin{equation}
    {\cal T}_{(ij)(mn)}^{\rm sn}  \coloneqq
    \left\{\begin{array}{ll}
        2\sigma_{\epsilon_1,\;i}^2\sigma_{\epsilon_1,\;j}^2 & \quad\text{if } i,j,m,n\in\text{source}\,,   \\
        \sigma_{\epsilon_1,\;i}^2 & \quad\text{if both  } (ij) \text{ and } (mn) \text{ contain a lens and source sample each} \,, \\
        1 & \quad\text{if } i,j,m,n\in\text{lens}\,. 
    \end{array}\right.
\end{equation}
We recall that `source' refers to the galaxy shape and `lens' to the galaxy position being used.
Furthermore, we note that the Latin indices carry a hidden label ‘lens’ or ‘source’ so that $\delta^\mathrm{K}_{ij} = 1$ only when both indices come from the same set. It should be noted that, while we use in the equations here the non-unique pairs, for any correlation measurement, however, only the unique pairs are relevant. This is accounted for in the covariance calculations. Since this often leads to confusion, let us use tomographic clustering with equi-populated bins as an example. The noise level will always be $\sim 1/N^2$, with $N$ being the number of galaxies in each bin. 
For auto-correlations, however, the Kronecker symbols in the brackets in  \Cref{eq:noise_covariance} make sure that only unique pairs are counted, effectively increases the covariance by a factor of two. In contrast, for cross-correlations, the $N^2$ is actually the number of unique pairs.

The mixed term is calculated in the same fashion as the sample variance contribution, \Cref{eq:cov_g_sva}:
\begin{equation}
    \label{eq:mixed_term}
    {\rm Cov}_{\rm G, mix} \bb{\Xi_\mu^{(ij)}\left(\bar{\theta}_p\right);\, \Xi_\nu^{(mn\vphantom{j})}\left(\bar{\theta}_q\right)} = \delta_{jn}\, \frac{{\cal T}_j^{\rm mix}}{2 \uppi n_{\rm eff}^{(j)}\, \mathrm{max}(A_{(ij)}A_{(mn)})  } \int \mathrm{d} \ell\, \ell\; {\cal K}_\mu \left(\ell \bar{\theta}_p\right) {\cal K}_\nu \left(\ell \bar{\theta}_q\right)\, \mathcal{P}_{im}(\ell) + \mbox{3 perm.}\,, 
\end{equation}
The noise level of the mixed term is defined as
\begin{equation}
    {\cal T}_j^{\rm mix} \coloneqq
    \left\{
\begin{array}{ll}
     \sigma^2_{\epsilon_1\,,j}& \quad \text{if } j\in\text{source}\,,  \\
     1 &\quad  \text{if } j\in\text{lens}\,. 
\end{array}
    \right.
\end{equation}
We note that this is an idealised setting for a rescaled full sky survey without a complicated mask. In \Cref{app:mixed_term} we revisit this term using the triplet counts of a catalogue to estimate the effect on the covariance and cosmological inference. 

The non-Gaussian covariance and SSC are
\begin{align}
    {\rm Cov}_{\rm NG/SSC} \bb{\Xi^{(ij)}_\mu\left(\bar{\theta}_p\right);\, \Xi^{(mn\vphantom{j})}_\nu\left(\bar{\theta}_q\right)} = &\; \frac{1}{4 \uppi^2 } \int \mathrm{d} \ell_1\, \ell_1\;  {\cal K}_\mu \left(\ell_1 \bar{\theta}_p\right) \int\mathrm{d} \ell_2\, \ell_2\;  {\cal K}_\nu \left(\ell_2 \bar{\theta}_q\right) \\ \nonumber \times & 
    \left\{
    \begin{array}{ll}
          \frac{1}{\mathrm{max}(A_{(ij)}A_{(mn)}) }\int_0^\uppi \frac{\mathrm{d} \phi_\ell}{\uppi} \; T^{(ij)(mn)}(\boldsymbol{\ell}_1,-\boldsymbol{\ell}_1,\boldsymbol{\ell}_2,-\boldsymbol{\ell}_2)  & \quad\mathrm{for\; nG}\\
        T^{(ij)(mn)}_\mathrm{SSC}({\ell_1,\ell_2})& \quad\mathrm{for\; SSC}
    \end{array}\right. \; .
\end{align}

\subsection{Bandpowers}
\label{sec:bandpowers}
\restructured{We define the bandpower signal their weights and further ingredients in \Cref{app:bandpowers} and focus on the covariance here.}
In contrast to the real space approach in \citet{joachimi_kids-1000_2021} we calculate the bandpower covariance from Fourier space directly using \Cref{eq:bp_cosmicshear1,eq:bp_cosmicshear2,eq:bp_cosmicshear3,eq:bp_cosmicshear4}. This is done to resemble the analytical calculations of the predicted signal and for numerical stability and speed.
In complete analogy to the real space correlation functions, we write
\begin{align}
 \label{eq:cov_g_bp_no_noise}
   &\nonumber {\rm Cov}_{\rm G, sva} \bb{\mathcal{C}^{(ij)}_{\mu}(L_1),  \mathcal{C}^{(mn\vphantom{j})}_{\nu}(L_2)}  =   \frac{2 \uppi}{\mathcal{N}_{L_1}\mathcal{N}_{L_2} \mathrm{max}(A_{(ij)}A_{(mn)}) }  \int \mathrm{d}\ell\, \ell\; {\cal W}^{L_1}_{\mu}(\ell) {\cal W}^{L_2}_{\nu}(\ell) \big[ \mathcal{P}_{im}(\ell) \mathcal{P}_{jn}(\ell)  + \mathcal{P}_{in}(\ell) \mathcal{P}_{jm}(\ell)\big] \\
  &  {\rm Cov}_{\rm G, mix} \bb{\mathcal{C}^{(ij)}_{\mu}(L_1),  \mathcal{C}^{(mn\vphantom{j})}_{\nu}(L_2)}  = \delta_{jn}\, \frac{2 \uppi \;{\cal T}_j^{\rm mix}}{\mathcal{N}_{L_1}\mathcal{N}_{L_2}n_{\rm eff}^{(j)}\, \mathrm{max}(A_{(ij)}A_{(mn)})  } \int \mathrm{d} \ell\, \ell\; {\cal W}^{L_1}_{\mu}(\ell) {\cal W}^{L_2}_{\nu}(\ell)\, \mathcal{P}_{im}(\ell) + \mbox{3 perm.} \,,
\end{align}
assuming that the cosmological $B$-mode signal vanishes.
The following shorthand notation for the weights was used:
\begin{equation}
        \mathcal{W}^L_\mu(\ell) = 
        \left\{
        \arraycolsep=1.pt\def\arraystretch{1.9}
        \begin{array}{ll}
            W^L_{EE}(\ell)/2 &  \quad\text{if $\mu = \epsilon\epsilon$E}\\ 
            W^L_{EE}(\ell) &  \quad\text{if $\mu = \;$nn}\\ 
                        W^L_{nE}(\ell) &  \quad\text{if $\mu = \mathrm{n}\epsilon$} \\
             W^L_{EB}(\ell) &  \quad\text{if $\mu = \epsilon\epsilon$B} 
        \end{array}\right.\,.
\end{equation}
To properly incorporate the pair counts the pure shot-noise contributions are calculated from real space:
\begin{align}
     {\rm Cov}_{\rm G, sn} \bb{\mathcal{C}^{(ij)}_{\mu}(L_1),  \mathcal{C}^{(mn\vphantom{j})}_{\nu}(L_2)} = &\;
    \frac{\uppi^2\delta^\mathrm{K}_{\mu\nu}}{\mathcal{N}_{L_1}\mathcal{N}_{L_2}} \left(\delta^\mathrm{K}_{im}\delta^\mathrm{K}_{jn} + \delta^\mathrm{K}_{in}\delta^\mathrm{K}_{jm}\right) \int\frac{\theta^2\mathrm{d}\theta}{n^{(ij)}_\mathrm{pair}(\theta)}\\ & \times\; \nonumber\left\{
      \arraycolsep=1.pt\def\arraystretch{1.9}
     \begin{array}{ll}
        {2\sigma^2_{\epsilon_1,\;i}\sigma^2_{\epsilon_1,\;j}}\left(g^{L_1}_+(\theta)g^{L_2}_+(\theta) + g^{L_1}_-(\theta)g^{L_2}_-(\theta)\right) &\; \text{if $\mu = \epsilon\epsilon$} \\
        {4\sigma^2_{\epsilon_1,\;i}} h^{L_1}(\theta)h^{L_2}(\theta) &\; \text{if $\mu  = \mathrm{n}\epsilon$} \\
        g^{L_1}_+(\theta)g^{L_2}_+(\theta)  &\; \text{if $\mu  = \mathrm{nn}$} \\
     \end{array}
     \right..
\end{align}
The differential pair counts, $n^{(ij)}_\mathrm{pair}(\theta)$, are defined such that the number of pairs, $N^{(ij)}_\mathrm{pair}$, in an angular bin $[\theta_\mathrm{l}, \theta_\mathrm{u}]$ is
\begin{equation}
    N^{(ij)}_\mathrm{pair} =  \int_{\theta_\mathrm{l}}^{\theta_\mathrm{u}}\mathrm{d}\theta^\prime\; n^{(ij)}_\mathrm{pair}(\theta^\prime)
\end{equation}
directly from the, possibly weighted, pair counts of the catalogue. Lastly, the nG and SSC terms can be calculated as follows:
\begin{align}
    {\rm Cov}_{\rm NG} \bb{\mathcal{C}^{(ij)}_{\mu}(L_1),  \mathcal{C}^{(mn\vphantom{j})}_{\nu}(L_2)} = &\; \frac{1}{\mathcal{N}_{L_1}\mathcal{N}_{L_2} \mathrm{max}(A_{(ij)}A_{(mn)}) } \int \mathrm{d}\ell_1\, \ell_1{\cal W}^{L_1}_{\mu}(\ell_1)\int \mathrm{d}\ell_2\, \ell_2\;  {\cal W}^{L_2}_{\nu}(\ell_2) \\ &\times \;\nonumber \int_0^\uppi \frac{\mathrm{d} \phi_\ell}{\uppi} \; T^{(ij)(mn)}(\boldsymbol{\ell}_1,-\boldsymbol{\ell}_1,\boldsymbol{\ell}_2,-\boldsymbol{\ell}_2)\,,  \\
     {\rm Cov}_{\rm SSC} \bb{\mathcal{C}^{(ij)}_{\mu}(L_1),  \mathcal{C}^{(mn\vphantom{j})}_{\nu}(L_2)} = &\; \frac{1}{\mathcal{N}_{L_1}\mathcal{N}_{L_2}} \int \mathrm{d}\ell_1\, \ell_1{\cal W}^{L_1}_{\mu}(\ell_1)\int \mathrm{d}\ell_2\, \ell_2\;  {\cal W}^{L_2}_{\nu}(\ell_2) T^{(ij)(mn)}_\mathrm{SSC}({\ell_1,\ell_2})\,.
\end{align}

\subsection{COSEBIs and $\Psi$ statistics}
\label{sec:cosebi}
Due to the similar structure of the COSEBIs we repeat the same calculation as for the band power covariance, i.e. projecting all terms from harmonic space except for the pure shot noise term. \restructured{With the definition of COSEBIs , see \Cref{app:cosebis}, this yields} the following:
\begin{align}
 \label{eq:cov_g_cosebi_no_noise1}
   & {\rm Cov}_{\rm G, sva} \bb{E^{(ij)}_{a},  E^{(mn\vphantom{j})}_{b}}  =\; \frac{1}{2\uppi \;\mathrm{max}(A_{(ij)}A_{(mn)}) } \int \mathrm{d}\ell\, \ell\; { W}_{a}(\ell) { W}_{b}(\ell) \bb{ \mathcal{P}_{im}(\ell) \mathcal{P}_{jn}(\ell) + \mathcal{P}_{in}(\ell) \mathcal{P}_{jm}(\ell) } \\
    \label{eq:cov_g_cosebi_no_noise2}
   & {\rm Cov}_{\rm G, mix} \bb{E^{(ij)}_{a},  E^{(mn\vphantom{j})}_{b}}  =\; \delta_{jn}\, \frac{\;{\cal T}_j^{\rm mix}}{2 \uppi\; n_{\rm eff}^{(j)}\, \mathrm{max}(A_{(ij)}A_{(mn)})  } \int \mathrm{d} \ell\, \ell\;  { W}_{a}(\ell) { W}_{b}(\ell)\, \mathcal{P}_{im}(\ell) + \mbox{3 perm.} \\
    \label{eq:cov_g_cosebi_no_noise3}
   & {\rm Cov}_{\rm NG} \bb{E^{(ij)}_{a},  E^{(mn\vphantom{j})}_{b}}  = \; \frac{1}{2\uppi\; \mathrm{max}(A_{(ij)}A_{(mn)}) } \int \mathrm{d}\ell_1\, \ell_1{ W}_{a}(\ell_1)\int \mathrm{d}\ell_2\, \ell_2\;  { W}_{b}(\ell_2)  \int_0^\uppi \frac{\mathrm{d} \phi_\ell}{\uppi} \; T^{(ij)(mn)}(\boldsymbol{\ell}_1,-\boldsymbol{\ell}_1,\boldsymbol{\ell}_2,-\boldsymbol{\ell}_2)\,,  \\
\label{eq:cov_g_cosebi_no_noise4}
   &  {\rm Cov}_{\rm SSC} \bb{E^{(ij)}_{a},  E^{(mn\vphantom{j})}_{b}}  = \; \frac{1}{2\uppi\;} \int \mathrm{d}\ell_1\, \ell_1{ W}_{a}(\ell_1)\int \mathrm{d}\ell_2\, \ell_2\;  { W}_{b}(\ell_2) T^{(ij)(mn)}_\mathrm{SSC}({\ell_1,\ell_2})\\
    \label{eq:cov_g_cosebi_no_noise5}
&     {\rm Cov}_{\rm G, sn} \bb{E^{(ij)}_{a},  E^{(mn\vphantom{j})}_{b}}  =\; 
    \frac{\sigma^2_{\epsilon_1,\;i}\sigma^2_{\epsilon_1,\;j} }{2} \left(\delta^\mathrm{K}_{im}\delta^\mathrm{K}_{jn} + \delta^\mathrm{K}_{in}\delta^\mathrm{K}_{jm}\right)\int\frac{\theta^2\mathrm{d}\theta}{n^{(ij)}_\mathrm{pair}(\theta)}\left[{T}_{+a}(\theta){ T}_{+b}(\theta) + {T}_{-a}(\theta){ T}_{-b}(\theta)\right]\,.
\end{align}
The expressions of the covariance of $\Psi$ statistics is in full analogy to \Cref{eq:cov_g_cosebi_no_noise1,eq:cov_g_cosebi_no_noise2,eq:cov_g_cosebi_no_noise3,eq:cov_g_cosebi_no_noise4,eq:cov_g_cosebi_no_noise5}.

\subsection{Stellar mass function}
To complement the halo model, we also implement the SMF covariance, \Cref{eq:smf}, which was already used in \citet{dvornik_kids-1000_2023}. For a flux-limited sample, we consider the standard $V_\mathrm{max}(M_\star)$-estimator, where $V_\mathrm{max}$ is the maximum volume out to which a galaxy with a given mass can be observed given the limiting magnitude of the survey. The estimator for the SMF is then \citep[see e.g.][]{smith_how_2012}:
\begin{equation}
    \hat\Phi ^{(i)}(M^\mu_\star) \coloneqq \frac{1}{\Delta M^\mu_\star} \frac{1}{V^{(i)}_\mathrm{max}(M^\mu_\star)}\sum_{a=1}^{N_\mathrm{tot}} \Theta_\mathrm{H}(z_\mathrm{photo ,a},z^{(i)}_\mathrm{photo}) \Theta_\mathrm{H}(M_{\star,a}, M^\mu_\star)\,.
\end{equation}
Here we assumed that the mapping from observed to real stellar mass is very tight and well approximated by a $\delta$ distribution. If the relation is less well known, this would amount to an additional integration. The indices $i,\mu, a$ label a possible splitting in tomographic bins, the stellar mass bin and the individual galaxy in each bin respectively. $\Theta_\mathrm{H}$ is the Heaviside function.
The auto-correlation consists of a noise term and an SSC contribution \citep{takada_probing_2007,smith_how_2012},
\begin{equation}
    \mathrm{Cov}\left[\hat\Phi ^{(i)}(M^\mu_\star),\hat\Phi ^{(j)}(M^\nu_\star)\right] = \mathrm{Cov}\left[\hat\Phi ^{(i)}(M^\mu_\star),\hat\Phi ^{(j)}(M^\nu_\star)\right]_\mathrm{sn} + \mathrm{Cov}\left[\hat\Phi ^{(i)}(M^\mu_\star),\hat\Phi ^{(j)}(M^\nu_\star)\right]_\mathrm{SSC}\,,
\end{equation}
with the halo occupation variance being neglected as it is subdominant \citep{smith_how_2012} and
\begin{align}
   & \mathrm{Cov}\left[\hat\Phi ^{(i)}(M^\mu_\star),\hat\Phi ^{(j)}(M^\nu_\star)\right]_\mathrm{sn} = \;  \delta^\mathrm{K}_{ij} \delta^\mathrm{K}_{\mu\nu}\frac{\Phi^{(i)}(M^\mu_{\star})}{\Delta M^\mu_{\star} V^{(i)}_{\mathrm{max},\mu}}\\
&    \mathrm{Cov}\left[\hat\Phi ^{(i)}(M^\mu_\star),\hat\Phi ^{(j)}(M^\nu_\star)\right]_\mathrm{SSC} = \; \frac{A^2 f^{(i)}f^{(j)}}{ V^{(i)}_{\mathrm{max},\mu} V^{(j)}_{\mathrm{max},\nu}}\int\mathrm{d}\chi\;\frac{p^{(i)}(\chi)p^{(j)}(\chi)}{p^2_\mathrm{tot}(\chi)} f^2_\mathrm{k}(\chi) \sigma^2_{\mathrm{bg,} A}(\chi) \Tilde\Phi_\mu(\chi) \Tilde\Phi_\nu(\chi)\,,
\end{align}
where $f^{(i)}$ is the fraction of galaxies in tomographic bin $i$, $A$ is the survey area, $p^{(i)}$ the redshift distribution of the $i$-th tomographic bin and $\sigma_{\mathrm{bg,}A}$ is given by \Cref{eq:sigma_bg} with the same footprint. Lastly, the quantity $\Tilde\Phi_\mu(\chi)$ is defined as
\begin{equation}
    \Tilde\Phi_\mu(\chi) \coloneqq \int \mathrm{d}M \;n_\mathrm{h}(M,z(\chi)) \Phi(M^\mu_\star|M)b_\mathrm{h}(M,z(\chi))\,.
\end{equation}
The SMF is also correlated with the LSS and their cross-variance includes a sample variance and SSC term:
\begin{equation}
    \mathrm{Cov}\left[\hat\Phi ^{(i)}(M^\mu_\star),\mathcal{O}_{a_1 a_2} (L)\right] = \mathrm{Cov}\left[\hat\Phi ^{(i)}(M^\mu_\star),\mathcal{O}_{a_1 a_2} (L)\right]_\mathrm{sva} +\; \mathrm{Cov}\left[\hat\Phi ^{(i)}(M^\mu_\star),\mathcal{O}_{a_1 a_2} (L)\right]_\mathrm{SSC}  \,.
\end{equation}
This is in analogy to \Cref{eq:linearmap_ell_to_obs} with the two contributions reading
\begin{equation}
    \mathrm{Cov}\left[\hat\Phi ^{(i)}(M^\mu_\star),\mathcal{O}_{a_1 a_2} (L)\right]_\mathrm{sva} = f^{(i)}\int\mathrm{d}\ell\; W_L(\ell) \int\mathrm{d}\chi\; \frac{p^{(i)}(\chi)}{p_\mathrm{tot}(\chi)}\frac{W_{A_1}(\chi)W_{A_2}(\chi)}{\chi^2} B_{\mathrm{cmm},\mu}\left(\frac{\ell +1/2}{\chi},z(\chi)\right)\,,
\end{equation}
with the bispectrum of counts and matter for a collapsed triangle configuration given by
\begin{align}
    B_{\mathrm{cmm},\mu}\left(k,z\right) = &\; \int\mathrm{d}M\; n_\mathrm{h}(M, z) \Phi(M^\mu_\star|M)\left(\frac{M}{\bar\rho_\mathrm{m}}\right)^2 \tilde u^2(k|M,z)
    \\ \nonumber & +\; 2P_\mathrm{lin}(k,z)\int\mathrm{d}M \; n_\mathrm{h}(M, z) \Phi(M^\mu_\star|M)\frac{M}{\bar\rho_\mathrm{m}} b_\mathrm{h}(M,z) \tilde u(k|M,z)
    \int\mathrm{d}M \; n_\mathrm{h}(M, z) b_\mathrm{h}(M,z) \tilde u(k|M,z)\,.
\end{align}
Finally, the SSC term is given by
\begin{equation}
    \mathrm{Cov}\left[\hat\Phi ^{(i)}(M^\mu_\star),\mathcal{O}_{a_1 a_2} (L)\right]_\mathrm{SSC} = A f^{(i)}\int\mathrm{d}\ell\; W_L(\ell) \int\mathrm{d}\chi \;\frac{p^{(i)}(\chi)}{p_\mathrm{tot}(\chi)}\frac{W_{A_1}(\chi)W_{A_2}(\chi)}{\chi^2}\frac{\partial P_{\mathrm{mm}}}{\partial \delta_\mathrm{bg}}\left(\frac{\ell+1/2}{\chi},z(\chi)\right)\sigma^2_{\mathrm{bg,} A}(\chi) \tilde\Phi_\mu(\chi)\,,
\end{equation}
with the power spectrum responses from \Cref{eq:responses}.

\section{The \mysc{OneCovariance} code}
\label{sec:basic_code}
This section aims to provide a brief rationale for the initial development of the code. For an overview of the code's general structure, please see \Cref{app:code_structure}. To enhance accessibility, a flowchart illustrating the typical workflow of the \mysc{OneCovariance} code \footnote{Available on \href{https://github.com/rreischke/OneCovariance}{https://github.com/rreischke/OneCovariance}.} is presented in \Cref{fig:flowchart}.

Several outstanding public codes are available for computing covariance matrices in harmonic space, notably packages like \mysc{CCL} \citep{chisari_core_2019} and its derived harmonic space covariance code \mysc{TJPCov}\footnote{\href{https://github.com/LSSTDESC/TJPCov}{https://github.com/LSSTDESC/TJPCov}}, \mysc{CosmoLike} \citep{krause_cosmolike_2017}, or \mysc{pySSC} \citep{lacasa_fast_2019}, offering comprehensive tools for constructing idealised harmonic space covariance matrices. For real space correlation functions, the excellent \mysc{COSMOCOV} \citep{fang_2d-fftlog_2020} utilises fast logarithmic Fourier transforms to compute real space covariance without the flat sky approximation adopted in this paper, rendering it extremely efficient for this purpose.

However, these tools are either focused on harmonic space or tailored to a specific observable or setup. Consequently, adapting the code to use different summary statistics, observables, or external inputs can be cumbersome. Integrating theory power spectra, trispectra, or power spectrum responses from files, or employing different weighting schemes such as the Bernardeau-Nishimichi-Taruya transformation \citep{bernardeau_cosmic_2014} for lensing efficiency, is not straightforward. Similarly, projecting an existing harmonic space covariance to a summary statistic presents challenges.

It is this need for input flexibility and user-friendliness that drove the development of the \mysc{OneCovariance} code. With this objective in mind, the code was designed to offer three key features:
\begin{enumerate}[(i)]
    \item \textit{Easy to use}: \mysc{OneCovariance} requires standard Python packages and includes its own \mysc{conda} environment to ensure stability. Running the code involves executing a single Python script, while code inputs are specified in a \mysc{.ini} (configuration) file read by the code using the \mysc{configparser} framework. The configuration file's design closely resembles that of \mysc{CLASS} \citep{lesgourgues_cosmic_2011,blas_cosmic_2011} or \mysc{CosmoSIS} \citep{zuntz_cosmosis_2015}. Sample configuration files, including a comprehensive \mysc{config.ini} file with detailed parameter explanations, are provided.
    
    \item \textit{Adaptability}: The \mysc{OneCovariance} code incorporates a default halo model and HOD-based prescription for biased tracer statistical properties. It communicates with \mysc{CAMB} \citep{lewis_efficient_2000, lewis_cosmological_2002,howlett_cmb_2012} for the matter power spectrum, providing both linear and nonlinear corrections. While these ingredients are modular and easily exchangeable, the code accepts almost all critical quantities as input files, enabling flexibility in various scenarios.
    \begin{enumerate}[(a)]
    \item Given an alternative HOD or a mass-concentration relation, various options are available. One can implement them in the \mysc{HOD-} and \mysc{halomodel-class}, or encode them and save them into a file to pass to the \mysc{OneCovariance} code. Alternatively, one can calculate the 3-dimensional power spectra, for instance via \mysc{CAMB}, \mysc{CLASS}, \mysc{HMCODE} and \mysc{baccoemu} \citep{arico_bacco_2021,angulo_bacco_2021} to name a few, to directly provide them to the code. It should be noted that the code itself communicates directly with \mysc{CAMB} and therefore provides all power spectra implemented there natively. 
    \item It is straightforward to provide harmonic space covariances, i.e., $C_\ell$, along with the SSC and NG contribution, for any tracer to the \mysc{OneCovariance} code and project them to an observable.
    \item Complete freedom is granted in choosing the summary statistics for each tracer. While four hard-coded cases are included: bandpowers, COSEBIs, real space correlation functions, and harmonic space covariance, any summary statistic can be passed as a file to the code, as long as it represents a linear transformation of the $C_\ell$. For example, measuring clustering with a real space correlation function, GGL with bandpowers and CS with COSEBIs.
    \item Consistency checks between different summary statistics are supported. Two summary statistics can be provided to any given tracer, enabling, for instance, the analysis of CS with both COSEBIs or bandpowers while accounting for their cross-covariance.
    \item Galaxy bias can be determined either by the default HOD prescription or by supplying a file containing the galaxy bias as a function of redshift for each lens bin considered, facilitating the incorporation of numerous (linear) bias models. For non-linear bias, one can pass the resulting power spectrum directly, as discussed in (b).

\end{enumerate}
This list only scratches the surface of the calculations achievable with the \mysc{OneCovariance} code. A comprehensive array of examples demonstrating the code's functionality can be found on the \mysc{readthedocs} webpage\footnote{\href{https://onecovariance.readthedocs.io/en/latest/index.html}{https://onecovariance.readthedocs.io/en/latest/index.html}}.

    \item \textit{Legacy}: In the KiDS collaboration, the steadfast aim has consistently been to utilise diverse sets of summary statistics and analytical methodologies to deduce cosmological parameters reliably, thereby attaining resilient constraints \citep[e.g.][]{asgari_kids-1000_2021}. Through the provision of a publicly accessible code, we empower the wider scientific community to conduct such analyses, not only with existing KiDS data but also with forthcoming datasets. Furthermore, our endeavour guarantees the applicability of the methodologies honed over years in the KiDS collaboration to future analyses.

\end{enumerate}
The code has been validated against the previously used covariance matrix code \citep{joachimi_kids-1000_2021} for all statistics \citep{asgari_kids-1000_2021} and tracers \citep{dvornik_unveiling_2018,dvornik_kids-1000_2023}, finding per cent agreement.
Furthermore, we compared the harmonic space lensing spectra against \mysc{CCL} for further cross-checks and found that they match below one per cent. We  show and discuss the results of these tests in \Cref{sec:comparison_codes}.

\begin{figure}
    \centering
    \includegraphics[width=0.42\textwidth]{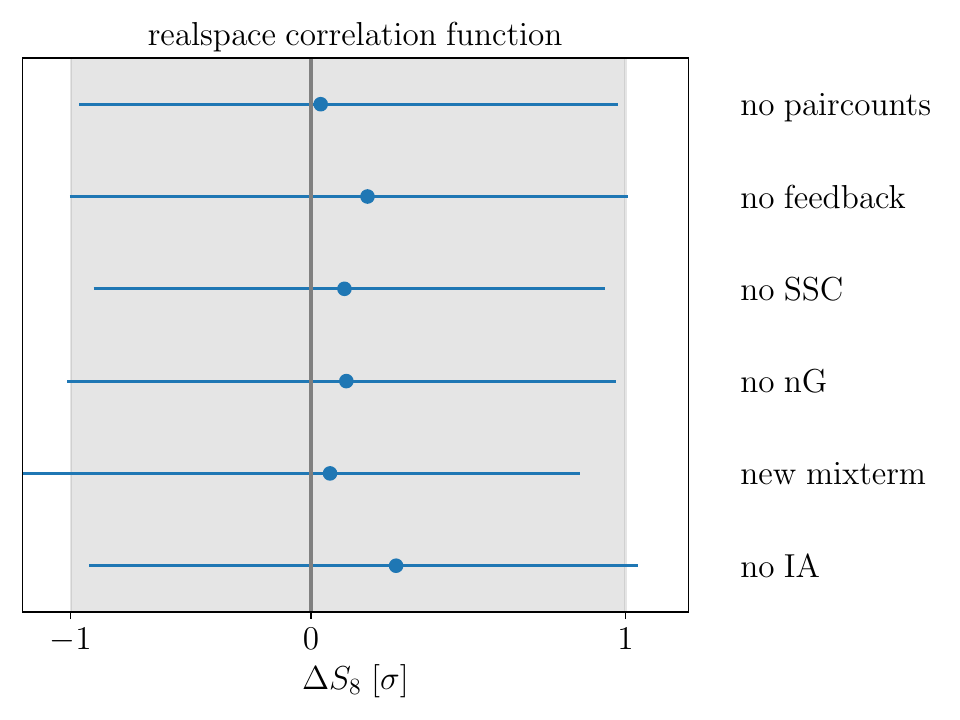}
    \includegraphics[width=0.42\textwidth]{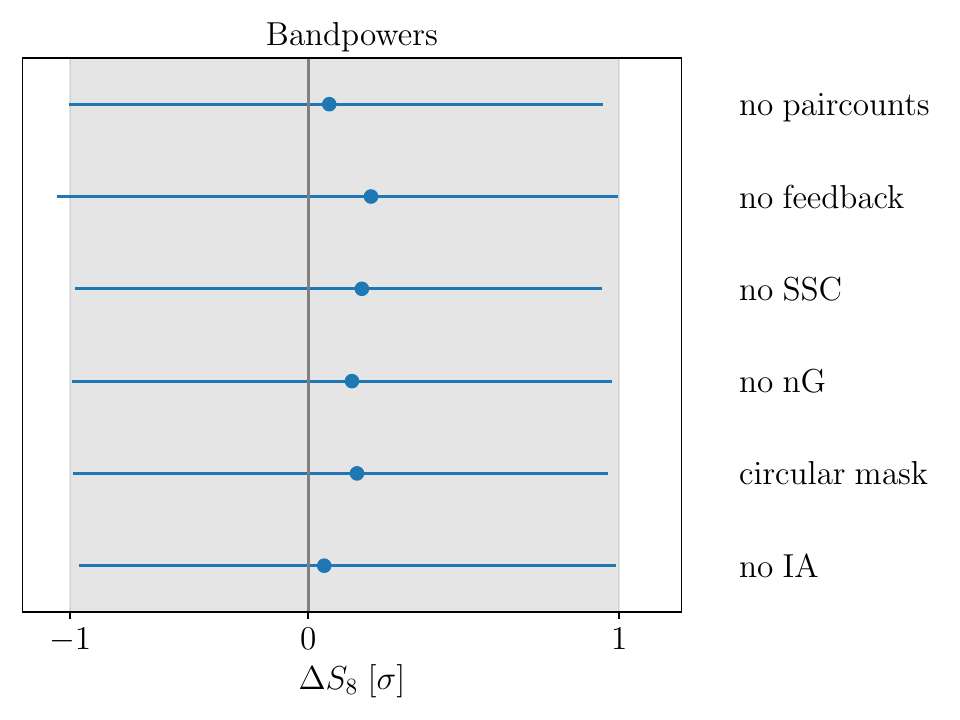}
    \caption{Effect of different choices for the covariance modelling for the inference of $S_8$ using the KiDS-Legacy data. Shown is the marginal maximum posterior and the corresponding 68$\%$ intervals (one $\sigma$) interval normalised to the fiducial settings assuming realistic pair counts and including feedback parametrised by $T_\mathrm{AGN}$ via \mysc{HMCODE2020} \citep{mead_hydrodynamical_2020}, the SSC and nG contributions, an idealised mixed term, a realistic survey mask, and the NLA model. Each blue data point replaces one of those assumptions at a time. \textit{Left}: Real space correlation functions. \textit{Right}: Bandpowers.}
    \label{fig:effect_covariance_on_inference}
\end{figure}

\section{The KiDS-Legacy covariance}
\label{sec:kids_legacy_sec}
\begin{figure}
    \centering
    \includegraphics[width =.9\textwidth]{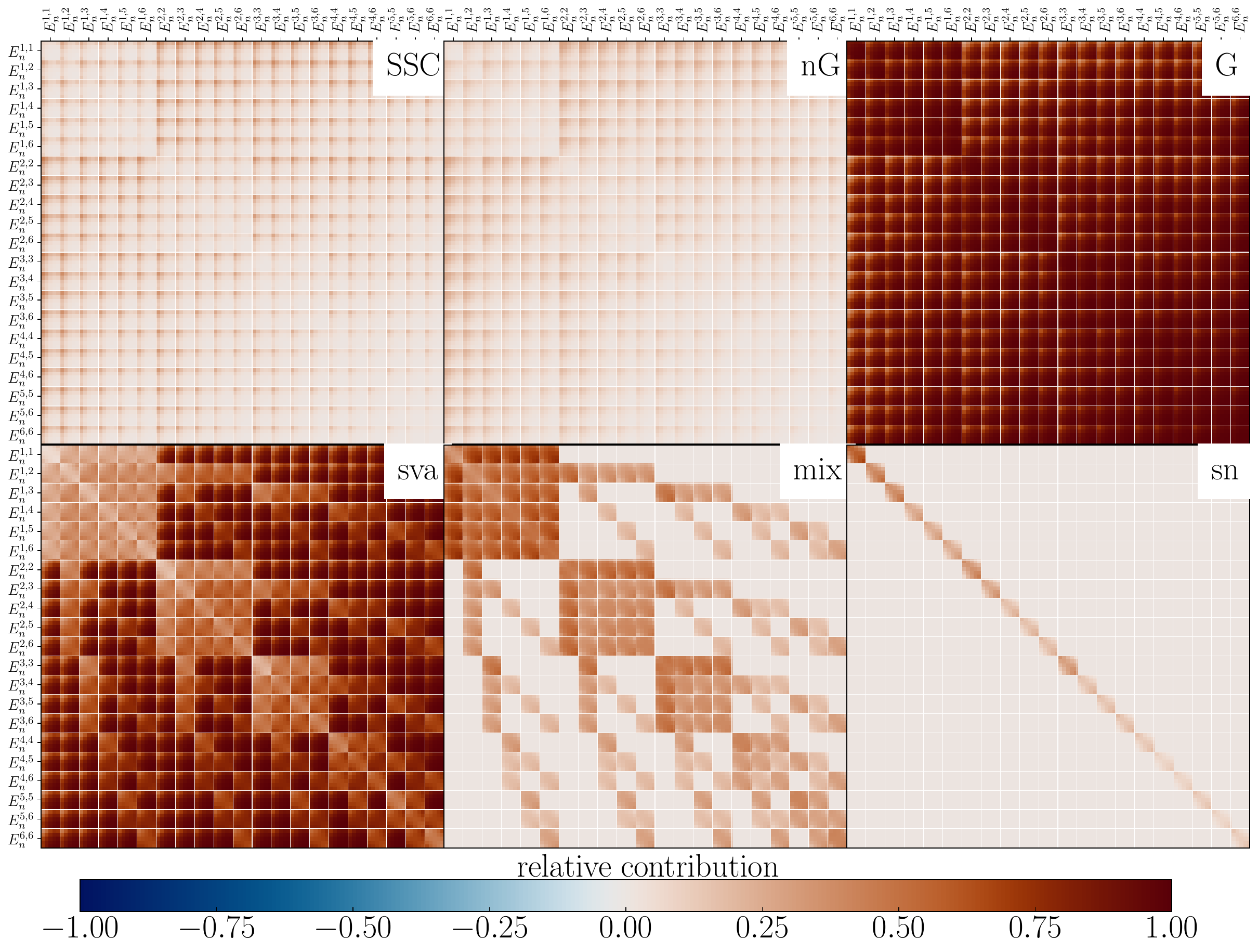}
    \caption{KiDS-Legacy-like covariance matrix for COSEBIs. The diagonals are in the following order $i\leq j$ with the tomographic bin indices $i$ and $j$ and in each little sub-block the order of the COSEBIs runs over $n=1,...,5$. Furthermore, we omit the $B$-mode signal here, since it is pure shape noise in the case of COSEBIs. The six different panels show the relative contribution of each term to the total covariance. In particular, we show the super-sample covariance (SSC), the non-Gaussian (nG), and the Gaussian (G) contribution in the upper three panels (compare \Cref{eq:covariance_general,eq:wichk}). The lower three panels show the three components of the Gaussian contribution, the sample variance (sva), the mixed term (mix), and the shape or shot noise term (sn), as described in \Cref{eq:Gaussian_split}.}
    \label{fig:terms_cosebis}
\end{figure}

\begin{figure}
    \centering
    \includegraphics[width = 0.42\textwidth]{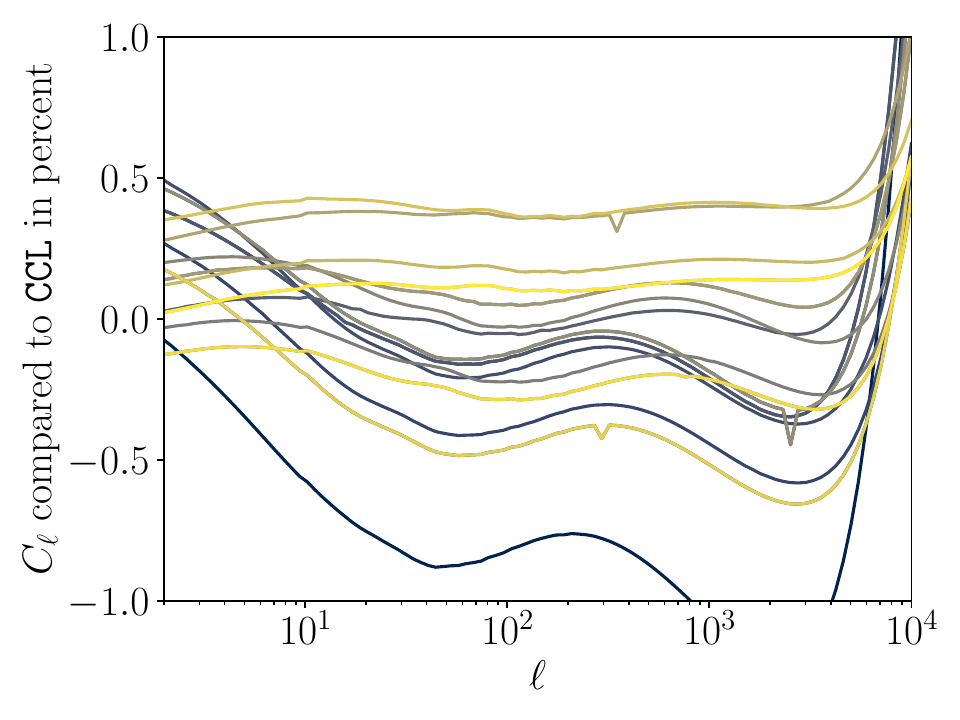}
    \includegraphics[width = 0.42\textwidth]{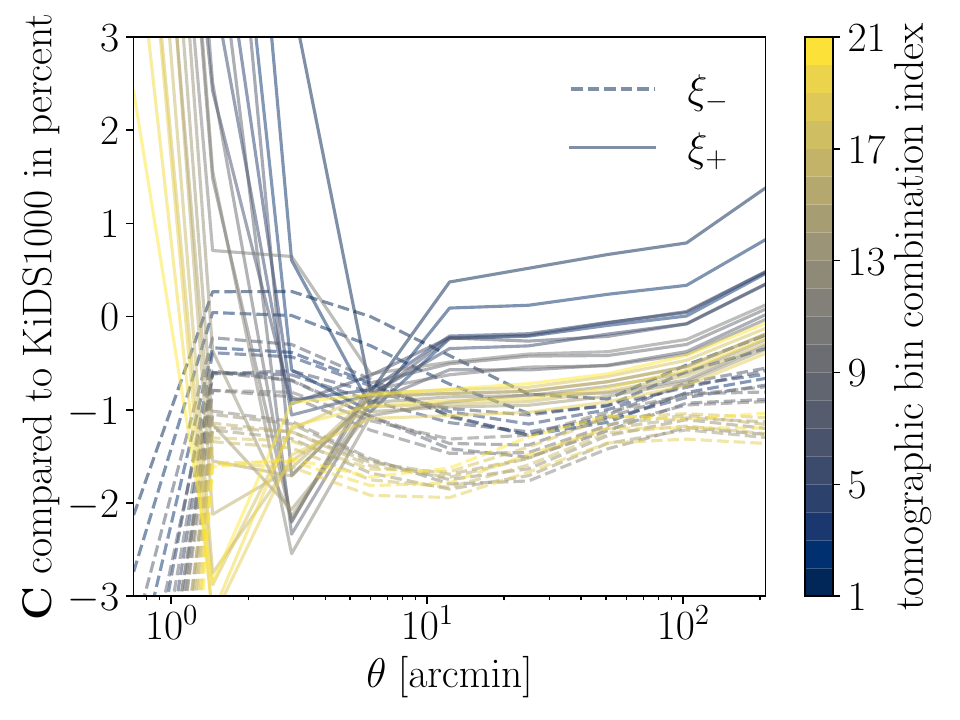}
    \caption{Cosmic shear setup with the six KiDS-Legacy tomographic bins. The colour bar shows the tomographic bin index combination, $I$ (i.e. for tomographic bins $i,\, j$ the {index} is $I =  \sum_\alpha^{i}\sum_{\beta=\alpha}^{j} $). \textit{Left}: Comparison between the \mysc{OneCovariance} code and \mysc{CCL}. The relative difference between the angular power spectrum calculation is shown as a percentage.  Matter power spectra have been modelled  using the \citet{takahashi_revising_2012} version of the halo model. \textit{Right}: Relative difference in per cent between the diagonal elements of the covariance matrix for $\xi_\pm$ calculated with the KiDS-1000 covariance code \citep{joachimi_kids-1000_2021} and the \mysc{OneCovariance} code. The covariance shown here does not contain shape noise.}
    \label{fig:comparison_to_previous_studies}
\end{figure}

In this section, we  provide a concise overview of a KiDS-Legacy-like CS sample and outline the error modelling strategies adopted for the associated CS legacy analysis. We  elucidate how these choices impact the inference of the structure growth parameter, $S_8 \coloneqq \sigma_8\sqrt{\Omega_\mathrm{m}/0.3}$, as illustrated in \Cref{fig:effect_covariance_on_inference}. Moreover, we   underscore the significance of the various terms comprising the covariance matrix, as demonstrated in \Cref{fig:terms_cosebis}. Finally, we   conduct a comparative analysis between the \mysc{OneCovariance} code and selected existing codes, as depicted in \Cref{fig:comparison_to_previous_studies}.

\subsection{KiDS-Legacy}
\label{sec:kids_legacy_definition}
The final data release of KiDS \citep{de_jong_kilo-degree_2013} and VIKING \citep[VISTA Kilo-degree INfrared Galaxy,][]{edge_vista_2013} is described in detail in \citet{wright_kids_dr5_2024} and is referred to as the fifth data release (\drfive). Covering a survey area of 1347$\;\mathrm{deg}^2$, it encompasses 9 photometric bands spanning from optical to near-infrared, with a $5\sigma$ limiting magnitude of 24.8 in the $r$-band. The footprint of the main survey includes a 4$\;\mathrm{deg}^2$ overlap with existing deep spectroscopic surveys. This data is complemented by an additional 23$\;\mathrm{deg}^2$ KiDS- and VIKING-like imaging over additional deep spectroscopic fields, yielding a total of about $126\,000$ sources with both spectroscopy and photometry, crucial for robust redshift calibration \citep{wright_kids_redshift_2024} across the {\drfive} footprint. Relative to the fourth data release \citep{kuijken_fourth_2019}, {\drfive} represents a roughly 30 per cent increase in area coverage and around 0.4 magnitude deepening in the $i$-band.

Significantly, the photometric redshift calibration has been refined, as outlined in \citet{wright_kids_redshift_2024}. Notably, a new matching algorithm enables the generation of highly realistic mock catalogues for both photometric and spectroscopic sources. Redshift calibration employs two techniques: colour-based self-organising maps (SOM) and a clustering redshift-based approach, leveraging extensively the spectroscopic samples from {\drfive}. The SOM method yields residual shifts in the mean redshift of each tomographic bin $\langle\delta_z\rangle \leq 0.01$, serving as a conservative error floor on the priors for the mean redshifts of each bin. These enhancements, combined with the {\drfive} data, facilitate the calibration of an additional tomographic bin with redshift up to $z\approx 2$, resulting in a total of six bins compared to five in the previous CS analysis in KiDS-1000. The KiDS-Legacy is then a subsample of the whole galaxy catalogue with certain quality requirements, for example for the shape measurements, as well as selection criteria such as masking or blended sources. This leads to a final catalogue of around $4\times 10^7$ galaxies for CS over an effective area of roughly 1000$\;\mathrm{deg}^2$.

\label{sec:existing_codes}

\subsection{KiDS-Legacy covariance}
\label{sec:klegacy_cov}
In this part, the covariance modelling choices in a KiDS-Legacy-like analysis as it will be carried out in \citet{wright_kids_legacy_2024} and \citet{stoelzner_kids_legacy_2024} are discussed. While the cosmology remains fixed to the fiducial values presented in \Cref{tab:KL_fid_cosm} for all plots, it is important to note that for the actual cosmological analysis, an iterative approach is adopted for covariance calculation. Initially, a fiducial covariance matrix is utilised, followed by maximising of the posterior and iterative updates to the covariance. This iterative process, typically converges after one iteration, as shown in \citet{van_uitert_kidsgama_2018}, resulting in final contours that exhibit negligible changes, as demonstrated in the subsequent results.

In \Cref{fig:effect_covariance_on_inference} we show how the inference of $S_8$ is affected by different modelling choices of the covariance for real space correlation functions and bandpowers (COSEBIs almost pick the same scales as bandpowers, so we do not show them here explicitly). The grey band corresponds to the 68 per cent interval together with the posterior maximum of the marginal distribution for the fiducial choice for the covariance modelling whose settings we are summarising in the following. We model the Gaussian term with an idealised mixed term (see \Cref{fig:mock_and_mix_term}), but shape noise is estimated from the catalogue. The covariance always includes a multiplicative shear bias uncertainty, since it is not marginalised over in the inference process via sampling. We include the non-Gaussian covariance and the SSC term with a non-binary mask with the realistic survey footprint to calculate its variance. We use the standard NLA model in the covariance as discussed in \Cref{sec:harmonic_space}, but stress again that the particular choice does not matter and that the code is flexible enough to deal with any alignment model on the Gaussian covariance level.
Lastly, the matter power spectrum is modelled using \mysc{HMCODE2020} \citep{mead_hydrodynamical_2020} with feedback controlled via $\log_{10} T_{\mathrm{AGN}}$. Each blue dot with error bars corresponds to an analysis where a specific attribute in the covariance modelling is varied. 

\Cref{fig:effect_covariance_on_inference} shows the influence of individual modelling choices on the marginal constraints on $S_8$  Nonetheless, it is important to acknowledge altering multiple fiducial assumptions about the covariance, e.g. ignoring the realistic pair counts, changing the IA model and removing the SSC contribution, could lead to a bias in the final value of $S_8$. However, ignoring the non-Gaussian and SSC contribution on its own has very little effect on the constraints on $S_8$. This is in particular reassuring for the non-Gaussian term, which is only accurate to 20 per cent \citep{joachimi_kids-1000_2021}. Especially the addition of the sixth tomographic bin in KiDS-Legacy compared to KiDS-1000 will make the nG contribution less pronounced as more linear scales with high signals are added to the analysis. Furthermore, due to the increase in effective area, the survey response is also smaller than in KiDS-1000. 

Interestingly, we find that the use of the idealised pair counts (i.e. not using the pair counts from the catalogue but instead the analytic formula) has little effect on the $S_8$ inference. This might be surprising at first, since using the analytic pair counts will change diagonal elements of the covariance matrix by up to 20 per cent \citep{troxel_des_2018}. However, we can explain this by the fact that the signal-to-noise ratio of the measurement in this case is almost unchanged. The reason for this is that changing the variance will also change the amount of correlation between the different data points. So if the analytic pair counts over (under) estimate the number of pairs, the variance in the fiducial setting will be larger (smaller), which makes the individual data point less significant for the inference. However, this effect also reduces (increases) correlations between the different data points, thus increasing the information content. Those two effects seem to balance each other such that the $S_8$ constraints do not change. Interestingly, we also find that using a circular mask instead of the real footprint has the same effect as ignoring the SSC term altogether. IA seems to have a larger effect on real space correlation functions than on bandpowers which could be indicative of the fact that this is driven by contributions from large multipoles to $\xi_\pm$. At the same time, however, we find that the case with no feedback changes both real space correlation functions and bandpowers in the same way. In general, we find the consistent trend that contributions decreasing the covariance will bias the constraint on $S_8$ to larger values, as expected by signal-to-noise considerations. It should also be highlighted that, while the marginal posterior maximum can shift, the upper limit of the 68 per cent interval is very robust. Here we find the largest effect from the new mixed term (compare \Cref{fig:mock_and_mix_term,app:mixed_term}), since it strongly changes the off-diagonal elements of the covariance matrix while leaving the diagonals untouched. We note that we do not include the new mixed term in the COSEBIs or bandpower covariance matrix at this stage, since they are mapped directly from harmonic space and not from real space to avoid discretisation effects or the computation of a very finely sampled real space covariance matrix. 

All those considerations are of course very specific to the KiDS-Legacy survey. They show, however, that our inference of $S_8$ and in particular the upper limit of the corresponding error is very robust when changing the covariance modelling. While the impact of the mixed term seems very daunting in the light of next-generation surveys it should be highlighted that KiDS is a ground-based, single-visit survey, thus being much more inhomogeneous than \textit{Euclid} or LSST, which are either in space or scan the same patch of the sky multiple times respectively. It is thus expected that the idealised mixed term is more accurate for these surveys.

In \Cref{fig:terms_cosebis}, we scrutinise the different contributions of the covariance matrix for the KiDS-Legacy analysis, focusing on the COSEBIs covariance. Since we have shown $\xi_\pm$ and bandpowers so far, we discuss this using the COSEBIs covariance. Specifically, we plot the Pearson correlation coefficient.
The structure of each of the six subplots is the following: the COSEBI mode, $n$, varies in each small square from $n = 1,\dots,5$ while the tomographic bin combination $i\leq j$, from one square to the next. We also do not show the covariance of the $B_n$ modes, as they are only given by the shape noise contribution. Lastly, we only include cosmological contributions and do not consider the multiplicative shear bias uncertainty. The six large subsquares of \Cref{fig:terms_cosebis} show the six terms of the covariance matrix discussed in \Cref{sec:general_considerations}. The covariance is dominated completely by the Gaussian term, with the SSC and nG contribution only making up to around 35 and 15 per cent each on the off-diagonals. The Gaussian contribution itself is dominated by the sample variance terms on the off-diagonals, while the shape noise and the mixed term dominate the diagonals, with relative importance increasing towards tomographic bins at higher redshift.

\subsection{Comparison with selected existing codes}
\label{sec:comparison_codes}
The analytical covariance utilised in KiDS-1000 has undergone extensive testing against mock data, as detailed in \citet{joachimi_kids-1000_2021}. While some of these tests are reiterated in the following sections using updated simulations, we   present a selection of results to validate the \mysc{OneCovariance} code.

Given that the \mysc{OneCovariance} code interfaces with \mysc{CAMB} for linear and non-linear matter power spectra, and other codes like \mysc{CCL} employ a modified version of our halo model trispectrum implementation, we do not validate these quantities directly. The first step of our validation is therefore to compare the CS angular power spectra, or in other words, the line-of-sight projection. On the left side of \Cref{fig:comparison_to_previous_studies} we demonstrate the relative accuracy of the \mysc{OneCovariance} code compared to \mysc{CCL}. A KiDS-Legacy-like setup with six source bins was used and the non-linear matter power spectrum has been calculated using \mysc{halofit} \citep{takahashi_revising_2012}. The tomographic bin combination index is colour-coded, that is labelling each bin combination, $i\leq j$ with a unique number starting at one.
We find that the agreement is well below one per cent, except for the very first tomographic bin. This difference originates from slightly different extrapolations used for the non-linear matter power spectrum (as can be seen by the steep rise at large multiples) as well as the kind of interpolation used for the source redshift distributions. \review{The small visible spikes are of order $10^{-3}$ and can be traced back to different accuracy settings when CAMB is called via \mysc{CCL} or the \mysc{OneCovariance} as well as slightly different interpolation schemes.} However, we find excellent agreement with \mysc{CCL}.

Another critical validation step involves comparing our code with the covariance code used in the KiDS-1000 analysis. The right panel of \Cref{fig:comparison_to_previous_studies} displays the percentage difference between the \mysc{OneCovariance} code and the previously employed code for the diagonal elements of the $\xi_\pm$ covariance matrix, excluding shape noise. We conducted this comparison across nine logarithmically spaced angular bins ranging from  0.5 and 300$\;\mathrm{arcmin}$. Once again, we observe excellent agreement, with differences within a few per cent. Slightly larger disparities are noted at small angular scales, primarily attributed to differences in integration accuracy and cut-off limits for the integrals, particularly for the non-Gaussian (nG) and super-sample covariance (SSC) contributions. However, these discrepancies, while reaching up to 15 per cent, are not significant as small scales are shot-noise dominated, thereby minimally impacting previous analyses. Similar validation procedures were repeated for bandpowers and COSEBIs, yielding agreement below a per cent with previous studies \citep{joachimi_kids-1000_2021, asgari_kids-1000_2021}. Notably, due to the more compact Fourier filters, the agreement for these cases is even better than for real space statistics. Additionally, we verified that we obtain identical results when passing Fourier filter functions externally to the code for different statistics (for how to pass external information, see \Cref{app:code_structure}).

\begin{table}
\renewcommand{\arraystretch}{1.15}
    \centering
    \begin{tabular}{lcc}
        \hline
        Parameter & Symbol & Value \\
        \hline\hline
        Hubble constant & $h$ & 0.670 \\
        Cold dark matter density & $\Omega_{\mathrm{c}}$ & 0.276 \\
        Baryonic matter density & $\Omega_{\mathrm{b}}$ & 0.050 \\
        Curvature density & $\Omega_{\mathrm{k}}$ & 0 \\
        Spectral index & $n_{\mathrm{s}}$ & 0.960 \\
        Matter fluctuation amplitude at 8 Mpc $h^{-1}$ & $\sigma_{8}$ & 0.786 \\
        Baryon feedback amplitude & $A_{\mathrm{bary}}$ & 3.1 \\ 
        Sum of the neutrino masses & $\sum m_{\nu}$ & 0.06 eV \\
    \end{tabular}\vspace{.1cm}
    \caption{Fiducial choice of cosmological parameters within $\Lambda$CDM used for the forward simulations to test the signal and noise modelling for KiDS-Legacy-like. The given cosmology is equivalent to $S_{8} \equiv \sigma_{8} [(\Omega_{\mathrm{b}}+\Omega_{\mathrm{c}})/0.3]^{1/2} =  0.82$. The parameters are set in accordance with the constraints from KiDS-1000 \citep{asgari_kids-1000_2021}, while $\sigma_{8}$ is set to be the mid-value between KiDS-1000 and Planck \citep{planck_collaboration_planck_2020}.}
    \label{tab:KL_fid_cosm}
\end{table}

\section{Validation against simulations}
\label{sec:sec_mock}
Testing the CS signal and noise modelling for the final release of the KiDS data, {\drfive}, necessitates a substantial number of realisations of the data to accurately characterise the covariance. Achieving this requires forward simulations that closely resemble real data, incorporating factors such as the survey mask and variable depth. In this section, we detail the construction of the forward simulations employed to validate the error modelling in KiDS-Legacy. \Cref{sec:glass_simulations} provides an overview of the log-normal mocks and \Cref{sec:legacy_mocks} briefly describes how the mocks are populated with galaxies and how the shape catalogues are created.
Lastly, in \Cref{sec:simulation_comparison} we compare the simulations against the analytical covariance. 

\begin{figure}
    \centering
    \includegraphics[width= 0.95\textwidth]{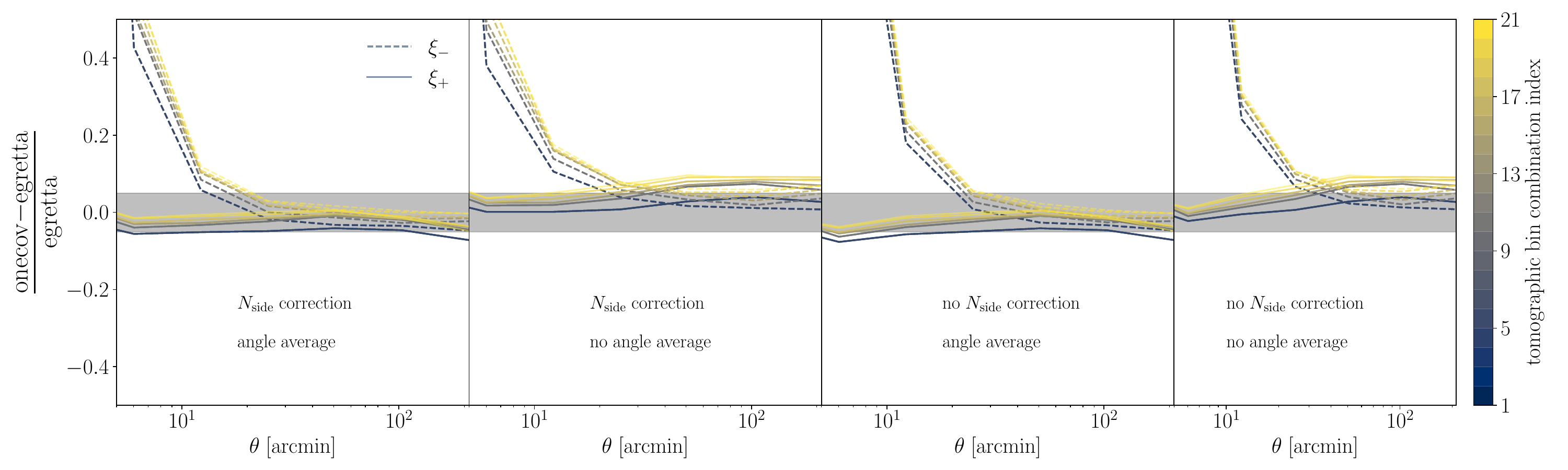}
    \caption{Relative difference between the signal measured in the 4224 \textit{Egretta} mocks (realistic mask and depth variations) with the signal prediction of \mysc{OneCovariance} code. The colour bar indicates the different unique tomographic bin combinations, the same as in \Cref{fig:comparison_to_previous_studies}. The dashed lines show $\xi_-$ and solid lines $\xi_+$. The grey band indicates a five per cent relative difference. The different plots show varying settings in the \mysc{OneCovariance} code. In particular, we distinguish whether the averaging over the $\theta$ bin (see Equation \ref{eq:bin_average_theta}) is carried out and if the pixel window due to the finite resolution of the \mysc{healpix} map is taken into account, i.e. damping power on small scales ($N_\mathrm{side}$ correction).}
    \label{fig:mock_onecov_signal}
\end{figure}

Central to our assessment are key comparison figures that directly juxtapose our most realistic mock with the \mysc{OneCovariance} code. As all summary statistics can be derived from the real space correlation function, we only measure those on the mocks.  \Cref{fig:mock_onecov_signal,fig:mock_and_mix_term,fig:diag_compare_egretta} show the comparison for the signal, the correlation coefficient and diagonal covariance matrix elements respectively.

\subsection{GLASS simulations}
\label{sec:glass_simulations}
The forward simulations for covariance validation in this study are founded on the KiDS-SBI\footnote{\href{https://github.com/mwiet/kids_sbi}{https://github.com/mwiet/kids\_sbi}} suite of simulations detailed in \citet{von_wietersheim-kramsta_kids-sbi_2024}. At the core of the pipeline lie log-normal mocks of the 3-dimensional matter distribution in radial shells, which are subsequently projected along the line of sight to create shear maps. This process is facilitated within the GLASS\footnote{\href{https://glass.readthedocs.io/stable/}{https://glass.readthedocs.io/stable/}} (Generator for Large-Scale Structure) framework \citep{tessore_glass_2023}. GLASS generates log-normal realisations to match the two-point statistics of the given input power spectrum. While higher-order statistics are not precisely recovered due to the inherently non-log-normal nature of the cosmic density field, this nonlinear transformation captures a significant portion of the trispectrum \citep{friedrich_dark_2020,hall_non-gaussian_2022} relevant for the covariance matrix.

Within this framework, $4224$ simulations are created for two distinct survey configurations, akin to the one selected in \citet{joachimi_kids-1000_2021}, a discussion of which is provided in further detail in \Cref{sec:legacy_mocks}. The cosmological parameters remain fixed at the values specified in \Cref{tab:KL_fid_cosm}. The 3-dimensional power spectrum of the input matter is computed using \mysc{CAMB} \citep{lewis_efficient_2000,lewis_cosmological_2002,howlett_cmb_2012}, with non-linear corrections computed from \mysc{HMCODE2020} \citep{mead_hydrodynamical_2020}. After computing the angular power spectra of the matter distribution in 22 shells of roughly equal thickness spanning the KiDS-Legacy redshift range, GLASS generates corresponding log-normal matter fields in each shell. The shell thickness ranges between 100 and 200$h^{-1}\mathrm{Mpc}$, ensuring a reasonably accurate description of the matter field as log-normal \citep{hall_non-gaussian_2022,piras_fast_2023}, while also being sufficiently finely sampled to accurately capture the matter overdensity along the line of sight.
The computationally intensive integration for the $C_\ell$ in the matter shells is executed using Levin's method \citep{levin_fast_1996,zieser_cross-correlation_2016,leonard_n5k_2022}. Subsequently, the concentric shells are integrated along the line of sight, weighted by the lensing efficiency function \citep[refer to][for detailed information]{tessore_glass_2023}, resulting in a CS field realisation covering the entire sky. It is noteworthy that intrinsic alignments are not included in the covariance matrix validation. However, this omission does not compromise the validity of our tests, given that intrinsic alignment exerts minimal influence on the covariance matrix (see \Cref{fig:effect_covariance_on_inference}) even if removing it completely and generally shares the same functional form as the lensing signal, thus being equally well accounted for.

\subsection{KiDS-Legacy-like mocks}
\label{sec:legacy_mocks}
Given a realisation of the density and shear fields, galaxies are sampled within those, including systematic effects such as the survey footprint and the effect of variable depth on the redshift distribution and the shape noise with the help of SALMO
\citep[Speedy Acquisition for Lensing and Matter Observables,][]{joachimi_kids-1000_2021}. 

To create a realistic number density of galaxies, we utilise organised randoms \citep[ORs, introduced in][]{johnston_organised_2021}. ORs employ self-organising-maps \citep[SOMs,][]{kohonen_self-organized_1982} to generate random samples that follow the spatial variability induced by a high-dimensional systematic space. SOMs are unsupervised neural network algorithms which are designed to map a high-dimensional data space onto a two-dimensional map while preserving topological features of the input data space. We define a systematic data vector including atmospheric seeing, sky background light, stellar number density, dust extinction and variations in the point-spread function. The SOM provides a map of Tiaogeng \citep[TG, see][for a detailed explanation]{yan_kidslegacy_2025} weights, $w_\mathrm{TG}$, as a function of pixel on the sky. The TG weights quantify the anisotropic selection function on the galaxy sample observed within KiDS as a function of the aforementioned observational systematics in a manner that is independent of the cosmological signal in the data. We found the TG weights to be a good indicator of the local seeing, atmospheric transparency and signal-to-noise, so it can be considered a good estimator of variable depth. In particular, the TG weights are uncorrelated with the $r$-band magnitude measurements of galaxies and their photometric redshift, while being highly correlated with the magnitude limit in the $r$-band which KiDS uses for shape measurements, the degree of background noise, and the PSF variations. Hence, $w_\mathrm{TG}$ can robustly predict the local observational depth independently of the galaxy position and shape measurements. 
Next, we partition each of the six tomographic bins into ten equi-populated bins in $w_\mathrm{TG}$.
For each $w^{i}_\mathrm{TG}$ we independently recompute the effective number density, $n_\mathrm{eff}$, and the total ellipticity dispersion $\sigma_\epsilon$, from the galaxy population. It is possible to linearly interpolate $n_\mathrm{eff}$ and $\sigma_\epsilon$ as a function of $w_\mathrm{TG}$ allowing the evaluations of both quantities at every pixel in the KiDS-Legacy-like footprint from the single spatial $w_\mathrm{TG}$ map. We use a footprint of approximately $1000\;\mathrm{deg}^2$ with an $N_\mathrm{side} = 1024$.

The redshift calibration itself uses the following photometric bin boundaries: $[0.1,\, 0.42,\, 0.58,\, 0.71,\, 0.9,\, 1.14,\, 2)$, adding a sixth tomographic bin in KiDS-Legacy compared to KiDS-1000 
(see \Cref{sec:kids_legacy_definition})
By applying this binning scheme to the mock catalogues, we obtain the redshift distributions discussed above from the SOM, mapping the observed photometric redshifts to spectroscopic redshifts based on a reference spectroscopic sample \citep{wright_photometric_2020,hildebrandt_kids-1000_2021}. 

Having set up the forward simulations, the two-point correlation functions $\xi_\pm$ are measured on each realisation using \mysc{TreeCorr} \citep{jarvis_skewness_2004}, following the same implementation described in \citet{giblin_kids-1000_2020, joachimi_kids-1000_2021}. Initially, $\xi_\pm$ are measured in the north and south separately using 300 logarithmically spaced bins in $\theta\in[0.5,\,300]\;\mathrm{arcmin}$. The estimated correlation functions, $\hat{\xi}_\pm$ are then combined as a weighted mean,
\begin{equation}
    \hat{\xi}_{\pm}^{(i j)} (\theta) = \frac{N^{(i j)}_{\mathrm{pairs}, \, \mathrm{N}} \, \hat{\xi}_{\pm, \, \mathrm{N}}^{(i j)}(\theta) + N^{(i j)}_{\mathrm{pairs}, \, \mathrm{S}} \, \hat{\xi}_{\pm, \, \mathrm{S}}^{(i j)}(\theta)}{N^{(i j)}_{\mathrm{pairs}, \, \mathrm{N}} + N^{(i j)}_{\mathrm{pairs}, \, \mathrm{S}}}\,,
\end{equation}
\noindent with the weighted galaxy pairs and measured correlation functions in tomographic bins $i$ and $j$ in the north (N) and south (S) fields respectively. Lastly, the finely binned correlation functions are then re-binned into the final nine logarithmically spaced $\theta$ bins covering the same range. 

In \citet{joachimi_kids-1000_2021}, three kinds of mocks, \textit{Buceros, Cygnus} and \textit{Egretta}, with increasing complexity were distinguished.
\textit{Buceros} is an idealised survey with a simple rectangular footprint and a homogeneous number density which serves as a baseline check of the covariance modelling, which was already found to agree within a few per cent with the analytical covariance \citep{joachimi_kids-1000_2021}. Given the agreement of the \mysc{OneCovariance} with the KiDS-1000 covariance matrix, see \Cref{fig:comparison_to_previous_studies}, this mock as such is not considered. In case of \textit{Cygnus}, we mask the simulations according to a realistic KiDS-Legacy-like mask. However, the survey is kept homogenous in $n_\mathrm{eff}$ and $\sigma_\epsilon$. In particular, the intrinsic ellipticities are sampled from a Gaussian distribution with fixed $\sigma_\epsilon$ and zero mean for each tomographic bin.
In the most complex mock, \textit{Egretta}, the homogeneity and isotropy assumption in $n(z)$, $n_\mathrm{eff}$ and $\sigma_\epsilon$ are dropped, and they are sampled from realisations in the $w_\mathrm{TG}$ maps. Therefore, \textit{Egretta} is a realistic representation of the real KiDS-Legacy-like catalogue, including all systematics discussed before. It will therefore be used as the benchmark for the analytical covariance matrix. For more details on the simulations, we refer the reader to \citet{joachimi_kids-1000_2021}.

\begin{figure}
    \centering
    \includegraphics[width =.45\textwidth]{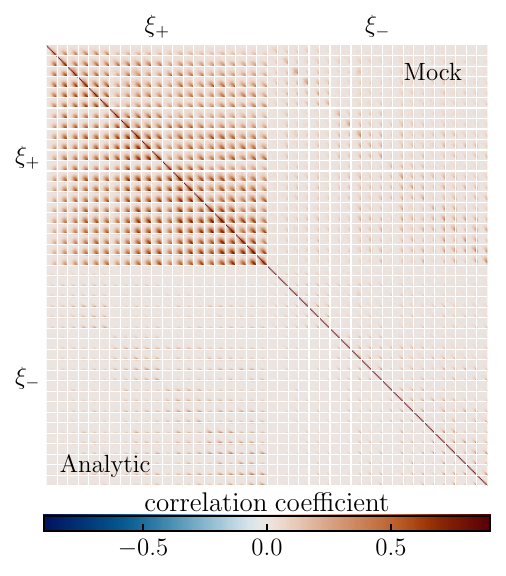}
    \includegraphics[width =.45\textwidth]{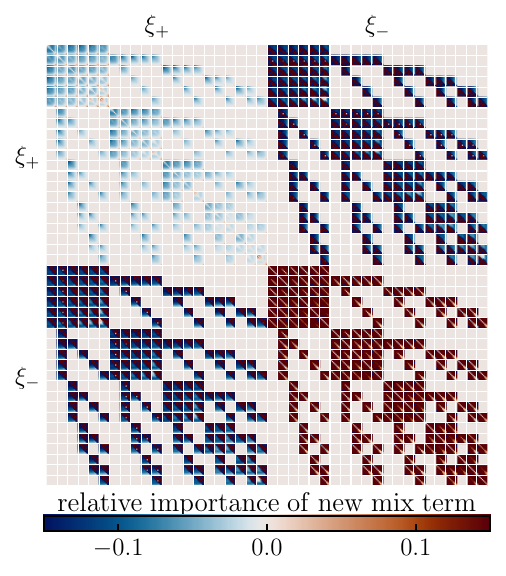}    
    \caption{KiDS-Legacy covariance matrix for real space correlation functions. Each square represents a unique combination of tomographic bins and contains the covariance for that combination in nine angular bins between 0.5 and 300$\;\mathrm{arcmin}$. The diagonals are in the order $i\leq j$ with the tomographic bin indices $i$ and $j$. \textit{Left}: Correlation coefficient of the full covariance matrix without multiplicative shear bias uncertainty. In the lower triangle, the analytic covariance is shown, while the upper triangle shows the covariance as measured in the \textit{Egretta} mocks (realistic mask and depth variations). \textit{Right}: Relative impact of the new mixed term (see \Cref{app:mixed_term}) on the Gaussian part covariance matrix compared to an idealised mixed term with a homogeneous pair and triplet counts. The latter were directly estimated from the KiDS-Legacy-like catalogue.}
    \label{fig:mock_and_mix_term}
\end{figure}

\subsection{Comparison with mocks}
\label{sec:simulation_comparison}

We   limit the discussion to the most realistic mock, \textit{Egretta}, since a general comparison between the different mock settings was already done in \citet{joachimi_kids-1000_2021} which the \mysc{OneCovariance} code was validated against. 
Having said that, we want to briefly summarise those results here again. For the \textit{Buceros} mock with an idealised footprint and homogeneous survey properties, we find agreement and the percentage with slightly larger difference at large separations. Due to the more complex survey mask in \textit{Cygnus}, those difference become more pronounced, leading to ten per cent differences at 300$\,\mathrm{arcmin}$ separation. At smaller scales, however, the difference is still well below five per cent.

Before delving into the covariance comparison with \textit{Egretta} we show the 
 relative difference between the signal measured in the simulations for $\xi_\pm$ and calculated from theory via the \mysc{OneCovariance} code is shown in \Cref{fig:mock_onecov_signal}. Dashed lines represent $\xi_-$ and solid lines $\xi_+$ while the colour bar indicates the tomographic bin index combination. The four panels show different settings of the theory calculation. In particular, the resolution of the simulation is partially mimicked by taking the pixel window into account which dampens the power on scales, $  \ell\leq N_\mathrm{side}$ and we explicitly average over angular bins in which the correlation function is measured. 
The grey band shows a 5 per cent difference between the simulation and the theory predictions. In general, we see good agreement between the theoretical predictions and the simulation, within 5 per cent. \review{Incorporating the damping of small-scale power into the theoretical predictions slightly enhances the agreement on small angular scales. However, because of the broad filter function in Fourier space of $\xi_\pm$, the small angular bins necessitate information from relatively small scales and are thus not accurately represented in the simulation. This effect is particularly noticeable for $\xi_-$ which requires integration up to $\ell > 10^4$ to reach convergence for $\theta< \mathrm{arcmin}$, although it is not evident for $\xi_+$ at the scales displayed in this figure for clarity. For $\xi_+$, this effect can be observed at scales below a few arcminutes.
Nevertheless, for the covariance comparison, this limitation should not present a significant issue, as the covariance will be predominantly influenced by shape noise on small scales.}
Additionally, we observe that bin-averaging introduces differences of a few per cent across all angular bins, with no significant variation as a function of $\theta$, consistent with expectations due to the logarithmic binning scheme.

\begin{figure}
    \centering
    \includegraphics[width=.85\textwidth]{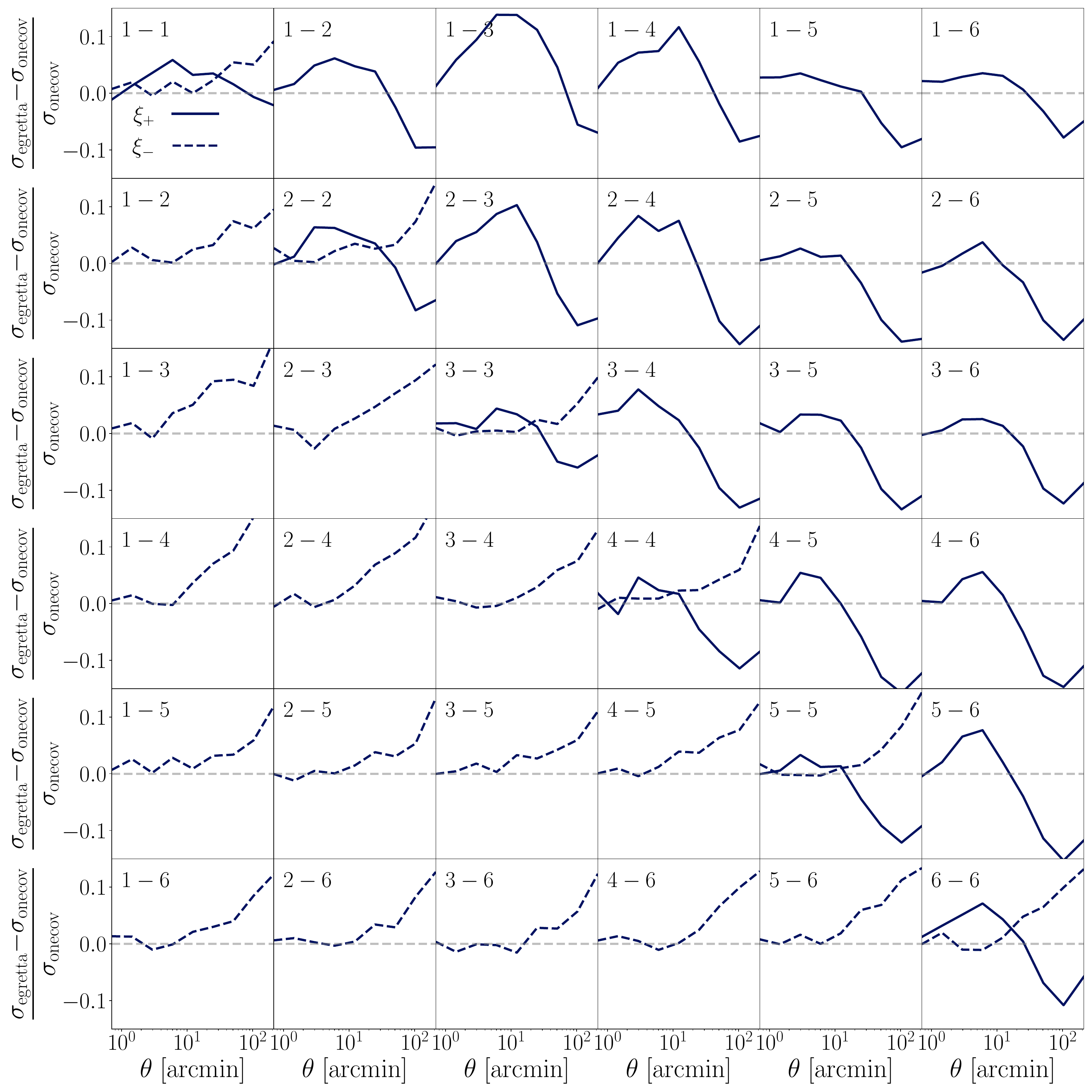}
    \caption{Relative difference between the standard deviation of the \textit{Egretta} mocks (realistic mask and depth variations) and that from the analytic covariance matrix from the \mysc{OneCovariance} code for the same setting as \Cref{fig:mock_and_mix_term}. The dashed lines show $\xi_-$ and solid lines depict $\xi_+$. The tomographic bin combination is shown in the top left corner of each  subplot.}
    \label{fig:diag_compare_egretta}
\end{figure}

On the left side of \Cref{fig:mock_and_mix_term}, we present the covariance for the setup outlined in \Cref{sec:legacy_mocks}. The lower triangle displays the analytic prediction, while the upper triangle shows the covariance estimated from the mock data. Visually, there is excellent agreement between the simulation and theory for the correlation coefficient. The elements are ordered in such a way that each square corresponds to a unique combination of tomographic bins, arranged in increasing order such that $i\leq j$.
Overall, it is evident that the covariance is dominated by shape noise. However, for the tomographic bins at higher redshifts, as discussed in more detail in \Cref{sec:klegacy_cov}, other contributions become more pronounced.

 The right panel of \Cref{fig:mock_and_mix_term} illustrates the impact of this modelling choice for the mixed term on the covariance matrix. Specifically, it shows the relative difference between these two cases for the Gaussian covariance matrix, i.e. considering an idealised mixed-term with homogeneous number density and no masking effects relative to using the triplet counts directly from the catalogue (compare \Cref{eq:Gaussian_split} and \Cref{app:mixed_term} respectively).
 Generally, we observe that the effect is most pronounced at large or small angular separations, while intermediate separations are less affected. This behaviour is expected because large scales are influenced by the extent of the survey, while very small scales can be affected by masking effects within the survey area. Additionally, the effect on the diagonal is less noticeable due to the contribution of shape noise.

For $\xi_+$, the effect of the updated mixed term reaches a maximum of 10 per cent. However, for $\xi_-$ and its cross-correlation with $\xi_+$, the effect can be significant for individual elements of the covariance matrix, exceeding 50 per cent. Nonetheless, since the signal strength for $\xi_-$ is relatively small, this effect is mitigated by the shape noise (its impact on inference is discussed in \Cref{sec:klegacy_cov}). It is worth noting that although there are individual points showing a substantial effect in the cross-covariance between $\xi_+$ and $\xi_-$, this is attributed to numerical noise and does not influence the results since the corresponding covariance matrix elements are very small.

We anticipate this effect to be less pronounced for larger and more homogeneous surveys than KiDS-Legacy. Surveys like \textit{Euclid} and LSST are expected to have fewer spatial variations in the number density of sources. This statement, however, bears the caveat that the other components of the covariance matrix will become more significant for these surveys as they are deeper, resulting in a stronger CS signal compared to the shape noise. Therefore, a concluding analysis of the impact of the mixed term in a stage 4 survey will be the subject of future work.

\begin{figure}
    \centering
    \includegraphics[trim={2.5cm 1cm 2.5cm 0},clip, width = .7\textwidth]{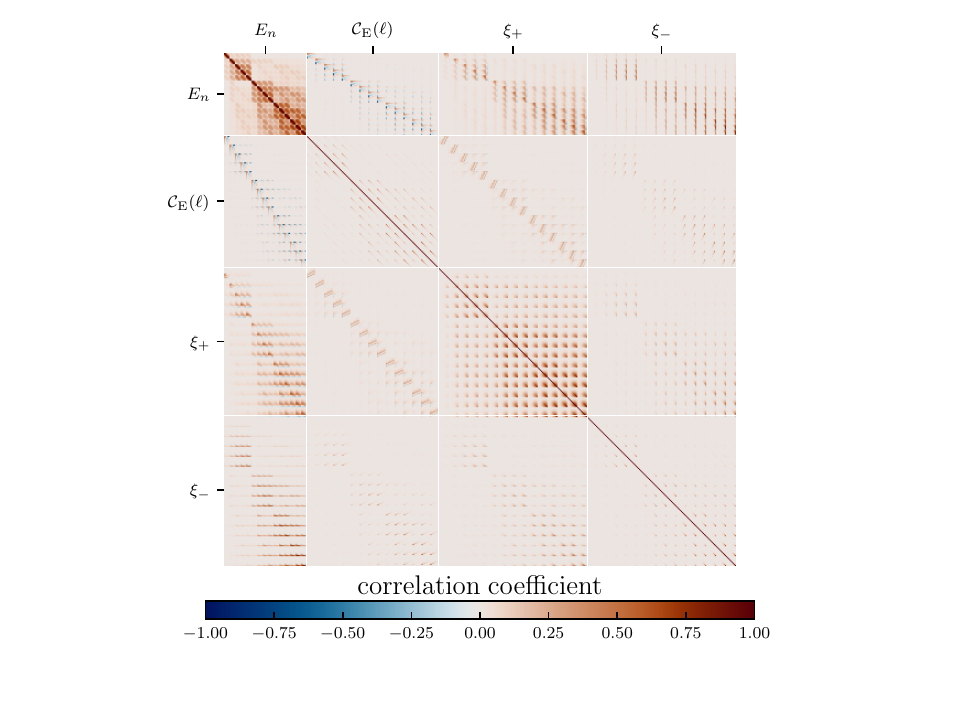}
    \caption{Showcase of the consistency tests possible with the \mysc{OneCovariance} code. This is a cosmic shear setting as in KiDS-1000 \citep{asgari_kids-1000_2021} (i.e. five tomographic bins, eight band power bins, five modes for COSEBIs, and nine bins for $\xi_\pm$. The code was used to create the full covariance matrix for any pair of those three statistics. Since these are all summary statistics for the same tracer and almost over the same physical scales, the matrix is close to being singular. The full matrix shown here has a single negative eigenvalue originating from numerical noise. However, the subcovariance matrices between each pair of summary statistics are still positive-definite. We do not show the B modes for COSEBIs and bandpowers since they mainly consist of shape noise (although they are still highly correlated with the shape noise of the other statistics).}
    \label{fig:consistency_test}
\end{figure}

 Finally, \Cref{fig:diag_compare_egretta} illustrates the relative difference in the standard deviation, i.e. the square root of the diagonal elements of the covariance matrix,  between \textit{Egretta} and the \mysc{OneCovariance} code. There is excellent agreement at small angular scales due to the contribution of shape noise, as the pair counts are matched to those measured in the \textit{Egretta} mocks. 
 It is worth noting that we address the most significant effect of variable depth, which generally changes the shape-noise in KiDS-Legacy compared to a uniform galaxy distribution. This effect is driven by the distribution of $w_{\mathrm{TG}}$, which skews above the mean, resulting in most pixels in the field being under-dense compared to the mean galaxy density. Generally, the agreement between the mocks and the analytical covariance is of a slightly better level than in previous analysis \citep{joachimi_kids-1000_2021}. \review{Due to the increased sensitivity of KiDS-Legacy compared to KiDS-1000 the question arises whether this agreement is good enough. As discussed above, the main effect is the survey geometry which we address in \Cref{app:mixed_term} reconsiderederation of the mix term and via the pair counts in \Cref{eq:noise_covariance}. In the `sva' term, however, the survey geometry is only accounted for via the $f_\mathrm{sky}$ approximation. To asses the accuracy we analysed mock data vectors with the fiducial KiDS-Legacy covariance (compare to the grey band in \Cref{fig:effect_covariance_on_inference}), including the new mixed term, and directly with \textit{Egretta}. We found that using the \textit{Egretta} covariance gives a shift in $\chi^2$ similar to the inclusion of the survey geometry in the mix term. From this we conclude that the $f_\mathrm{sky}$ approximation in the `sva' term is still accurate enough for KiDS-Legacy. If the sensitivity with upcoming surveys is further increased, this approximation might need to be reconsidered. This point is discussed a bit more in the conclusions.}

\section{\mysc{OneCovariance} applications}
\label{sec:examples}
\begin{figure}
    \label{fig:clusteringz_sample}
    \centering
    \includegraphics[width = .45\textwidth]{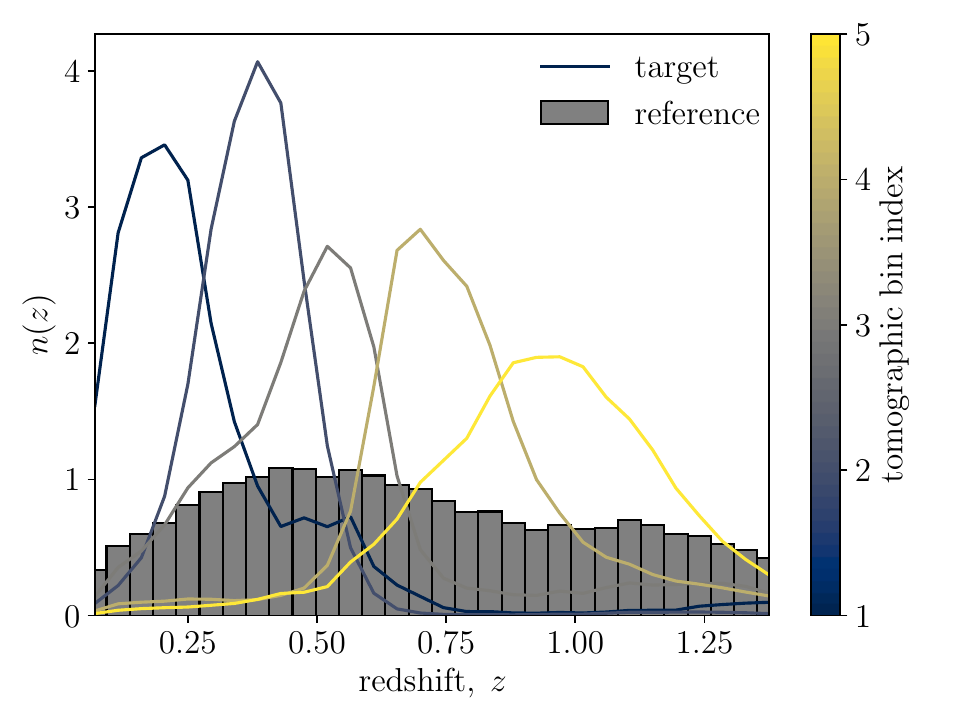}
    \includegraphics[width = .45\textwidth]{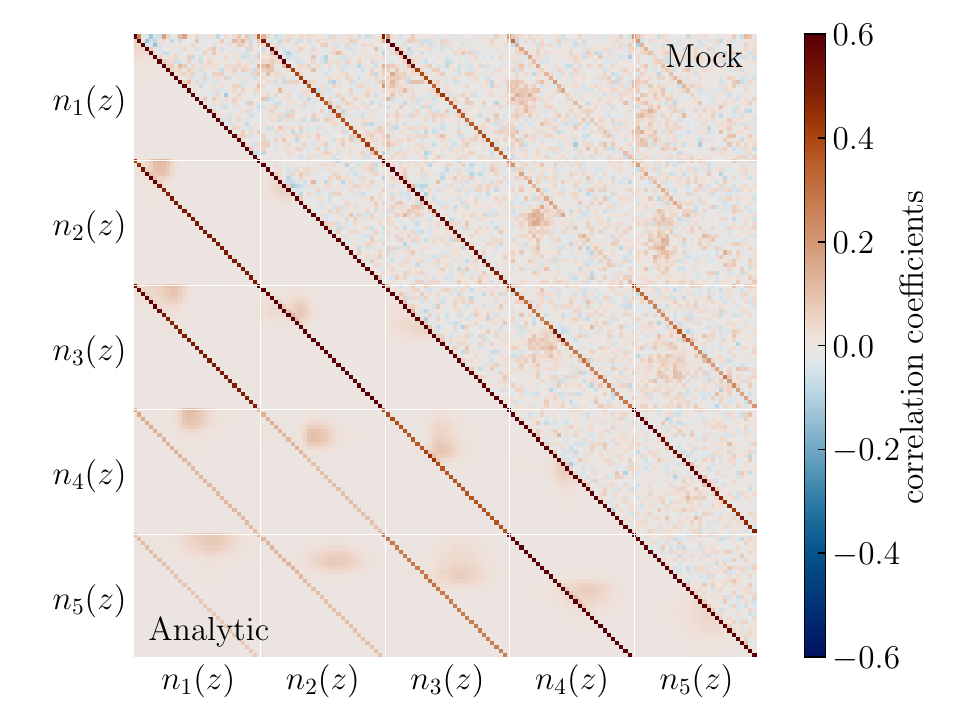}
    \caption{\textit{Left}: Redshift distribution of the target sample for five tomographic bins of KiDS-1000 as solid lines. The grey bars indicate the reference sample. \textit{Right}: Pearson correlation coefficient with the lower triangle showing the predictions of the analytical prescription, \Cref{eq:clustering_z_cov}, while the upper triangle shows the results obtained by jackknife resampling from \citet{hildebrandt_kids-1000_2021} (compare their Figure 3.}
\end{figure}

To demonstrate the versatility of the \mysc{OneCovariance} code, we employ it to address two additional key aspects of the KiDS-Legacy analysis: consistency tests among different summary statistics in \Cref{sec:consistency} and clustering redshifts in \Cref{sec:clustering_redshifts}. These aspects are integral to the studies presented in \citet{stoelzner_kids_legacy_2024} and \citet{wright_kids_legacy_2024}, respectively.
In \Cref{fig:consistency_test}, we present the correlation coefficient between the three main summary statistics utilised in KiDS-Legacy. For the clustering redshift covariance, our primary findings are depicted in \Cref{fig:clusteringz_sample} and \Cref{fig:contributions_clusteringz}. These illustrations offer a comparison with mocks and validate the fundamental structure of the jackknife covariance employed for the redshift calibration.

\subsection{Consistency tests}
\label{sec:consistency}
As an initial showcase of the \mysc{OneCovariance} code's capabilities for KiDS-Legacy CS, we present the covariance matrix between various summary statistics. This calculation, facilitated by the \mysc{ARBsummary} class of the code (see \Cref{app:code_structure}), serves as a foundational component for the consistency checks (or cross-validation) that will be carried out in \cite{stoelzner_kids_legacy_2024}, a process that hinges not only on calculating the covariance of each summary statistic but also on their cross-covariance. Given that the different summary statistics probe nearly identical physical scales, the resulting covariance matrix exhibits significant cross-correlation and can be close to singular.

As a specific example and to demonstrate the versatility of the \mysc{OneCovariance} code, we undertake the covariance matrix calculation for a KiDS-1000 setup. This pairwise computation encompasses our principal CS summary statistics. It is, however, available for arbitrary summary statistics of all tracers supported by the code. 

 The resulting correlation coefficient is depicted in \Cref{fig:consistency_test}, showing large off-diagonal elements between different summary statistics originating both from sample variance terms as well as shape noise. Moreover, the pronounced correlations highlight that the diverse summary statistics employed in the KiDS-Legacy analysis are almost completely degenerate.

The \mysc{OneCovariance} code can straightforwardly accommodate any combination of summary statistics envisioned by the user since it propagates the relation between summary statistics and the harmonic covariance through simple integral transformations (compare Equation \ref{eq:linearmap_ell_to_obs}). Consequently, users can seamlessly integrate different summary statistics for various tracers or explore the impact of different scale cuts for any summary statistic, as done for COSEBIs in \citet{dark_energy_survey_and_kilo-degree_survey_collaboration_y3_2023}.

\subsection{Clustering redshifts}
\label{sec:clustering_redshifts}

As the final application of the \mysc{OneCovariance} code in this study, we investigate the covariance matrix of clustering redshift estimation (see e.g. \citealt{lima_estimating_2008, bonnett_redshift_2016, van_den_busch_testing_2020,hildebrandt_kids-1000_2021}) which utilises a spectroscopic reference sample (s) and a target sample with noisy redshift estimates (p). In \citet{hildebrandt_kids-1000_2021} a covariance using jackknife resampling was used. So far, however, it was never compared to analytical expectation. In this section we are filling this gap.
By measuring the correlation function, \Cref{eq:gc_correlation_function}, between these two samples at fixed physical scale $r$ at different redshifts, the redshift distribution $n(z)$ can be recovered up to an irrelevant multiplicative constant (as the redshift distribution will be normalised in the end) and bias evolution of the target sample,
\begin{equation}
    n_\mathrm{p}(z) = \frac{w_{\mathrm{sp}}(\theta(r),z)}{\Delta z\sqrt{ w_{\mathrm{ss}}(\theta(r),z)}}\,,
\end{equation}
where $\Delta z$ is the redshift width of each cross-correlation measurement.
By labelling tomographic bin indices of the target sample with Greek letters and reference sample redshift bins with Latin letters, the covariance for this estimator at first order is
\begin{equation}
\label{eq:clustering_z_cov}
\begin{split}
    \mathrm{Cov}\left[n_\alpha({z_i}),n_\beta({z_j})\right] \equiv \; &  \mathrm{Cov}\left[\frac{w_{i\alpha}}{\Delta z_i\sqrt{ w_{ii}}}\frac{ w_{j\beta}}{\Delta z_j \sqrt{w_{jj}}}\right] \\
    \approx &\;\frac{1}{\Delta z_i \Delta z_j\sqrt{w_{ii}w_{jj}}}\left[C_{i\alpha j \beta} - \frac{w_{i\alpha}}{2w_{ii}}C_{iij\beta}  - \frac{w_{j\beta}}{2w_{jj}}C_{i\alpha jj} +  \frac{w_{i\alpha}w_{j\beta}}{4w_{ii}w_{jj}}C_{iijj}\right]\,.
\end{split}
\end{equation}
Here $C_{ijkm}$ denotes the covariance between clustering measurement $w_{ij}$ and $w_{km}$ at fixed scale $r$. The last three terms are only non-zero if $i=j$, with some small off-diagonal elements from the full non-Limber expression. Only the first term can have important off-diagonal elements, in particular where the redshift of the reference sample $i$ approaches the support of the target sample $\beta$ (and the same for $j$ and $\alpha$. 

The performance evaluation of the analytic covariance for clustering redshifts is conducted using the synthetic data outlined in \citet{hildebrandt_kids-1000_2021} and \citet{van_den_busch_testing_2020}. This dataset comprises mock galaxy samples closely resembling the {KiDS+VIKING-450} (KV450) CS sample. The basis of these mock samples is the {MICE} simulation \citep{fosalba_mice_2015}. The cosmological model adopted in these simulations adheres to a $\Lambda$CDM model with parameters: $\Omega_\mathrm{m} = 0.25$, $\Omega_\Lambda = 1 -\Omega_\mathrm{m}$, $\Omega_\mathrm{b} = 0.044$, $\sigma_8 = 0.8$, and $h = 0.7$. These simulations are accompanied by the corresponding galaxy catalogues \citep{crocce_mice_2015}, serving as the foundation for the publicly accessible pipeline for generating KiDS mock samples\footnote{Accessible at \href{https://github.com/KiDS-WL/MICE2_mocks}{https://github.com/KiDS-WL/MICE2\textunderscore mocks}}.
Due to the nature of the estimator, the covariance retains a residual dependence on the bias evolution of the target sample. \citet{davis_cross-correlation_2018} characterised this bias using the parametrisation $\mathcal{B}_\alpha(z) = (1+z)^\alpha$, we use the values from \citet{van_den_busch_testing_2020,hildebrandt_kids-1000_2021} for the residual bias evaluation in the covariance modelling. 

\begin{figure}
    \centering
    \includegraphics[width= 0.42\textwidth,clip]{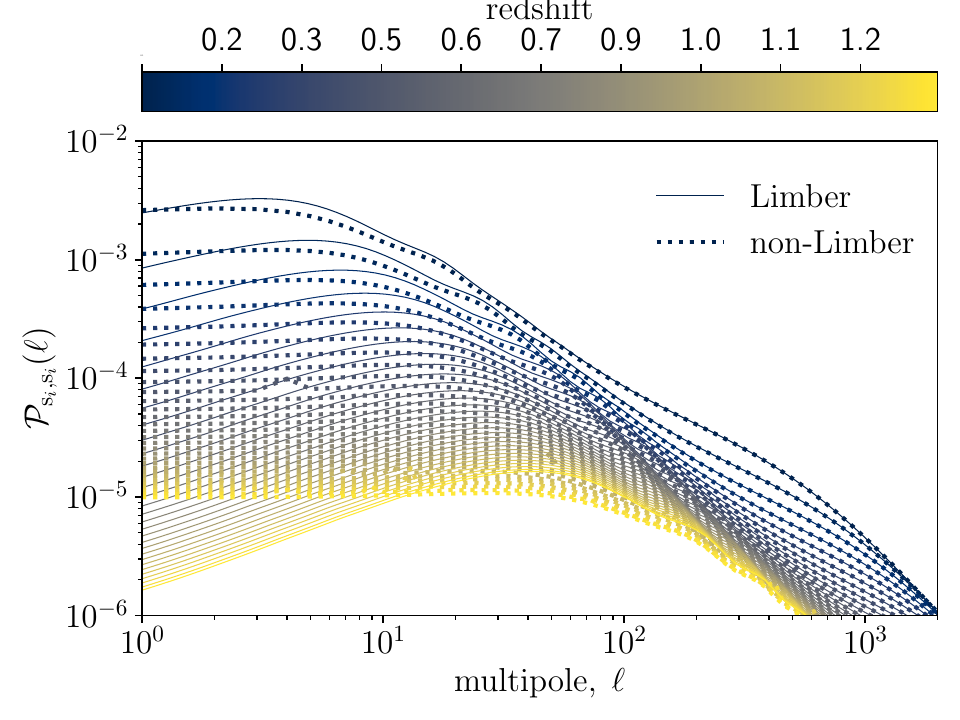}
    \includegraphics[width= 0.42\textwidth]{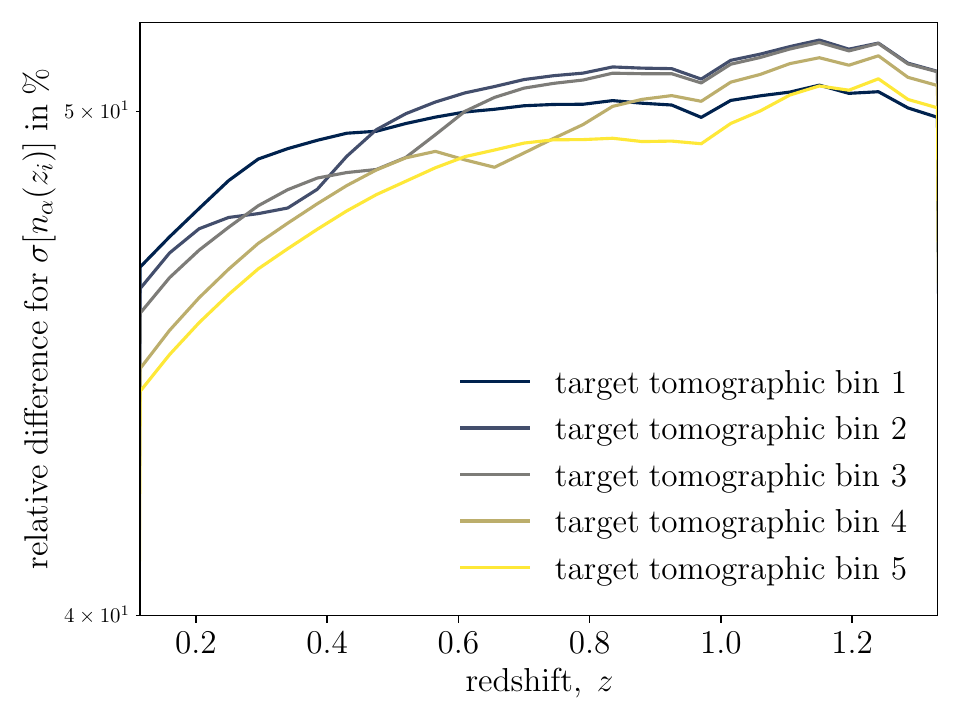}
    \caption{\textit{Left}: Auto-correlation angular power spectra of the reference sample as a function of spectroscopic redshift (colour bar). The solid lines use the Limber approximation, \Cref{eq:limber}, while the dashed lines make use of the full expression, \Cref{eq:powerspectrum_general}. \textit{Right}: Fractional difference as a percentage of the Gaussian covariance term when using Limber vs non-Limber.
    }
    \label{fig:nonlimber_effect}
\end{figure}

\Cref{fig:clusteringz_sample} illustrates the resulting redshift distributions for the five tomographic bins of KiDS1000, represented by solid colour-coded lines on the left side, denoting the target sample. The redshift distribution of the reference sample is depicted as a bar chart. On the right, the Pearson correlation coefficient of the corresponding covariance matrix is displayed for scales $r\in[0.1,1]h^{-1}\mathrm{Mpc}$. In the upper triangle, we show the analytic prediction of \Cref{eq:clustering_z_cov}, while the lower triangle uses the mock catalogue and jackknife re-sampling over the individual subsamples of the simulation.  The general structure is very similar, in particular, one can see that the previously mentioned non-vanishing off-diagonals at $i\neq j$ are in the correct position. Moreover, the diagonal of each block matrix exhibits comparable structures, providing valuable qualitative validation of the jackknife covariance outlined in \citet{hildebrandt_kids-1000_2021}.

With the narrow redshift bins employed in cross-correlation measurements with spectroscopic surveys, the applicability range of the Limber approximation, as denoted by \Cref{eq:limber}, is significantly constrained. As discussed in \citet{loverde_extended_2008}, the next-to-leading order term in the expansion becomes subdominant for $\ell > \chi/\Delta\chi$, where $\Delta\chi$ is the width of a Gaussian redshift bin at co-moving distance $\chi$. For the present situation, this requires the non-Limber calculation up to $\ell \sim 10^3$.
In the left plot of \Cref{fig:nonlimber_effect}, we depict the angular power spectrum of spectroscopic sample auto-correlations. Dashed lines represent the complete integral, as described in \Cref{eq:powerspectrum_general}, while solid lines utilise the Limber approximation. All lines are colour-coded with respect to the spectroscopic redshift bins (as illustrated in the bar plot of \Cref{fig:clusteringz_sample}). It is evident from the figure that the non-Limber projection significantly impacts the integrand of $w_{\mathrm{s}_i\mathrm{s}_j}$.
On the right side of \Cref{fig:nonlimber_effect}, we illustrate the impact of the non-Limber approximation on the Gaussian error of the covariance, revealing approximately 50 per cent effects for all tomographic bins. 

\begin{figure}
    \centering
    \includegraphics[width= 0.42\textwidth,clip]{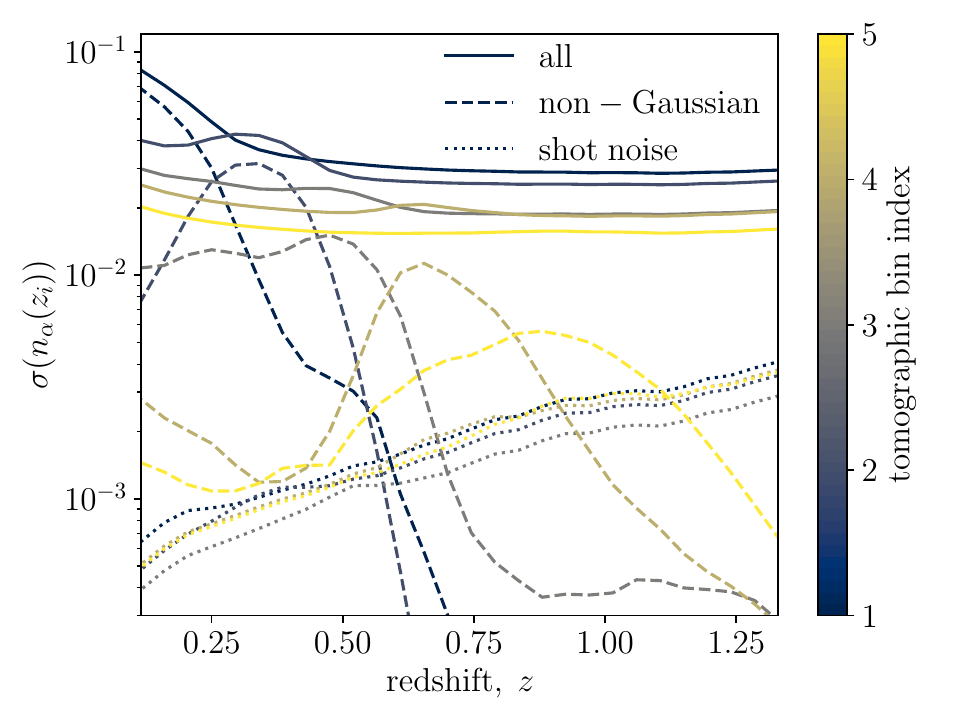}
    \includegraphics[width= 0.42\textwidth]{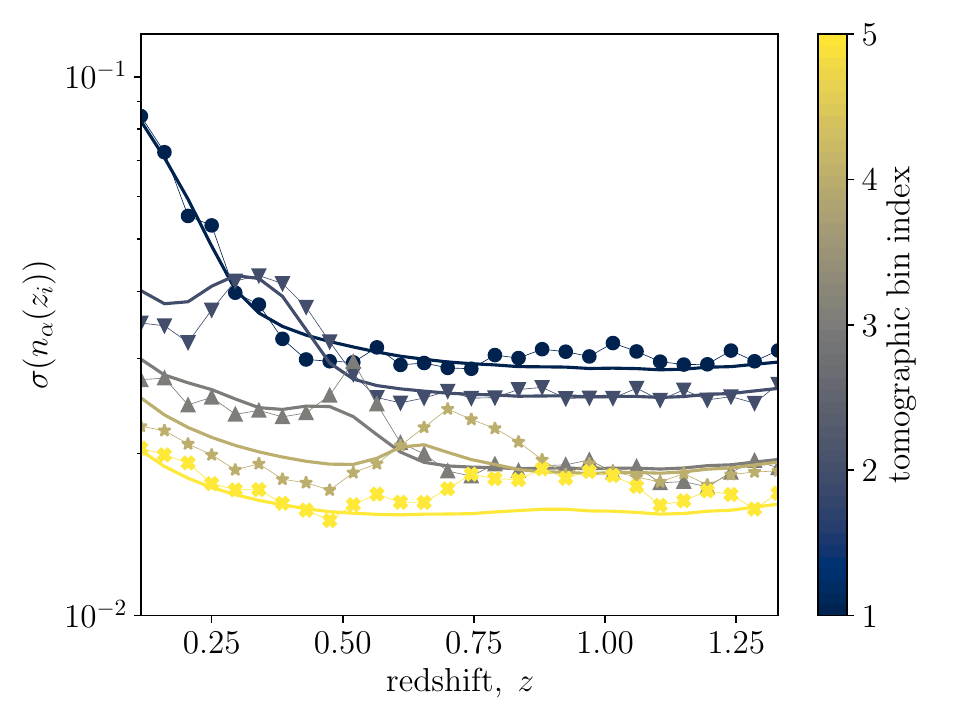}
    \caption{\textit{Left}: Contribution to the standard deviation of the clustering redshift measurements. The solid lines correspond to the total variance, the dashed lines to the non-Gaussian contribution, and the dotted lines to the shot noise. For all tests, we ignored the SSC term since it is suppressed for galaxy clustering. 
    \textit{Right}: Comparison of the analytical standard deviation (solid lines) with the simulations (symbols). 
    }
    \label{fig:contributions_clusteringz}
\end{figure}

\Cref{fig:contributions_clusteringz} presents, on the left side, the contributions to the covariance as a function of redshift, colour-coded according to the tomographic bin index. It is evident that low redshifts are significantly influenced by the non-Gaussian covariance (as indicated by dashed lines), while the shot noise contribution remains negligible for all redshifts.
On the right side of \Cref{fig:contributions_clusteringz}, we juxtapose the analytical covariance with simulations (indicated by symbols with thin lines). We observe a good agreement for the lower tomographic bins, with the increase attributed to the non-Gaussian contribution clearly visible. However, this increase is less pronounced in the fourth and fifth tomographic bins for the analytical covariance, whereas it remains evident in the mock data. This difference is likely due to uncertainties in the non-linear modelling of the trispectrum, especially as we measure signals deep within the non-linear regime, $k>1h/\mathrm{Mpc}$. 

It is worth noting that while the non-Limber effect is easily detectable for the Gaussian covariance, its influence on the non-Gaussian term presents a more challenging task and has thus far only been explored in \citet{lee_cosmological_2020} for trispectra using specific kernels from perturbation theory and the galaxy bias expansion. Their analysis suggests significant effects on the trispectrum. However, a similar analysis for general trispectra, such as those from the halo model, remains unexplored and is beyond the scope of this section for several reasons. Firstly, while the \mysc{OneCovariance} can compute the non-Limber projection of power spectra, its primary utility lies in photometric surveys with broader window functions, where Limber's approximation is more readily applicable. Secondly, spectroscopic samples, necessitating non-Limber calculations, are typically conducted on quasi-linear scales, unlike the one presented here. On those scales, the non-Gaussian contribution is generally less significant. Lastly, this application serves as a comprehensive test and sanity check for clustering redshift covariance, where jackknife resampling is readily available. Furthermore, realistic samples are likely to be more influenced by noise, as galaxy density tends to be substantially lower, at least by a factor of ten.

\section{Present and future of the \mysc{OneCovariance} code}
\label{sec:summary_onecov}
In this section, we offer a summary of the capabilities of the code, delineate the modelling choices incorporated, underscore its flexibility, and contemplate potential avenues for future enhancements. 

\subsection{Current status of \mysc{OneCovariance} code}
We have provided a comprehensive review of the covariance matrix modelling in KiDS so far in \Cref{sec:correlators_in_fourier_space,sec:harmonic_space}. Employing a halo model approach, we have consolidated all relevant equations used in various analyses, encompassing CS \citep{hildebrandt_kidsviking-450_2020,asgari_kids-1000_2021}, GGL, GC, and the CSMF \citep{heymans_kids-1000_2020, dvornik_unveiling_2018, dvornik_kids-1000_2023}. This effort culminates in establishing the legacy of analytic covariance modelling within KiDS. While our primary focus lies on CS for the direct validation of the code in this study, we have implemented all pertinent quantities into the new \mysc{OneCovariance} code from scratch and cross-validated them against previous results. Moreover, we have reviewed and incorporated all summary statistics utilised within KiDS in \Cref{sec:harmonic_to_real}, encompassing real space correlation functions, COSEBIs, and bandpowers, each applicable across all three tracers.
Through the development of a unified code for covariance matrices, we have streamlined the calculation process, now directly linking harmonic space statistics to observables via integral transformations. This approach differs from that of KiDS-1000, where the bandpowers covariance was derived from the $\xi_\pm$ covariance. This modification brings the analytic covariance for bandpowers closer to the theoretical modules employed in KiDS, enhancing the consistency and efficiency of our analyses.

The \mysc{OneCovariance} offers users the capability to effortlessly download and compute simulation-validated covariance matrices for different traces across various statistical representations including harmonic space, real space, COSEBIs, and bandpowers, all under the flat sky approximation. While we focused on the comparison using CS in this paper, the code has been validated against all previous covariance codes employed in KiDS also for the other tracers.
Minimal user input is required beyond specifying the redshift distributions, as the code operates out of the box.
Moreover, the code's flexibility allows for extensive customisation. Users can readily adjust inputs, such as modifying the HOD prescription or directly altering galaxy power spectra or bias parameters. Additionally, it offers the flexibility to alter line-of-sight weight functions through input, facilitating the definition of entirely new tracers. For instance, by defining a new line-of-sight weight function and adjusting the HOD prescription, the code can compute the covariance matrix for the Cosmic Infrared Background or any other intensity mapping quantity.
Alternatively, users can provide harmonic-space quantities directly, allowing the code to perform the necessary projections to observables. Furthermore, the code includes scripts to compute the weight functions used for various summary statistics in KiDS-Legacy. 

The overarching philosophy of the \mysc{OneCovariance} code is centred on the replaceability of all essential quantities by file inputs. If a file input is provided, it supersedes the standard configuration of the code, utilising the file input instead. This approach permeates the entirety of the code, from fundamental quantities required for theory calculations, such as line-of-sight weights and bias prescriptions, to higher-level quantities including power spectra, their responses, and trispectra. The final elements of the covariance are effectively arrays for each tracer combination with the general shape
\begin{equation*}
    (\texttt{spatial}_{t_1,t_2},\;\texttt{spatial}_{t_3,t_4}, \;\texttt{mass}_{t_1,t_2}\;, \texttt{mass}_{t_3,t_4}\;, \texttt{tomo}_{i,t_1}\;, \texttt{tomo}_{j,t_2}\;,  \texttt{tomo}_{i,t_3}\;, \texttt{tomo}_{j,t_4})\;.
\end{equation*}
Here, $\texttt{spatial}_{t_1,t_2}$ labels the spatial index of the tracer pair $t_1, t_2$ over which its two-point function is measured. This can be for example the $\theta$ bin, the band power mode, the order of the COSEBIs, or any other label of an arbitrary summary statistic passed to the code. $\texttt{mass}_{t_1,t_2}$ is the corresponding stellar mass bin (this could be also modified to cluster number counts, for example). For CS, this always has a length of one. Lastly, $\texttt{tomo}_{i,t_1}$ and $\texttt{tomo}_{j,t_2}$ is the tomographic bin combination for the pair $(t_1, t_2)$. With this structured approach, the \mysc{OneCovariance} code possesses the versatility required to handle various scenarios in cosmology, allowing users to specify inputs via file paths in a configuration file. This means that users can specify the necessary inputs, thereby solving the often cumbersome task of ordering covariance matrix elements and performing the projection to observables. In essence, the \mysc{OneCovariance} code provides a comprehensive solution for covariance estimation whenever a HOD prescription-based halo model is employed for the signal, offering a valuable resource for the cosmological community.

\subsection{Future developments of the code}
The future development roadmap for the \mysc{OneCovariance} code includes both incremental improvements and significant expansions to its functionality. Here is an outline of the planned enhancements:
\begin{enumerate}[i)]
    \item Currently, the code employs the flat sky approximation for defining observables. In the next version, full sky transformations will be implemented, especially for the Gaussian terms, as they dominate the covariance on scales where curved sky effects become significant.
    \item The code already supports calculating the non-Limber covariance for Gaussian contributions. The next step involves extending this capability to cover connected terms as well, enhancing the accuracy of covariance estimates.
    \item Incorporating survey window effects into the harmonic space covariance and propagating them to observables, particularly for sample variance, will be a crucial addition to enhancing the accuracy of covariance estimates. This could be done by streamlining the code with \mysc{NaMaster} \citep{alonso_unified_2019}.
    \item A longer list of pre-implemented models will be added to reduce the number of required input quantities, including alternative HOD and bias prescriptions, streamlining the user experience and expanding the code's usability.
    \item Interfacing CCL \citep{chisari_core_2019} will be implemented, enabling the use of pre-defined quantities within CCL and leveraging its extensive functionalities for enhanced accuracy and efficiency in covariance calculations.
\end{enumerate}

\section{Conclusion}
\label{sec:summary}
In this study, we have provided a comprehensive account of the covariance methodology utilised in and up to the latest data release of KiDS. This encompasses the validation of the CS covariance matrix utilised in \citet{wright_kids_legacy_2024} and \citet{stoelzner_kids_legacy_2024}, as well as its application to clustering redshifts as employed in \citet{wright_kids_redshift_2024}. Accompanying this paper is the \mysc{OneCovariance} code, a unified covariance tool tailored for photometric LSS surveys.
 We have conducted a thorough validation of the \mysc{OneCovariance} code for analysing KiDS-Legacy data, with a specific focus on CS. Our validation efforts have yielded several noteworthy conclusions.
We have demonstrated that the theoretical predictions generated by the \mysc{OneCovariance} code align closely with those obtained from external libraries, such as \mysc{CCL}, at the per cent level. This underscores the robustness and accuracy of the code's theoretical predictions.
Furthermore, comparisons with the independent implementation of the covariance matrix calculations from KiDS-1000, as detailed in \citet{joachimi_kids-1000_2021}, being typically at the per cent level, reinforces the accuracy of the \mysc{OneCovariance} code. Any larger discrepancies identified were traced back to differences in accuracy settings or extrapolation schemes used for integrations when transitioning from harmonic space to observables.
Importantly, we have verified that these discrepancies would not have significantly impacted any previous analyses conducted on the KiDS data. This ensures the consistency and reliability of previous findings derived from KiDS data analyses.

Our validation efforts provide strong evidence supporting the efficacy and accuracy of the \mysc{OneCovariance} code for CS analyses within the KiDS framework. This instils confidence in the code's utility for future cosmological investigations, ensuring robust and reliable covariance estimation in the analysis of LSS data.
Next, we established a simulation pipeline to further validate the covariance matrix. We used log-normal realisations of the density field with both a realistic mask and homogeneous sources (\textit{Cygnus}), as well as variations in source density (\textit{Egretta}). On these mocks, we measured the real space correlation functions of the shear, $\xi_\pm$, and found excellent agreement between the theoretical signal and the signal measured in the mocks. This agreement was observed across nine logarithmically spaced angular bins ranging from 0.5 to 300$\;\mathrm{arcmin}$.
 By leveraging the exact pair counts from the \textit{Egretta} mocks, we also found similar agreement between the simulation and the analytic covariance as in KiDS-1000. Particularly, we observed agreement well within ten per cent on most scales, with only the largest angular scales exhibiting more significant deviations due to survey boundary effects. Importantly, these differences do not significantly impact the final constraints derived from the analysis.

A significant deviation from KiDS-1000 was the re-evaluation of the mixed term, namely the cross-correlation between shape noise and two-point statistics. This term is influenced by both the response of the signal to the survey geometry and the number of pairs used to estimate the signal itself, owing to the complex survey mask of KiDS.
We found that assuming a homogeneous survey misestimates the Gaussian term by roughly 10 per cent in the case of $\xi_+$ for off-diagonal elements, with this effect potentially being both positive and negative. For $\xi_-$ and its cross-correlation with $\xi_+$, the deviation can be substantial on off-diagonal elements. This is attributed to the broad support of the Fourier filter for $\xi_-$ and its sensitivity to many scales, thus being affected by the inhomogeneity of the KiDS-Legacy sample and the intricate survey footprint. However, despite the potentially large effects on individual covariance matrix elements, the low signal of $\xi_-$ and high shape noise mitigate these discrepancies.
For upcoming surveys, the situation is less clear. On the one hand, shape noise effects will become less important and the surveys will become more homogeneous as well as larger, reducing the importance of edge effects and variable depth. On the other hand, however, due to the increasing importance of sample variance and the increased overall sensitivity, the usage of an ideal mixed term might still lead to more significant effects on parameter inference. \review{In future work, we plan an investigation of this issue along with  the inclusion of the survey geometry in the `sva' term. This can be done by using the pseudo-$C_\ell$ covariance directly to incorporate  the effect of the mask.}

We examined the influence of various modelling choices for the covariance on the final inference of $S_8$, demonstrating their robustness. These choices included the baryonic feedback prescription, the Super Sample Covariance (SSC) and non-Gaussian (nG) terms, the updated mixed term, and the intrinsic alignment model, among others. The upper bound of the 68 per cent interval of $S_8$ remained largely unaffected by these modelling choices.
Furthermore, we compared the impact of different terms on the covariance matrix and found that the Gaussian term overwhelmingly dominated on almost all scales and tomographic bins. The non-Gaussian and SSC contributions were of minor importance, with the SSC term slightly more significant across tomographic bins. One contributing factor to the diminished impact of these terms in KiDS-Legacy is the inclusion of the sixth tomographic bin, which enhances sensitivity to more linear scales, especially given the substantial CS signal in the sixth bin.

Furthermore, we demonstrated that the \mysc{OneCovariance} code can serve as a valuable tool for conducting consistency tests between different summary statistics or different surveys. That is, we calculate the covariance matrix for the same tracer (CS in this case), but two different summary statistics. We confirmed that the covariance matrices required for the KiDS-Legacy consistency tests \citep{stoelzner_kids_legacy_2024} are invertible and well-behaved for pairs of summary statistics.
This functionality holds significant importance when performing consistency tests between large-scale and small-scale analyses of CS data, particularly in addressing tensions such as the $S_8$ tension. By enabling such tests, the \mysc{OneCovariance} code enhances the capability to probe and validate cosmological models across different scales and datasets.

In our final application of the \mysc{OneCovariance} code, we turn our attention to the estimation of the covariance matrix for clustering redshifts, a critical component utilised in \citet{wright_kids_legacy_2024} to validate the jackknife covariance employed for $n(z)$ calibration. Our approach involved applying the code to a KiDS-1000 redshift distribution, slated for calibration against an idealised reference sample derived from the MICE2 simulations, approximately ten times denser than the actual dataset.
Our analysis revealed encouraging results; there was a generally good agreement observed for the correlation coefficient between the analytical and jackknife covariance derived directly from the mocks. However, the narrow redshift shells over which the signal is averaged at a given physical scale posed a challenge, rendering the Limber expression inapplicable over a broad range of scales. Consequently, the covariance experienced deviations of up to 50 per cent.
To address this issue, we incorporated the full non-Limber expression in the Gaussian covariance, while assuming the Limber expression for the non-Gaussian component. However, we recognise that a comprehensive treatment necessitates a full non-Limber modelling of the non-Gaussian covariance, a task beyond the scope of this paper. Nevertheless, our analytical covariance matrix effectively captured the salient features of the jackknife covariance, thereby serving as a valuable cross-check for the clustering redshift calibration process.

In summary, our study confirms the robustness of the KiDS-Legacy covariance in delivering accurate results for the $S_8$ parameter. By providing the cosmological community with a versatile tool capable of calculating covariance matrices for a wide array of photometric(-like) LSS observables in both current and future surveys, we hope to facilitate further advancements in cosmological research.

\begin{acknowledgements}
The authors would like to thank an anonymous referee for valuable comments.
RR, AD, JLvdB, HH and CM are supported by a European Research Council Consolidator Grant (No. 770935). 
SU, BS and ZY acknowledges support from the Max Planck Society and the Alexander von Humboldt Foundation in the framework of the Max Planck-Humboldt Research Award endowed by the Federal Ministry of Education and Research.
HH is supported by a DFG Heisenberg grant (Hi 1495/5-1), the DFG Collaborative Research Center SFB1491 and the DLR project 50QE2305.
BJ acknowledges support from the ERC-selected UKRI Frontier Research Grant EP/Y03015X/1 and by STFC Consolidated Grant ST/V000780/1.
MA is supported by the UK Science and Technology Facilities Council (STFC) under grant number ST/Y002652/1 and the Royal Society under grant numbers RGSR2222268 and ICAR1231094.
MvWK acknowledges the support by the Science and Technology Facilities Council. AHW is supported by the Deutsches Zentrum für Luft- und Raumfahrt (DLR), made possible by the Bundesministerium für Wirtschaft und Klimaschutz, and acknowledges funding from the German Science Foundation DFG, via the Collaborative Research Center SFB1491 `Cosmic Interacting Matters -- From Source to Signal'.
MB and PJ are supported by the Polish National Science Center through grant no. 2020/38/E/ST9/00395. MB is supported by the Polish National Science Center through grants no. 2018/30/E/ST9/00698, 2018/31/G/ST9/03388 and 2020/39/B/ST9/03494.
PB acknowledges financial support from the Canadian Space Agency (Grant No. 23EXPROSS1) and the Waterloo Centre for Astrophysics.
JHD acknowledges support from an STFC
Ernest Rutherford Fellowship (project reference ST/S004858/1).
CG acknowledges support from the project `A rising tide: Galaxy intrinsic alignments as a new probe of cosmology and galaxy evolution' (with project number 1393VI.Vidi.203.011) of the Talent programme Vidi which is (partly) financed by the Dutch Research Council (NWO).
SJ acknowledges the Dennis Sciama Fellowship at the University of Portsmouth and the Ram\'{o}n y Cajal Fellowship from the Spanish Ministry of Science.
KK acknowledges support from the Royal Society and Imperial College.
LL is supported by the Austrian Science Fund (FWF) [ESP 357-N].
CM acknowledges support under the grant number PID2021-128338NB-I00 from the Spanish Ministry of Science.
TT acknowledges funding from the Swiss National Science Foundation under the Ambizione project PZ00P2\_193352 
MY acknowledges funding from the European Research Council (ERC) under the European Union’s Horizon 2020 research and innovation program (Grant agreement No. 101053992). \newline

{\it Software:} The figures in this work were created with {\sc matplotlib} \citep{Hunter:2007}, making use of the {\sc NumPy} \citep{Numpy}, {\sc SciPy} \citep{Scipy}, {\sc Cosmosis} \citep{zuntz_cosmosis_2015}, {\sc Nautilus} \citep{Lange23} and {\sc CosmoPower} \citep{SpurioMancini22}  software packages. \newline
\textit{Author contributions}: All authors contributed to the development and writing of this paper. The authorship list is given in three groups: the lead authors (RR, SU) followed by two alphabetical groups. The first alphabetical group includes those who are key contributors to both the scientific analysis and the data products. The second group covers those who have either made a significant contribution to the data products, or to the scientific analysis.
\end{acknowledgements}

\bibliographystyle{aa}
\bibliography{MyLibrary_single}

\begin{appendix}

\section{Code structure}
\label{app:code_structure}
\begin{figure}
    \centering
    \includegraphics[width = 1.\textwidth, trim={1.5cm 10cm 1.5cm 0},clip]{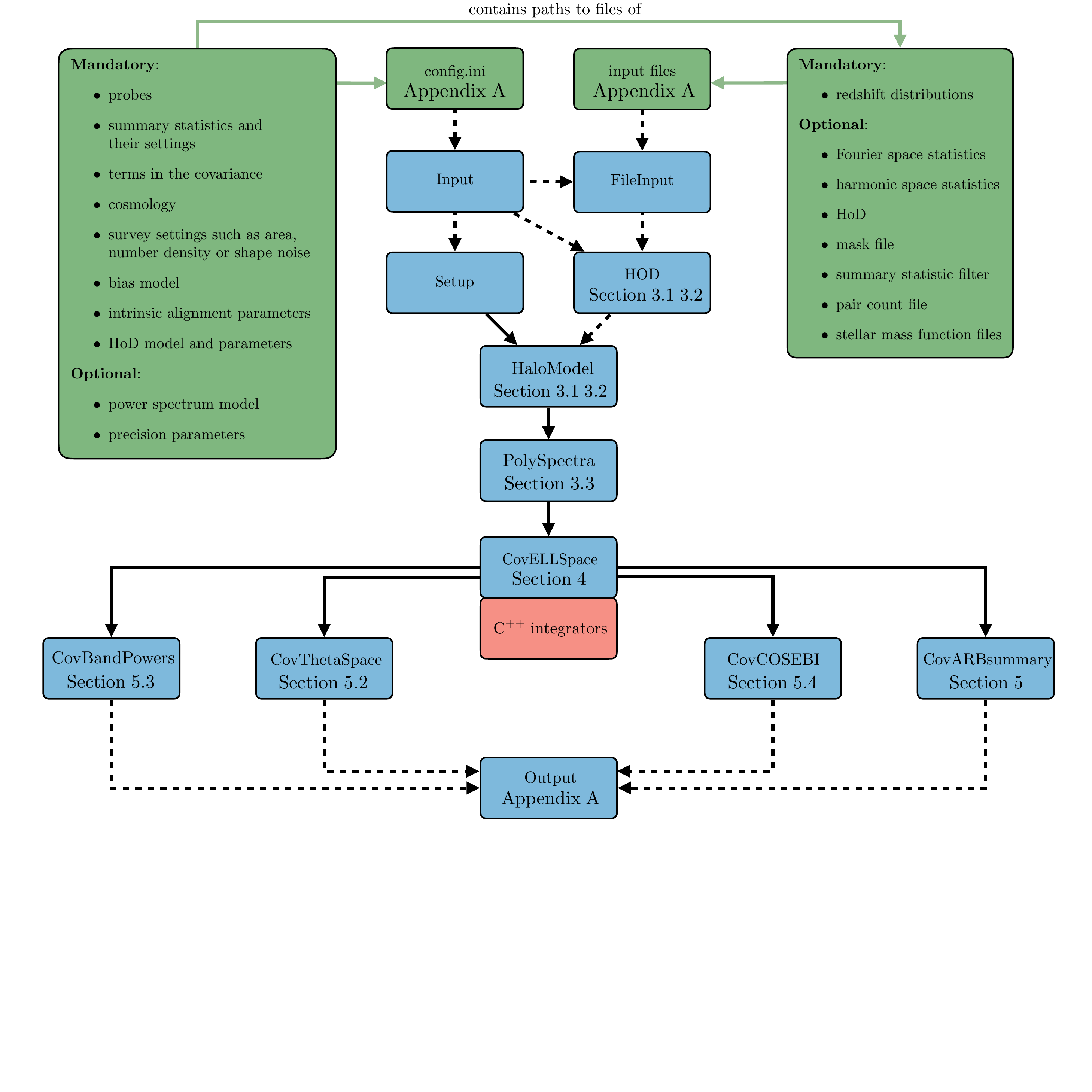}
    \caption{General flowchart of the \mysc{OneCovariance} code. The green boxes indicate files that are fed to the code (visit the \mysc{readthedocs} webpage, \href{https://onecovariance.readthedocs.io/en/latest/index.html}{https://onecovariance.readthedocs.io/en/latest/index.html}, for a more detailed discussion). The dashed lines with arrows indicate that files or functionalities are included, but not inherited. The solid arrows instead indicate inheritance. Each blue box indicates a \mysc{python} module in the form of a class with the corresponding name. The red box is a \mysc{C++} class wrapped into \mysc{python} with \mysc{pybind} to carry out some of the heavy lifting in the code. Section numbers indicate where the corresponding equations and description of the content of each module can be found in the paper.}
    \label{fig:flowchart}
\end{figure}
The general code structure is depicted in the flowchart of \Cref{fig:flowchart}. Dashed black arrows provide input, while solid black arrows indicate inheritance. Each blue square is a class object in the code, while green boxes indicate input files and their description. 
Here we   give a brief explanation of each of the different modules and inputs to provide an idea of the philosophy of the code.

\begin{enumerate}
    \item[] \textbf{Input, FileInput, and configuration files}\\ In the code, users specify all parameters through the configuration file, encompassing cosmological, HOD, bias, precision, and other parameters, as well as settings for considered summary statistics or survey specifications. Certain inputs, such as the redshift distribution of sources and lenses, are handled through input files, which are mandatory. While some settings are mandatory, others have default values.
The \mysc{Input} and \mysc{FileInput} classes play a crucial role in managing user input. They communicate with the user to ensure that all mandatory information is provided and that settings are compatible. Upon reading all information from the configuration file via the \mysc{Input} class, variables are stored in dictionaries and passed to the \mysc{FileInput} class. Subsequently, the \mysc{FileInput} class reads in all specified files and stores them in a specified dictionary. These dictionaries containing all necessary information are then passed through the code.
It is worth noting that any input-related messages or warnings include factors of $h$ in the units, where applicable.\vspace{.2cm}

    \item[] \textbf{Setup}\\ 
    The \mysc{Setup} class sets the cosmology via the \mysc{astropy} library \citep{astropy_collaboration_astropy_2022} and contains a lot of auxiliary functions which are called later in the code, such as the mode calculation of the survey footprint or the binning of the number of pairs. Furthermore, it contains a lot of the functionality to perform consistency checks of the input parameters, for example whether the $k$ range is large enough to support a required multipole range.\vspace{.2cm}

    \item[] \textbf{HOD and HaloModel}\\
    In the \mysc{HOD} class, all the components of the HOD and CSMF (\Cref{sec:csmf}) are defined. This class is then instantiated in the \mysc{HaloModel} class, where all the halo model quantities are implemented (see \Cref{sec:halo_mod_ingredients}). For the halo mass function calculations, the \mysc{hmf} package \citep{murray_hmfcalc:_2013} is used. All quantities are calculated on a grid in $k$ at a single redshift.\vspace{.2cm}

    \item[] \textbf{PolySpectra}\\
    Uses the functionality of the \mysc{HaloModel} class together with the perturbation theory kernels to calculate all power spectra, their responses and trispectra in Fourier space (\Cref{sec:polysectra}) for all observables considered at a single redshift. Up to this point, all calculations are completely independent of any survey specifications. Quantities in this class live on a $k$-grid ($k_1$ and $k_2$ in the case of trispectra) and are eventually divided into stellar mass bins if required.\vspace{.2cm}

     \item[] \textbf{CovELLSpace}\\
    Carries out the projections from three-dimensional Fourier space to harmonic space as described in \Cref{sec:harmonic_space}. The code inherits the functionality from the \mysc{PolySpectra} class, updates all the quantities that depend on redshift, and creates splines for the line-of-sight integration. It is here where the survey specifications enter the stage via the redshift distribution, survey mask, for example. The \mysc{CovELLSpace} is also the first class that can produce a covariance matrix. All covariance matrix elements are stored in \mysc{numpy} arrays. \vspace{.2cm}

    \item[] \textbf{Integrators}\\
    The red box in the middle of \Cref{fig:flowchart} is an external \mysc{C++} code which has been wrapped into \mysc{python} for the numerical integration of the highly oscillatory integrals. 
For the Bessel functions occurring in some of the Fourier weights, we use either \mysc{cquad} from the Gnu Scientific Library (\mysc{GSL}) or Levin's method \citep{levin_fast_1996, zieser_cross-correlation_2016,spurio_mancini_3d_2018,leonard_n5k_2022}. The latter is in particular used for large $\theta$, i.e. when many oscillations of the Bessel function are crossed. In particular, we use \mysc{cquad} only when there are less than 100 extrema of the Bessel functions in the integral domain.

For most of the Fourier weights, however, Levin's method is not applicable as it requires an analytic relation between the oscillatory part of the integral and its derivative. This is only given via the recurrence relations of the Bessel functions, but not for the Fourier filter functions. Therefore, the integration domain is simply split into different intervals, each covering roughly a specified (in the configuration file) number of extrema of the Filter functions. These integrals are then again evaluated using \mysc{cquad}. The integrator is parallelised over all combinations of tomographic bins and can therefore benefit greatly from a large number of cores. Convergence of the integrals is reached if the final result does not change by a user-specified relative amount. \vspace{.2cm}

    \item[] \textbf{CovBandPowers, CovThetaSpace, and CovCOSEBI}\\
    The \mysc{BandPowers}, \mysc{ThetaSpace}, and \mysc{COSEBI} classes all work in the same way. They use the harmonic space expressions and map them into the desired $L$, $\theta$ or $n$-range respectively (\Cref{sec:bandpowers}, \Cref{sec:real_space_correlation_functions} and \Cref{sec:cosebi}) as specified in the configuration file with all Fourier and real space filter functions computed internally. The main method in this class is always the \mysc{calc\_covbandpowers} method (and similarly for the other classes). They provide lists of 10 arrays containing all unique combinations of probes for a standard $3\times2$ analysis\footnote{Ten because it includes B modes for lensing. Instead, for pure $C_\ell$, there are only six.} with the different covariance matrix components.\vspace{.2cm}

    \item[] \textbf{ARBsummary}\\
    This class works in principle in the same way as the previous classes for the covariance calculation, however, it does not calculate any filters (see Equations \ref{eq:linearmap_ell_to_obs} and \ref{eq:linearmap_theta_to_obs}) for the summary statistics internally but requires them to be passed as file inputs via the configuration file. It is therefore completely agnostic to the kind of summary statistic which is passed. In particular, you can combine different summary statistics and also check the consistency between summary statistics of different probes via this class by just passing two kinds of Fourier and real space filters for each summary statistic.\vspace{.2cm}

    \item[] \textbf{Output}\\
    The last class of the code just takes the list of covariance matrix entries computed by one of the previous four classes and does three things. It writes this long list of all components into a file, depending on the setting they can be split into ‘sva’, ‘mix’, ‘sn’, ‘SSC’, ‘nG’ and ‘cov’ (the sum of all) or just ‘gauss’, ‘SSC’, ‘nG’ and ‘cov’. The order of the elements can be specified so that either the spatial index is the fastest or the slowest varying index. Secondly, the long list is brought into a matrix format, which is also written into a file. Here the order of the indices is always the probe, the tomographic bin index, and the spatial index labelled from the slowest to the fastest. Lastly, this matrix is also used to create a plot of the correlation coefficient.
    
\end{enumerate}
The code is executed by just running the main script \mysc{python covariance.py your\_config.ini}. It will then go through the flowchart depending on the settings you have chosen and communicate via the terminal until it concludes with the output.

\section{Spherical harmonic decomposition}
\label{app:full_sky_fluctuations}
A two-dimensional field $a(\boldsymbol{\hat{n}})$ can be expanded in a spherical harmonic basis:
\begin{equation}
\label{eq:spherical_harmonics}
    a(\boldsymbol{\hat{n}}) = \sum_{\ell, m} a^{\mathstrut}_{\ell m} \yellmarb{}{\mathstrut}{\ell m}(\boldsymbol{\hat{n}})\,.
\end{equation}
The spherical harmonics, $ \yellmarb{}{}{\ell m}(\boldsymbol{\hat{n}})$, satisfy the orthogonality relation
\begin{equation}
    \int\mathrm{d}\Omega_{n}  \yellmarb{}{\mathstrut}{\ell_1 m_1}(\boldsymbol{\hat{n}}) \yellmarb{}{*}{\ell_2 m_2}(\boldsymbol{\hat{n}}) = \delta^\mathrm{K}_{\ell_1\ell_2}\delta^\mathrm{K}_{m_1 m_2}\,,
\end{equation}
with the solid angle element d$\Omega_n$ associated with $\boldsymbol{\hat{n}}$. Furthermore, we   use the following Fourier space convention:
\begin{equation}
\label{eq:Fourier_convention}
    A(\boldsymbol{x}) = \int_{\boldsymbol{k}} A(\boldsymbol{k})\mathrm{e}^{-\mathrm{i}\boldsymbol{k}\cdot\boldsymbol{x}}\coloneqq \int\frac{\mathrm{d}^3k}{(2\uppi)^3}A(\boldsymbol{k})\mathrm{e}^{-\mathrm{i}\boldsymbol{k}\cdot\boldsymbol{x}}\,.
\end{equation}
With this definition, the Fourier components of statistically homogeneous fields have the following $n$-point statistics
\begin{equation}
\label{eq:3dpoly_spectra}
    \langle A_1(\boldsymbol{k}_1,\chi_1)\dots A_1(\boldsymbol{k}_n,\chi_n) \rangle =(2\uppi)^3 P_{A_1\dots A_n}(\boldsymbol{k}_1,\chi_1,\dots, \boldsymbol{k}_n,\chi_1,\dots,\chi_n) \delta^{(3)}_\mathrm{D}(\boldsymbol{k}_{1\dots n})\,,
\end{equation}
where $\boldsymbol{k}_{1\dots n} \coloneqq \boldsymbol{k}_{1} + \dots + \boldsymbol{k}_{n}$. 
Using the Rayleigh expansion, the exponential can be written as
\begin{equation}
    \label{eq:rayleigh_expansion}
    \mathrm{e}^{\mathrm{i}\boldsymbol{k}\cdot\boldsymbol{x}} = 4\uppi\sum_{\ell, m}\mathrm{i}^\ell j^{\mathstrut}_\ell(kx)\yellmarb{}{\mathstrut}{\ell m}(\boldsymbol{\hat{k}})\yellmarb{}{*}{\ell m}(\boldsymbol{\hat{n}})\,,
\end{equation}
where $x = |\boldsymbol{x}|$ is the absolute value and $\boldsymbol{\hat{n}}$ the angular part of the vector $\boldsymbol{x}$, likewise for $\boldsymbol{k}$.

\section{Polyspectra on the sphere}
\label{app:higher_order_correlators_sphere}
Expanding the expression using \Cref{eq:spherical_harmonics}, \Cref{eq:3dpoly_spectra} and \Cref{eq:rayleigh_expansion} yields
\begin{equation}
\label{eq:n_point_angular}
\begin{split}
\left\langle    a_{1_{\ell_1 m_1}}    \dots a_{n_{\ell_n m_n}}\right\rangle= & \; (\mathrm{i})^{\ell_{1\cdot n}}\ ({2\uppi})^{3}\prod_{\alpha=1}^n\left(4\uppi\int\mathrm{d}\chi_\alpha\; W_{A_\alpha}(\chi_\alpha)\int\frac{ k_\alpha^2\mathrm{d}k_\alpha}{(2\uppi)^3}\mathrm{d}\Omega_{k_\alpha} j_{\ell_\alpha}(k_\alpha\chi_\alpha)  Y^*_{\ell_\alpha m_\alpha}(\boldsymbol{\hat{k}}_\alpha)\right)
\; \\ & \times P_{A_1\dots A_n}(\boldsymbol{k}_1,\dots,\boldsymbol{k}_n)\delta^{(3)}_{\mathrm{D}}(\boldsymbol{k}_{1\dots n}) \,,
\end{split}
\end{equation}
where we carried out the angular integrals $\mathrm{d}\Omega_{n_i}$ and used the orthonormality of the spherical harmonics. Angular integrations over spherical harmonics are trivially evaluated casting \Cref{eq:n_point_angular} into a form respecting rotational symmetries. For \Cref{eq:powerspectrum_general} this is for example realised by the diagonality in $\ell$ and $m$. 
To make further progress, we expand the delta distribution into plane waves as well:
\begin{equation}
\begin{split}
\label{eq:angular_polyspectrum}    
\left\langle    a_{1_{\ell_1 m_1}}    \dots a_{n_{\ell_n m_n}}\right\rangle= & \; (\mathrm{i})^{\ell_{1\cdot n}} (2\uppi)^3\prod_{\alpha=1}^n\left((4\uppi)^2\int\mathrm{d}\chi_\alpha\; W_{A_\alpha}(\chi_\alpha)\int \frac{ k_\alpha^2\mathrm{d}k_\alpha}{(2\uppi)^3}\mathrm{d}\Omega_{k_\alpha} j_{\ell_\alpha}(k_\alpha\chi_\alpha)j_{\ell_\alpha}(k_\alpha x)Y^*_{\ell_\alpha m_\alpha}(\boldsymbol{\hat{k}}_\alpha)\right)
\;   \\
& \times P_{A_1\dots A_n}(\boldsymbol{k}_1,\dots,\boldsymbol{k}_n)\; \prod_{\beta = 1}^n\sum_{\tilde\ell_\beta \tilde m_\beta}(\mathrm{i})^{\tilde\ell_\beta}\mathcal{H}^{\tilde m_1\dots \tilde m_n}_{\ell_1\dots\tilde\ell_n}\int x^2\mathrm{d}x \; j_{\tilde\ell_\beta}(k_\beta x) Y_{\tilde\ell_\beta \tilde m_\beta}(\hat{\boldsymbol{k}}_\beta)  \,.    
\end{split}
\end{equation}
Here we defined the integral over spherical harmonics,
\begin{equation}
    \mathcal{H}^{m_1\dots m_n}_{\ell_1\dots\ell_n} \equiv \int\mathrm{d}\Omega_x \prod_{\alpha = 1}^n Y_{\ell_\alpha m_\alpha}(\hat{\boldsymbol{x}})\,,
\end{equation}
which can be decomposed into integrals over two and three spherical harmonics using 
\begin{equation}
    Y_{\ell_1 m_1}(\hat{\boldsymbol{x}})    Y_{\ell_2 m_2}(\hat{\boldsymbol{x}}) = \sum_{\ell m} c_{\ell m}(\ell_1,\ell_2,m_1,m_2) Y_{\ell m}(\hat{\boldsymbol{x}})\,,
\end{equation}
with the Klebsch-Gordon coefficients
\begin{equation}
    c_{\ell m}(\ell_1,\ell_2,m_1,m_2) = (-1)^m\mathcal{G}^{m_1m_2-m}_{\ell_1\ell_2\ell}
\end{equation}
and the Gaunt integral ${G}^{m_1m_2m_3}_{\ell_1\ell_2\ell_3}$:
\begin{equation}
    \mathcal{G}^{m_1m_2m_3}_{\ell_1\ell_2\ell_3} = \int\mathrm{d}{\Omega}_xY_{\ell_1m_1}(\boldsymbol{\hat{x}})Y_{\ell_2m_2}(\boldsymbol{\hat{x}}) Y_{\ell_3m_3}(\boldsymbol{\hat{x}}) = \sqrt{\frac{(2\ell_1+1)(2\ell_2+1)(2\ell_3+1)}{4\uppi}}
\begin{pmatrix}
    \ell_1 & \ell_2 & \ell_3 \\
    0 & 0  & 0
\end{pmatrix}
\begin{pmatrix}
    \ell_1 & \ell_2 & \ell_3 \\
    m_1 & m_2  & m_3
\end{pmatrix}
    \,. 
\end{equation}
The angular polyspectrum, $\mathcal{P}^{a_1\dots a_n}_{\ell_1\dots\ell_n}$, is then defined via
\begin{equation}
\label{eq:angular_n_point_function}
    \left\langle    a_{1_{\ell_1 m_1}}    \dots a_{n_{\ell_n m_n}}\right\rangle=   \mathcal{H}^{m_1\dots m_n}_{\ell_1\dots\ell_n} \mathcal{P}^{a_1\dots a_n}_{\ell_1\dots\ell_n}\,.
\end{equation}
For $n =2,\;3$ the functional form of the angular polyspectrum is completely fixed by rotational symmetry since the integral over $\Omega_k$ can always be evaluated. 

\section{Flat sky approximation}
\label{sec:flat_sky}
For computational simplicity, one often relies not on a spherical harmonic decomposition but rather a flat two-dimensional discrete Fourier transform by replacing the directional vector $\hat{\boldsymbol{n}}$ by a unit vector in the flat sky ${\boldsymbol{\theta}} \perp \hat{\boldsymbol{e}}_z$. In this flat sky, one can decompose the projected field $a$ into Fourier modes $\tilde a$
\begin{equation}
    a(\boldsymbol{\theta}) = \frac{1}{A_\mathrm{s}}\sum_{\boldsymbol{\ell}} \tilde{a}_{\boldsymbol{\ell}}\;\mathrm{e}^{-\mathrm{i}\boldsymbol{\ell}\cdot\boldsymbol{\theta}}\,,
\end{equation}
where $A_\mathrm{s}$ is the area of the flat sky patch.
The sum runs over all $\boldsymbol{\ell}^\mathrm{T} = 2\uppi/\sqrt{A_\mathrm{s}}(i_x,i_y)$, where $i_x,i_y\in \mathbb{N}$. For $A_\mathrm{s}\to\infty$ we can replace the discrete by a continuous Fourier transformation, such that
\begin{equation}
    a(\boldsymbol{\theta}) = \int\mathrm{d}^2\ell \tilde{\boldsymbol{a}}_{\boldsymbol{\ell}}\;\mathrm{e}^{-\mathrm{i}\boldsymbol{\ell}\cdot\boldsymbol{\theta}}\,.
\end{equation}
The corresponding angular power spectra are defined via
\begin{equation}
\label{eq:flat_sky_power}
    \langle a_{1\boldsymbol{\ell}_1}a_{2\boldsymbol{\ell}_2}\rangle = A_\mathrm{s}\delta^\mathrm{K}_{\boldsymbol{\ell}_1 - \boldsymbol{\ell}_2} \mathcal{P}_{a_1a_2}(\ell_1)\,,\quad \langle a_{1\boldsymbol{\ell}_1}a_{2\boldsymbol{\ell}_2}\rangle = (2\uppi)^2\delta^\mathrm{D}(\boldsymbol{\ell}_1 - \boldsymbol{\ell}_2)\mathcal{P}_{a_1a_2}(\ell_1)
\end{equation}
for the discrete and continuous cases respectively, where $\delta^\mathrm{D}(\boldsymbol{x} - \boldsymbol{y})$ is the Dirac $\delta$-distribution. Higher-order correlators follow analogously.

\section{Halo profiles}
\label{app:halo_profiles}
For an NFW profile \citep{navarro_universal_1997} the Fourier transform assumes the  form
\begin{equation}
\tilde u(k|M,z) = \cos(kr_\mathrm{s})\left[\mathrm{Ci}(k(1+c)r_\mathrm{s}) - \mathrm{Ci}(kr_\mathrm{s})\right] -\frac{\sin(ckr_\mathrm{s})}{kr_\mathrm{s}(1+c)} + \frac{\sin(kr_\mathrm{s})\left(\mathrm{Si}(kr_\mathrm{s}(1+c))-\mathrm{Si}(kr_\mathrm{s})\right)}{\frac{1}{1+c} +\ln(1+c) -1}\,,
\end{equation}
where $\mathrm{Si}(x)$ and $\mathrm{Ci}(x)$ are sine and cosine integrals. The scaling radius
\begin{equation}
r_\mathrm{s} = \frac{r_\mathrm{vir}}{c} = \left(\frac{3M}{4\uppi \Delta_\mathrm{V} \bar\rho_\mathrm{m}c^3}\right)^{1/3}\,
\end{equation}
defines the empirical concentration relation $c(M)$; here $\Delta_\mathrm{V}$ is the virial overdensity. It should be noted that \Cref{eq:halomod_integral} can, in principle, include an integral over the concentration. However, assuming a mean relation between $c$ and $M$, allows for analytic integration.
For the baseline functional form, we assume the relation provided in \citet{duffy_dark_2008}:
\begin{equation}
\label{eq:cmrelation}
    c(M,z) = 10.14\left[\frac{M}{2\times 10^{12}h^{-1}M_\odot}\right]^{-0.81} (1+z)^{-1.01}\,.
\end{equation}

\section{Mixed term reconsidered}
\label{app:mixed_term}
It should be noted that the equations in Sect \ref{sec:real_space_correlation_functions} for the mixed term \Cref{eq:mixed_term} and the non-Gaussian term assume the tracers to be uniformly distributed. This will not be fulfilled when estimating the statistics in a realistic survey with a nontrivial angular mask and depth variations across the survey footprint. \citet{schneider_analysis_2002} have derived the corresponding equations for the shear correlation functions from which we reproduce the mixed term of $\hat{\xi}_+$ for a tomographic setup,
\begin{align}
\label{eq:XipMixedTermGeneral1}
    \mathrm{Cov}_{\mathrm{G,mix}} \big[\hat{\xi}_+^{(a_2a_2)}(\bar{\theta}_1);\hat{\xi}_+^{(a_3a_4)}(\bar{\theta}_2)\big]
    \equiv & \;
    \frac{\sigma^2_{\epsilon_1}}{N_\mathrm{pair}^{(a_1a_2)}(\bar{\theta}_1) \ \ N_\mathrm{pair}^{(a_3a_4)}(\bar{\theta}_2)}
    \ \\&  \times   \;
    \Bigg[
    \sum_{i,j,k}^{N_{a_1},N_{a_2},N_{a_3}} w^2_iw_jw_k \ \xi_+^{(a_2a_3)}(\theta_{jk}) \ \Delta_{\theta_1}(\theta_{ij})\Delta_{\theta_2}(\theta_{ik}) \ \delta^{\mathcal{K}}_{a_1a_4}
    + \ 3 \ \mathrm{perm.}\Bigg] \ 
    ,
\end{align}
where the indices $i,j,k$ stand for the individual galaxies having weights $w_i$, the summation runs over all $N_{a_i}$ galaxies residing in the $a_i$th tomographic redshift bin and $\Delta_{\theta_1}(\theta_{ij})$ denotes the angular bin selection function which is unity iff $\theta_{ij}\equiv |\boldsymbol{\theta}_i-\boldsymbol{\theta}_j|\in \theta$.

For an efficient evaluation of the triplet sums we proceed along the lines of \citet{slepian_computing_2015} and decompose the three-point correlator in its multipoles:
\begin{align}
    M_{++}^{(a_1a_2a_3)}(\bar{\theta}_1,\bar{\theta}_2, \phi) &\equiv
     \sum_{i,j,k}^{N_{a_1}, N_{a_2},N_{a_3}} w_i^2w_jw_k \ \Delta_{\theta_1}(ij) \ \Delta_{\theta_2}(ik) \ \delta^{\mathcal{K}}_{\varphi_{kj}\phi}
     \nonumber \\ &=
     \frac{1}{2\uppi} \sum_{n=-\infty}^{\infty} M_{++,n}^{(a_1a_2a_3)}(\bar{\theta}_1,\bar{\theta}_2) \ \mathrm{e}^{-\mathrm{i} n \phi} \ ,
\end{align}
where $\varphi_{ji}$ denotes the polar angle of $\boldsymbol{\theta}_{ij}$, $\phi\equiv \varphi_{ki}-\varphi_{ji}$ and the multipole components can be calculated as 
\begin{align}
    M_{++,n}^{(a_1a_2a_3)}(\bar{\theta}_1,\bar{\theta}_2)
    &=
    \sum_{i}^{N_{a_1}}w_i^2 \ c^{(a_1a_2)}_{i,n}(\bar{\theta}_1) \ \br{c^{(a_1a_3)}_{i,n}(\bar{\theta}_2)}^* \ 
    \nonumber \ ; \\
    c^{(a_1a_2)}_{i,n}(\bar{\theta}_1) 
    &\equiv
    \sum_{j=1}^{N_{a_2}} w_j \ \mathrm{e}^{in\varphi_{ji}} \ \Delta_{\theta_1}(\theta_{ij}) \ .
\end{align}
To further speed up the computation we make use of the combined estimator method introduced in \citet{porth_roadmap_2023} that amends the equations presented here with grid-based methods for which the $c_{i,n}^{(a_1a_2)}$ can be computed via FFTs.

Once the multipoles are computed, we can perform the numerical integral to obtain
\begin{align}
    T_{++}^{(a_1a_2a_3)}(\bar{\theta}_1, \bar{\theta}_2)
    &\equiv
    \int_0^{2\uppi} \mathrm{d} \phi \ M_{++}^{(a_1a_2a_3)}(\bar{\theta}_1, \bar{\theta}_2,\phi) \ \xi_+^{(a_2a_3)}(\bar{\theta}_3) \ ,
\end{align}
where we evaluate the shear correlation function at $\bar{\theta}_3 = \sqrt{\bar{\theta}_1^2 + \bar{\theta}_2^2 - 2\bar{\theta}_1\bar{\theta}_2\mathrm{cos(\phi)}}$. Plugging everything together we arrive at an approximation of \Cref{eq:XipMixedTermGeneral1} in which $\xi_+$ is not evaluated for every triplet, but instead the mean distance across the bin is taken\footnote{For ensuring a higher accuracy one could also divide the $\theta_i$ into sub-intervals which are then put into the corresponding $(\bar{\theta}_1,\bar{\theta}_2)$ combination.},
\begin{align}
    \label{eq:XipMixedTermGeneral2}
    \mathrm{Cov}_{\mathrm{G,mix}} \left[\hat{\xi}_+^{(a_2a_2)}(\bar{\theta}_1);\hat{\xi}_+^{(a_3a_4)}(\bar{\theta}_2)\right]
    &\approx
    \frac{\sigma^2_{\epsilon_1}}{N_\mathrm{pair}^{(a_1a_2)}(\bar{\theta}_1) \ \ N_\mathrm{pair}^{(a_3a_4)}(\bar{\theta}_2)}
    \ \
    \left[ T_{++}^{(a_1a_2a_3)}(\bar{\theta}_1, \bar{\theta}_2) \ \delta^{\mathcal{K}}_{a_1a_4} + \ 3 \ \mathrm{perm.}\right] \ 
    .
\end{align}
Performing similar computations we can also work out a multipole-based decomposition for the mixed covariance of $\hat{\xi}_-$:
\begin{align}
    \label{eq:XipMixedTermGeneral3}
    \mathrm{Cov}_{\mathrm{G,mix}} \left[\hat{\xi}_-^{(a_2a_2)}(\bar{\theta}_1);\hat{\xi}_-^{(a_3a_4)}(\bar{\theta}_2)\right]
    &\approx
    \frac{\sigma^2_{\epsilon_1}}{N_\mathrm{pair}^{(a_1a_2)}(\bar{\theta}_1) \ \ N_\mathrm{pair}^{(a_3a_4)}(\bar{\theta}_2)}
    \ \
    \left[ T_{--}^{(a_1a_2a_3)}(\bar{\theta}_1, \bar{\theta}_2) \ \delta^{\mathcal{K}}_{a_1a_4} + \ 3 \ \mathrm{perm.}\right] \ 
    ,
\end{align}
where the $T_{--}$ are given as
\begin{align}
    T_{--}^{(a_1a_2a_3)}(\bar{\theta}_1, \bar{\theta}_2)
    &\equiv
    \int_0^{2\uppi} \mathrm{d} \phi \ M_{--}^{(a_1a_2a_3)}(\bar{\theta}_1, \bar{\theta}_2,\phi) \ \xi_+^{(a_2a_3)}(\bar{\theta}_3) \ ; \nonumber \\
    M_{--}^{(a_1a_2a_3)}(\bar{\theta}_1, \bar{\theta}_2,\phi) &\equiv
    \frac{1}{2} \sum_{i=1}^{N_{a_1}}w_i^2\left\{c_{i,4+n}^{(a_1,a_2)}(\bar{\theta}_1)\left[c_{i,4+n}^{(a_1,a_3)}(\bar{\theta}_2)\right]^* + c_{i,n-4}^{(a_1,a_2)}(\bar{\theta}_1)\left[c_{i,n-4}^{(a_1,a_3)}(\bar{\theta}_2)\right]^*\right\} \ .
\end{align}

\section{Bandpowers}
\label{app:bandpowers}
Band powers are angular averages over the angular power spectrum, \Cref{eq:powerspectrum_general}, and can as such make more use of the easier calculations in harmonic space, while including effects of survey mask and finite coverage of scales more accessible. Generally, they are defined as
\begin{equation}
\label{eq:bpdef}
    \mathcal{C}_{(a_1 a_2)}({L}) \coloneqq \frac{1}{\mathcal{N}_L}\int \ell \mathrm{d}\ell \; S_L(\ell) \;\mathcal{P}_{a_1a_2}({\ell})\,,
\end{equation}
where $S_L$ is the band power response function and $\mathcal{N}_L$ is the normalisation
\begin{equation}
\mathcal{N}_L     = \int \frac{\mathrm{d\ell}}{\ell} S_L(\ell)\,,
\end{equation}
such that the band powers trace the angular power $\ell^2 \mathcal{P}^{a_1a_2}_{\ell} $ at the log-centered bins in $L$.
Due to the orthogonality of the Bessel functions, \Cref{eq:hankelshear1} and \Cref{eq:hankelshear2} can be inverted to express the band powers in terms of the real space correlation functions $w$, $ \gamma_\mathrm{t}$ and $\xi_\pm$. Thus, by using \Cref{eq:bpdef}, one finds \citep{schneider_analysis_2002}
\begin{align}
        \label{eq:bpestimate_ideal}
&{\mathcal C}_{\mathrm{n}_i\mathrm{n}_j}({L}) = \; \frac{\uppi}{{\cal N}_L}\; \int  \theta \mathrm{d}\theta\; T(\theta) w^{(ij)}(\theta) g_+^L(\theta)  \,,\\
&{\mathcal C}_{\mathrm{n}_i\epsilon_j}({L}) = \; \frac{2\uppi}{{\cal N}_L}\; \int  \theta \mathrm{d}\theta\; T(\theta)  \gamma^{(ij)}_\mathrm{t}(\theta) h^L(\theta) \,,\\
&{\mathcal C}_{\epsilon_i\epsilon_j{\rm E/B}}({L}) = \; \frac{\uppi}{{\cal N}_L}\; \int  \theta \mathrm{d}\theta\; T(\theta) \left[ \xi_+^{(ij)}(\theta)\; g_+^L(\theta) \pm \xi_-^{(ij)}(\theta)\; g_-^L(\theta)  \right]\,,
\end{align}
for GC, GGL, and CS respectively. Where the E-mode corresponds to the sum and the B-mode to the difference.
With `+' corresponding to ${\rm J}_0$, the kernels are given by
\begin{align}
\label{eq:gplusminus}
g_\pm^L(\theta) = &\; \int  \ell\;\mathrm{d} \ell\, S_L(\ell)\; {\rm J}_{0/4}(\ell \theta)\,, \\
h^L(\theta) = &\; \int  \ell\;\mathrm{d} \ell\, S_L(\ell)\; {\rm J}_{2}(\ell \theta)\,.
\end{align}
If the real-space correlation functions are only accessible over a finite range of angular separations, one apodises the kernels with a Hann window \cite[see][]{joachimi_kids-1000_2021}:
\begin{equation}
    T(\theta) =
    \begin{cases}
    \;\cos^2 \left(\frac{\uppi[x- (x_{\rm lo}+\Delta_x/2)]}{2\Delta_x} \right)  &\quad \text{if} \quad x_{\rm lo} -\frac{\Delta_x}{2} \leq x <  x_{\rm lo} +\frac{\Delta_x}{2} \\ 
    \;1  &\quad \text{if}\quad  x_{\rm lo} +\frac{\Delta_x}{2} \leq x <  x_{\rm up} -\frac{\Delta_x}{2} \\
    \;\cos^2 \left(\frac{\uppi[x-(x_{\rm up} - \Delta_x/2)]}{2\Delta_x} \right)  &\quad \text{if} \quad x_{\rm up} -\frac{\Delta_x}{2}  \leq x <  x_{\rm up} +\frac{\Delta_x}{2}\\ 
    \; 0  &\quad\quad\quad\quad\quad\quad\text{else} 
    \end{cases} \,,
\end{equation}
with $x= \log\theta$ and $\Delta x$ the logarithmic bandwidth.
The apodisation is such that the lower and upper limit is centred on $x_{\rm lo} = \log \theta_{\rm lo}$ and $x_{\rm up} = \log \theta_{\rm up}$, respectively. 

Here we chose the band power response to be a top hat between $\ell_{{\rm lo},l}$ and $\ell_{{\rm up},l}$ leading to a normalisation ${\cal N}_l = \ln \big(\ell_{{\rm up},i}/\ell_{{\rm lo},i}\big)$
and a closed form for $g^L_\pm$ and $h^L$:
\begin{align}
\label{eq:bp_kernel_cosmicshear}
&g_+^L(\theta) = \frac{1}{\theta^2} \left[\theta \ell_{{\rm up},L}\;
  {\rm J}_1(\theta \ell_{{\rm up},L}) - \theta \ell_{{\rm lo},L}\;   {\rm J}_1(\theta \ell_{{\rm lo},L})\right] \,, \\ 
&g_-^L(\theta) = \frac{1}{\theta^2} \left[ {\cal G}_-(\theta \ell_{{\rm
      up},L})  - {\cal G}_-(\theta \ell_{{\rm lo},L}) \right]\,,\\ 
      &h^L  = -\frac{1}{\theta^2}\left[\theta\ell_{{\rm up},L} \mathrm{J}_1(\theta\ell_{{\rm up},L}) - \theta\ell_{{\rm lo},L}\mathrm{J}_1(\theta\ell_{{\rm lo},L}) + 2\mathrm{J}_0(\theta\ell_{{\rm up},L}) - 2\mathrm{J}_0(\theta\ell_{{\rm lo},L})\right]\,,
\end{align}
where
\begin{equation}
{\cal G}_-(x) = \left(x - \frac{8}{x}\right) {\rm J}_1(x) - 8 {\rm J}_2(x)\,.
\end{equation}
Linking the band powers directly to the angular power spectra yields the final expressions
\begin{align}
\label{eq:bp_cosmicshear1}
{\cal C}_{{\mathrm{n}_i\mathrm{n}_j}} (L) &=  \frac{1}{{\cal N}_L} \int\ell\;\mathrm{d}
\ell\,  W^L_{\rm EE}(\ell)\; \mathcal{P}_{\mathrm{n}_i\mathrm{n}_j}(\ell) \,, \\ 
\label{eq:bp_cosmicshear2}
{\cal C}_{{\mathrm{n}_i\epsilon_j}} (L) &=  \frac{1}{{\cal N}_L} \int \ell\;\mathrm{d}
\ell\,  W^L_{\rm nE}(\ell)\; \mathcal{P}{\mathrm{n}_i\epsilon_j}(\ell) \,, \\ 
    \label{eq:bp_cosmicshear3}
{\cal C}_{\epsilon_i\epsilon_j{\rm E}}(L) &=  \frac{1}{{2\cal N}_L} \int \ell\;\mathrm{d}
\ell\,  \left[ W^L_{\rm EE}(\ell)\; \mathcal{P}_{\epsilon_i\epsilon_j,\rm
    E}(\ell) + W^L_{\rm EB}(\ell)\; \mathcal{P}_{\epsilon_i\epsilon_j,\rm B}(\ell) \right] \,, \\ 
    \label{eq:bp_cosmicshear4}
{\cal C}_{\epsilon_i\epsilon_j{\rm B}} (L)&=  \frac{1}{2 {\cal N}_L} \int\ell\;\mathrm{d}
\ell\,  \left[ W^L_{\rm BE}(\ell)\; \mathcal{P}_{\epsilon_i\epsilon_j,\rm
    E}(\ell) + W^L_{\rm BB}(\ell)\; \mathcal{P}_{\epsilon_i\epsilon_j,\rm B}(\ell) \right]\,,
    \end{align}
with kernels given by 
\begin{align}
    \label{eq:kernelshear1}
W^L_{\rm EE}(\ell) &= W^L_{\rm BB}(\ell) = \!\! \int \!\! \theta\;\mathrm{d} \theta\, T(\theta) \left[ {\rm J}_0(\ell \theta)\; g_+^L(\theta) + {\rm J}_4(\ell \theta)\; g_-^L(\theta) \right]\,,\\
\label{eq:kernelshear2}
W^L_{\rm EB}(\ell) &= W^L_{\rm BE}(\ell) = \!\! \int \!\! \theta\;\mathrm{d} \theta\, T(\theta) \left[ {\rm J}_0(\ell \theta)\; g_+^L(\theta) - {\rm J}_4(\ell \theta)\; g_-^L(\theta) \right] \,, \\
\label{eq:kernelshear3}
W^L_{\rm nE}(\ell) & = \int \!\! \theta\;\mathrm{d} \theta\, T(\theta) \;{\rm J}_2(\ell \theta)\; h^L(\theta)\,.
\end{align}

\section{COSEBIs and $\Psi$ statistics}
\label{app:cosebis}
The complete orthogonal sets of E/B-integrals \citep[COSEBIs,][]{schneider_cosebis_2010} are summary statistics for CS, avoiding leakage between E- and B-modes on a finite range of angular scales. These are discrete values and can be directly linked to the two-point correlation functions:
\begin{align}
\label{eq:COSEBIsReal}
 E^{(ij)}_n &= \frac{1}{2} \int_{\theta_{\rm min}}^{\theta_{\rm max}}
 \theta\;\mathrm{d}\theta\: 
 [T_{+n}(\theta)\,\xi^{(ij)}_+(\theta) +
 T_{-n}(\theta)\,\xi^{(ij)}_-(\theta)]\,, \\ 
 B^{(ij)}_n &= \frac{1}{2} \int_{\theta_{\rm min}}^{\theta_{\rm
     max}}\theta\;\mathrm{d}\theta\: 
 [T_{+n}(\theta)\,\xi^{(ij)}_+(\theta) -
 T_{-n}(\theta)\,\xi^{(ij)}_-(\theta)]\,.
 \end{align}
The filter functions $T_{\pm n}(\theta)$ are defined for a given angular range, $\theta\in[\theta_{\rm min}$, $\theta_{\rm max}]$.  Two families of COSEBIs have been used in the past, linear-COSEBIs and log-COSEBIs, defining whether the oscillations of $T_\pm(\theta)$ are linearly or logarithmically spaced respectively. The COSEBIs are labelled by $n$ natural numbers, with filters having $n+1$ roots in their range of support.
Here, we employ log-COSEBIs, but the code can use any filter function.
The theoretical model prediction for COSEBIs is given by
\begin{align}
\label{eq:EnBnFourier}
E^{(ij)}_n &= \int_0^{\infty}
\frac{\ell\;\mathrm{d}\ell}{2\uppi}\mathcal{P}_{\epsilon_i\epsilon_j,\rm E}(\ell)\,W_n(\ell)\,,\\ 
B^{(ij)}_n &= \int_0^{\infty}
\frac{\ell\;\mathrm{d}\ell}{2\uppi}\mathcal{P}_{\epsilon_i\epsilon_j,\rm B}(\ell)\,W_n(\ell)\,,
\end{align} 
where the weight functions, $W_n(\ell)$, are Hankel transforms of $T_\pm(\theta)$  
\begin{align}
\label{eq:Wn}
W_n(\ell) & =  \int_{\theta_{\rm{min}}}^{\theta_{\rm{max}}}\theta\;\mathrm{d}\theta\:
T_{+n}(\theta) \rm{J}_0(\ell\theta)  = \int_{\theta_{\rm{min}}}^{\theta_{\rm{max}}}\theta\;\mathrm{d}
\theta\:T_{-n} (\theta) \rm{J}_4(\ell\theta)\,.
\end{align} 
A very similar set of observables can be constructed for GGL and GC \citep{buddendiek_rcslens_2016, asgari_minimizing_2021} to avoid including physical scales outside $[\theta_\mathrm{min}, \theta_\mathrm{max}]$. Similarly to the COSEBIs, they are defined as
\begin{align}
    \Psi^{\mathrm{n}_i\mathrm{n}_j}_m = &\;  \int_{\theta_{\rm min}}^{\theta_{\rm max}}
 \theta\;\mathrm{d}\theta\;  U_m(\theta) w^{(ij)}(\theta)\\
  \Psi^{\mathrm{n}_i\epsilon_j}_m = &\;  \int_{\theta_{\rm min}}^{\theta_{\rm max}}
 \theta\;\mathrm{d}\theta\; Q_m(\theta) \gamma^{(ij)}_\mathrm{t}(\theta)\,,
\end{align}
so that $U_m(\theta)$ is compensated and orthogonal and $Q_m(\theta)$ given by the expression
\begin{equation}
    Q_m(\theta) = \frac{2}{\theta^2}\int^\theta_0 \theta^\prime\mathrm{d}\theta^\prime U_m(\theta^\prime) - U_m(\theta)\,.
\end{equation}
Their respective Fourier counterparts are calculated in the same manner as the COSEBIs $E$-mode, \Cref{eq:EnBnFourier}, with a Fourier weight:
\begin{equation}
    W^\Psi_m(\ell) = \int_{\theta_{\rm min}}^{\theta_{\rm max}}
 \theta\;\mathrm{d}\theta\; U_m(\theta) \mathrm{J}_0(\ell\theta) =\int_{\theta_{\rm min}}^{\theta_{\rm max}}
 \theta\;\mathrm{d}\theta\; Q_m(\theta) \mathrm{J}_2(\ell\theta)\,.
\end{equation}

\section{Definition of Fourier and real space filters for summary statistics}
\label{app:mapping_to_summary}
As an example, we define the linear maps $\boldsymbol{\tens{W}}_L (\ell)$ and $\boldsymbol{\tens{R}}_L(\theta)$ for the three summary statistics considered in \Cref{sec:harmonic_to_real}. For real space correlation functions, one finds the following:
\begin{equation}
    \boldsymbol{\tens{W}}^{\mathrm{2pcf}}_L(\ell) = 
    \begin{pmatrix}
        {\cal K}_0(\ell L) & 0 & 0 & 0 \\
        0 & {\cal K}_2(\ell L) & 0 & 0 \\
        0 & 0 & {\cal K}_0(\ell L) & {\cal K}_0(\ell L) \\
        0 & 0 & {\cal K}_4(\ell L)  &-{\cal K}_4(\ell L)
    \end{pmatrix}
    \quad\text{and}\quad \boldsymbol{\tens{R}}^{\mathrm{2pcf}}_L(\theta) =\boldsymbol{\mathds{1}}\frac{1}{\theta}\uppi\left(\frac{\theta - L}{\Delta L}\right)\,.
\end{equation}
Here $\uppi$ is a top-hat filter centred around $L$ with bandwidth $\Delta L$.
This includes the averaging over $\theta$ bins in the weights. For bandpowers one finds
\begin{equation}
    \boldsymbol{\tens{W}}^\mathrm{BP}_L(\ell) =
    \frac{\uppi}{\mathcal{N}_L}
    \begin{pmatrix}
        2W^L_\mathrm{EE}(\ell) & 0 & 0 & 0\\
        0 & 2W^L_\mathrm{nE}(\ell) & 0 & 0 \\
        0 & 0 & W^L_\mathrm{EE}(\ell) & W^L_\mathrm{EB}(\ell)\\
        0 & 0& W^L_\mathrm{EB}(\ell) &W^L_\mathrm{BB}(\ell)
    \end{pmatrix}
    \quad\text{and}\quad \boldsymbol{\tens{R}}^{\mathrm{BP}}_L(\theta) =  \frac{\uppi}{\mathcal{N}_L} T(\theta)\begin{pmatrix}
       g^L_+(\theta) & 0 & 0 & 0\\
       0 & 2h^L(\theta) &  0 & 0 \\
       0 &  0 & g^L_+(\theta) & g^L_-(\theta)\\
       0 &  0 & g^L_+(\theta) & -g^L_-(\theta) 
    \end{pmatrix}\,,
\end{equation}
and for COSEBIs:
\begin{equation}
    \boldsymbol{\tens{W}}^\mathrm{COSEBI}_L(\ell) =
    \begin{pmatrix}
        W^\Psi_L(\ell)  & 0 & 0 & 0\\
        0 & W^\Psi_L(\ell)& 0 & 0 \\
        0 & 0 & W_L(\ell) & 0 \\
        0 & 0& 0 &W_L(\ell)
    \end{pmatrix}
    \quad\text{and}\quad \boldsymbol{\tens{R}}^{\mathrm{COSEBI}}_L(\theta) =
    \frac{1}{2}\begin{pmatrix}
       2U_L(\theta) & 0 & 0 & 0\\
       0 & 2Q_L(\theta) &  0 & 0 \\
       0 &  0 & T_{+L}(\theta) & T_{-L}(\theta)\\
       0 &  0 & T_{+L}(\theta) & -T_{-L}(\theta) 
    \end{pmatrix}\,,
\end{equation}
We now label components of $\mathrm{vec}(\boldsymbol{C})(\ell)$ with Latin indices and similar for $\boldsymbol{\tens{W}}_L(\ell)$ and the summary statistic  $\mathrm{vec}(\boldsymbol{\mathcal{O}})(L)$. 
Using \Cref{eq:linearmap_ell_to_obs} and \Cref{eq:Gaussian_covariance}, Gaussian covariances which are not pure shot noise terms can be expressed as
\begin{align}
    \mathrm{Cov}_\mathrm{G, sva}\left[\mathcal{O}^{ij}_{p_1,L_1}\mathcal{O}^{mn\vphantom{j}}_{p_2,L_2}\right] = & \;\frac{1}{2 \uppi\; \mathrm{max}(A_{(ij)}A_{(mn)}) } \sum_{q_1,q_2} \int\ell\mathrm{d}\ell \left(\boldsymbol{\tens{W}}_{L_1}(\ell)\right)_{p_1q_1}\left(\boldsymbol{ \tens{W}}_{L_2}(\ell) \right)_{p_2q_2}\\  & \times  \left[ \mathcal{P}^{(im)}(\ell) \mathcal{P}^{(jn)}(\ell) + \mathcal{P}^{(in)}(\ell) \mathcal{P}^{(jm)}(\ell) \right]_{q_1q_2}
\end{align}
where it is understood that the summary statistic $i$ is the one identified with the tracer combination $a_1, a_2$ and the same of $j$ and $a_3, a_4$. The shot noise term can then be calculated as
\begin{align}
    \mathrm{Cov}_\mathrm{G, sn}\left[\mathcal{O}^{ij}_{p_1,L_1}\mathcal{O}^{mn\vphantom{j}}_{p_2,L_2}\right] = \sum_q N^{ij}_q\int\frac{\theta^2\mathrm{d}\theta}{n^{ij}_q(\theta)} \left(\boldsymbol{\tens{R}}_{L_1}(\theta)\right)_{p_1q}\left(\boldsymbol{\tens{R}}_{L_2}(\theta)\right)_{p_2q} \left[\delta^\mathrm{K}_{im}\delta^\mathrm{K}_{jn} + \delta^\mathrm{K}_{in}\delta^\mathrm{K}_{jm}\right]\,,
\end{align}
where $n^{ij}_q(\theta)$ is the differential pair-count density and $N^{ij}_q$ the noise level in the $q$-th summary statistic and the $i,j$ tomographic bin combination. Lastly, connected terms such as NG and SSC are given by
    \begin{align}
    \mathrm{Cov}_\mathrm{NG, sn}\left[\mathcal{O}^{ij}_{p_1,L_1}\mathcal{O}^{mn\vphantom{j}}_{p_2,L_2}\right] = &\; \frac{1}{4 \uppi^2 \mathrm{max}(A_{(ij)}A_{(mn)}) } \sum_{q_1,q_2} \int \mathrm{d} \ell_1\, \ell_1\;  \int\mathrm{d} \ell_2\, \ell_2\; \left(\boldsymbol{\tens{W}}_{L_1}(\ell_1)\right)_{p_1 q_1 }\left(\boldsymbol{\tens{W}}_{L_2}(\ell_2)\right)_{p_2 q_2}  \\ \nonumber &\times \int_0^\uppi \frac{\mathrm{d} \phi_\ell}{\uppi} \; T^{(ij)(mn)}_{q_1 q_2}(\boldsymbol{\ell}_1,-\boldsymbol{\ell}_1,\boldsymbol{\ell}_2,-\boldsymbol{\ell}_2)\,,  \\
    \mathrm{Cov}_\mathrm{SSC, sn}\left[\mathcal{O}^{ij}_{p_1,L_1}\mathcal{O}^{mn\vphantom{j}}_{p_2,L_2}\right] = &\;\frac{1}{4 \uppi^2} \int\mathrm{d} \ell_1\, \ell_1\;  \int\mathrm{d} \ell_2\, \ell_2\; \left(\boldsymbol{\tens{W}}_{L_1}(\ell_1)\right)_{p_1 q_1 }\left(\boldsymbol{\tens{W}}_{L_2}(\ell_2)\right)_{p_2 q_2} \left(T^{(ij)(mn)}_\mathrm{SSC}({\ell_1,\ell_2})\right)_{q_1 q_2}\,.
    \end{align}

   \end{appendix} 
\end{document}